\documentclass[12pt]{iopart}
\usepackage{bm}
\usepackage{graphicx}
\usepackage{amssymb}
\def\ra{\rightarrow}

\def\be{\begin{equation}}
\def\ee{\end{equation}}
\def\bea{\begin{eqnarray}}
\def\eea{\end{eqnarray}}
\def\jac{x}
\def\jacb{{\bf x}}
\def\hypfi{\varphi}

\def\qb{{\bf q}}
\def\a{\alpha}
\def\r{{\bf r}}
\def\htm{{\hbar^2\over m}}
\def\tri{{{}^3{\rm H}}}
\def\het{{{}^3{\rm He}}}
\def\hel{{{}^3{\rm He}}}
\def\heq{{}^4{\rm He}}

\def\n{\phantom{0}}
\def\e{\phantom{(0)}}
\newcommand{\bfx}{{\bm x}}

\newcommand{\bfk}{{\bm k}}

\begin{document}

\maketitle

\title[A High-Precision Variational Approach $\ldots$]
{A High-Precision Variational Approach to Three- and 
Four-Nucleon Bound and Zero-Energy Scattering States}

\author{A. Kievsky$^1$, S. Rosati$^{2,1}$, 
M. Viviani$^1$, L.E. Marcucci$^{2,1}$, and L. Girlanda$^1$}

\address{$^1$  Istituto Nazionale di Fisica Nucleare, Sezione di Pisa, 
Largo B. Pontecorvo 3, I-56127, Pisa, Italy \\
$^2$ Department of Physics ``E. Fermi'', University of Pisa, 
Largo B. Pontecorvo 3, I-56127, Pisa, Italy \\
 }

\ead{sergio.rosati@df.unipi.it.}

\begin{abstract}

The hyperspherical harmonic (HH) method has been widely
applied in recent times to the study of the bound states, 
using the  Rayleigh-Ritz variational principle,
and of low-energy scattering processes, using
the Kohn variational principle, of 
$A=3$ and 4 nuclear systems. When the wave function of the system is
expanded over a sufficiently large set of HH basis functions,
containing or not correlation factors, 
quite accurate results can be obtained for the observables
of interest. In this paper, the main aspects of the method are
discussed together with its application to
the $A=3$ and 4 nuclear bound and zero-energy scattering states. Results for 
a variety of nucleon-nucleon (NN) and three-nucleon (3N) 
local or non-local interactions are reported. 
In particular, NN and 3N interactions derived in the
framework of the chiral effective field theory and NN potentials from which 
the high momentum components have been removed, as recently presented
in the literature, are considered for the first time within 
the context of the HH method. The purpose of this paper is two-fold.
First, to present a complete description of the HH method for bound and
scattering states, including also detailed formulas for the
computation of the matrix elements of the NN and 3N interactions.
Second, to report 
accurate results for bound  and zero-energy scattering states
obtained with the most commonly used interaction models. These results 
can be useful for comparison with those obtained by other
techniques and are a significant test for different future
approaches to such problems.

\end{abstract}



\section{Introduction}
\label{sec:intro}

The study of nuclear physics started just in the first few years
after the introduction of quantum mechanics.  From the very beginning, 
the non-relativistic Schr\"odinger 
equation was introduced, written as
\begin{equation}
H\Psi=(T+V)\Psi=E\Psi \ .
\label{eq:sch}
\end{equation}
Here $E$ and $\Psi$ are the energy and the wave 
function of the considered nuclear system, and 
$H$ is the Hamiltonian operator, given as the sum of the kinetic ($T$)
and potential ($V$) energy operators.
Correspondingly, two important and difficult problems 
immediately were raised: the first one was the
determination of the nuclear interaction, i.e. the potential energy operator, 
the second was the solution of Eq.~(\ref{eq:sch}) to calculate the 
wave function $\Psi$. These two aspects are reviewed 
in the following sections.

\subsection{The Nuclear Interaction}
\label{subsec:nuclint}

Significant progress on the problem of determining the nuclear 
interaction was made
after the fundamental paper of Yukawa~\cite{yukawa}
and the subsequent
experimental discovery of the pion.
However, the very first field theoretical approaches
focusing on pion-exchanges~\cite{brueckner:a,brueckner:b}
were not successful, due to the non-renormalizability of meson
field theory. As a consequence, semi-phenomenological approaches were
then pursued. They were based on the assumption that 
the nuclear interaction is mainly a nucleon-nucleon (NN) interaction, 
whose long-range part can be described 
as due to a one-pion exchange (OPE). For the 
medium- and shorter-range parts, 
empirical forms were used, with the inclusion, in some cases, 
of a hard-core repulsion~\cite{hamada,reid}.

The discovery of heavy mesons gave rise to other attempts to 
describe the NN interaction as due to a OPE, heavy meson and 
multi-meson exchanges, with the short-range part still treated in a
phenomenological way. Moreover, a vast theoretical effort was made to
calculate the two-pion exchange contribution producing the intermediate
range attraction of the NN force. The Paris~\cite{cottingham}, 
the Nijmegen~\cite{nagels}
and the Bonn~\cite{machleidt:a} models are the most representative ones 
of such effort.

In recent times, accurate calculations 
on few-nucleon systems have clearly indicated that, to perform a
meaningful comparison with the experimental data, the input NN
potential must reproduce, with a $\chi^{2}/\mathrm{datum}$
close to unity, the deuteron binding energy and NN scattering
data up to the pion threshold, namely  
$\mathrm{E}_\mathrm{lab} \lesssim  350\,\mathrm{MeV}$, 
$\mathrm{E}_\mathrm{lab}$ being the NN laboratory kinetic 
energy~\cite{stoks,wiringa:a,machleidt:b}. 
Consequently, a new generation of interaction models, as the so-called  
Nijmegen~\cite{stoks}, Argonne $v_{18}$ (AV18)~\cite{wiringa:a}, 
and charge-dependent Bonn (CDBonn)~\cite{machleidt:b} potentials, 
has been derived. These models
explicitly include charge symmetry breaking
(CSB) terms in the nuclear interaction in order to reproduce
equally well the $np$ and the $pp$ data. Furthermore, they 
can be local, as the first two models, or non-local, as the last one.
However, it has been shown that none of these models 
reproduces the $A>2$ dynamics in a satisfactory way. 
For example, they strongly underpredict the $A>2$ binding energies. 
A possible cure is the inclusion of three-nucleon (3N) 
interaction terms in the potential energy operator. 
The very first model for a 3N interaction was 
proposed by Fujita and Miyazawa~\cite{fujita} 
and was again based on meson theory. In fact, the 3N interaction 
was derived from the exchange of two pions 
between three nucleons, with the intermediate 
excitation of a $\Delta$ resonance. This two-pion-exchange 3N interaction
mechanism is the starting point of the 3N interaction models 
derived by the Urbana group~\cite{wiringa:b,pudliner:a}. 
The Urbana models are written as the sum of the two-pion-exchange 
3N interaction plus a phenomenological repulsive  
term. The strengths of the two contributions are adjusted 
to reproduce the triton binding energy and the nuclear matter equilibrium 
density, in conjunction with one of the Argonne NN potentials. In particular, 
the so-called Urbana IX 3N interaction~\cite{pudliner:a} is often used 
with the Argonne AV18 NN interaction. 
A more recent and sophisticated 3N interaction model~\cite{pieper:a}, still 
derived by the Urbana group, contains two-pion-exchange terms due 
to pion-nucleon scattering in $S$- and $P$-waves, 
three-pion-exchange terms due to ring diagrams with one $\Delta$ resonance
in the intermediate states, and again a phenomenological repulsive term. 
The model has five parameters which are fitted  to the light nuclei 
mass spectrum. This more recent model has a rather complicated 
operatorial structure and is currently object of study.

Another family of models for the 3N interaction is known as the 
Tucson-Melbourne~\cite{tucson} (TM) potential, which arises from 
an off-mass-shell model for the pion-nucleon scattering based upon 
current algebra and a dispersion-theoretical axial vector amplitude 
dominated by the $\Delta$ resonance. The model contains monopole 
form factors, whose cutoff is adjusted to reproduce the triton 
binding energy. More recently, the model has been revisited 
within a chiral symmetry approach~\cite{friar:a}, 
and it has been demonstrated that the contact term present in the TM model
should be dropped. This new TM potential, known as TM$'$, has been 
subsequently readjusted~\cite{tmp}. The final operatorial structure 
coincides with that one
given in the 3N Brazil interaction already derived many years 
ago~\cite{brazil}.

The main difficulty associated with the aforementioned approach 
is the following. The potential models considered 
fit the experimental NN scattering
data up to $\mathrm{E}_\mathrm{lab} \lesssim  350\,\mathrm{MeV}$,
and the $A=3$ nuclei binding energies, 
but they treat each one in its peculiar way the higher momentum contributions.
More in general, the problem is the connection between these
NN plus 3N potentials and the fundamental theory of strong
interactions, namely quantum chromodynamics (QCD). The approach
of Weinberg~\cite{weinberg:a,weinberg:b}, when applied to low-energy processes,
suggests how the problem can be overcome,
at least partially. The starting point
is to consider the most general Lagrangian that incorporates the
assumed symmetry principles, in particular the (broken) chiral
symmetry of QCD, constructed in terms of the pion and nucleon
fields, which are the effective degrees of freedom at low energy,
and their covariant derivatives. The old-fashioned perturbation
theory can be applied to NN scattering, resulting in an infinite
number of Feynman diagrams.  Weinberg has shown~\cite{weinberg:b} that
a systematic expansion of the nuclear potential can be made in
terms of $(Q/\Lambda)^{\nu}$, $Q$ being the pion momentum, 
$\Lambda \approx 1\,\mathrm{GeV}$
the chiral breaking scale and $\nu$ a non-negative integer number. 
For each value of
$\nu$, i.e. the order of the expansion, the number of corresponding
Feynman diagrams is finite and can be calculated.
Such an approach is model independent and is known as
the chiral perturbation theory ($\chi$PT). Following the first
approach of Weinberg, a number of NN potentials
has been obtained by many 
authors~\cite{ordonez,kolck,kaiser,epelbaum:a,epelbaum:b,epelbaum:c,entem} 
increasing the
order of the expansion. For $\nu = 1, 2, \ldots$ the notation NLO,
next-to leading order, N2LO, next-to-next, and so on, is
commonly used. 

It is important to notice that $\chi$PT makes specific
predictions for many-nucleon forces, too. Three-body forces, 
for instance, appear
for $\nu \geq 2$, i.e. firstly at N2LO. This explains why the contributions 
of the 3N interaction to the description of the $A>2$ dynamics
are rather small if compared with the ones arising from the 
NN interaction.

More recently a new class of NN interactions has been
obtained. With the purpose of eliminating from the
semi-phenomenological high precision potentials the strong
high momentum parts, the Hilbert space is separated into
low and  high momentum regions and the renormalization
group method~\cite{rgm} is used to integrate out the 
high-momentum components
above a cutoff $\Lambda$. In this way a model independent
potential $V_{low-k}$ can be 
obtained~\cite{bogner:a,bogner:b,bogner:c,schwenk,bogner:d}.
An important issue of this procedure is what value should be used for 
$\Lambda$. As already discussed in Refs.~\cite{bogner:a,nogga:vlowk}, 
$\Lambda$ should be large enough to include all the relevant 
degrees of freedom. For instance, if $\Lambda<2\,m_\pi$, $m_\pi$ being the 
pion mass, the two-pion exchange part of the bare NN interaction is 
integrated out. As a consequence, the model-dependence of triton 
binding energy is remarkably reduced~\cite{nogga:vlowk}. On the 
other hand, such model-dependence becomes very strong
for $\Lambda>2.0-2.5$ fm$^{-1}$, where the differences between 
the various semi-phenomenological NN potentials 
in their short-distance structure start 
to be important.

\subsection{Calculations on Three- and Four-Nucleon Systems}
\label{subsec:calc}

Different methods can be used to
study few-body bound and scattering states. In nuclear physics
many calculations have been made on systems with $A=3$ and 4
by solving the Schr\"odinger equation using various NN local 
or non-local interactions with and without the inclusion of 3N
forces. For $A=3$, the Faddeev equations~\cite{faddeev} (FE) technique 
has been widely applied both in coordinate and in momentum space. 
A detailed discussion of the method and the results for
bound and scattering states can be 
found in Ref.~\cite{glockle:a}. The Faddeev-Yakubovsky (FY)
equations~\cite{yakubovsky} 
are the generalization of the Faddeev equations to the $A=4$ case. They
can be solved in configuration space~\cite{lazauskas} as well as 
in momentum space~\cite{nogga:a,deltuva:a,deltuva:b}. 

Alternative approaches to the FE and FY equations techniques make use 
of variational principles. Among them we can mention the Rayleigh-Ritz
variational principle used for studying bound states and the Kohn 
variational principle~\cite{kohn}  used in the case of scattering
states. A few of the most important variational approaches for 
studying few-nucleon systems are briefly recalled here.

The quantum Monte Carlo variational methods have been applied in two
different ways. In the first one, known as the variational Monte Carlo,
flexible functions
containing a number of trial parameters are conveniently chosen.
The mean value of the Hamiltonian is calculated by a Monte
Carlo technique and the parameters are varied so as to obtain the
minimum mean value~\cite{wiringa:c}. The second approach 
is the so-called
Green's function Monte Carlo method. It is based on a sophisticated
algorithm  for evaluating appropriate path integrals in order to obtain 
the properties of the bound states of the 
Hamiltonian~\cite{pudliner:b, wiringa:d}. 
Further refinements of the 
algorithm have made it possible to push the calculation up to 
$A=10$~\cite{pieper:b}.

The coupled-rearrangement-channel Gaussian (CRCG) basis proposed by 
Kamimura~\cite{kamika:a} has been applied successfully to various 
systems, in particular to $A=3$ and 4 nuclei~\cite{kamika:b}.
A further development of this approach is the so-called stochastic
variational method (SVM) in which the Gaussian-type basis 
is optimized by means of a stochastic procedure~\cite{suzuki:a,usukura:a}.

The no-core shell model (NCSM) is an approach which has been successfully
applied both to s-shell and p-shell 
nuclei~\cite{navratil:a,navratil:b,navratil:c,caurier,navratil:d}.
The calculation is done with a finite harmonic-oscillator basis.
In the model space spanned by these states,  an $A$-body operator is 
approximated to a two(three)-body effective interaction, 
by applying the Suzuki-Lee method~\cite{suzuki:b}, 
and therefore an 
effective Hamiltonian is constructed. By extending 
the basis, the NCSM calculation should converge to
the exact solution.

The hyperspherical harmonic (HH) method has a rather long history.
In principle the HH basis functions can be used for expanding the wave 
function of a generic $A$-body system~\cite{fabre,avery}. The
problem with such a basis is its quite large degeneracy which 
increases very rapidly with $A$ and makes necessary the inclusion 
of a very high number of basis elements in the expansion. Exceptions
to this general rule arise when the system interacts through very
soft interactions or when the interaction has an hyperradial
character as the harmonic oscillator potential.
However, this is not the case of the strong state-dependent 
nuclear potential
making the application of the HH technique to nuclear systems
a difficult task. One possibility to tackle this problem
is to introduce an effective interaction in the Hilbert
space spanned by the finite basis considered~\cite{barnea:a},
as has been done in the applications of the NCSM mentioned above.

During the last few years, the authors of this paper started
a strong collaboration
in the application of the HH technique to describe bound states of
light nuclear systems and to low-energy scattering processes involving few 
nucleons. In the first approaches, initiated by some of the authors,
the HH functions were multiplied by suitable correlation factors
in order to strongly improve the convergence of the 
basis~\cite{kievsky:a,kievsky:e,viviani:a}. 
Subsequently, thanks to the improvement in the
computational facilities, it has been possible to
consider very large sets of basis functions in order to 
accurately describe all the details of the wave function of the 
system~\cite{kievsky:b,viviani:b}.
A significant test of the accuracy reached in the calculation of
the wave function of the $\alpha$-particle bound state 
using the HH expansion is given in Ref.~\cite{kamada:a}. In that
work, the HH results for the binding
energy and other quantities of interest are compared to those ones 
obtained by the methods quoted above. In the production of this
benchmark, the Argonne $v_8'$ (AV8$'$) 
NN potential~\cite{pudliner:b} has been used. 
Very recently, the HH technique has been 
further developed making it possible to use 
for the first time non-local potentials in the 
description of the $A=3,4$ systems~\cite{viviani:c,viviani:d}.

The HH method is a powerful tool for investigating scattering processes
in the $A=3$~\cite{kievsky:c,kievsky:d,kievsky:f} 
and 4~\cite{viviani:e} systems. The scattering
wave function is expanded in the HH basis taking into
account explicitly its asymptotic part describing
the relative motion of the incident nucleon and the target. The
Kohn variational principle~\cite{kohn} 
is then used to calculate the $S$-matrix of
the reaction from which the observables of interest can be computed.
To be noticed that in the application to scattering states the 
electromagnetic long range potential has to be taken into account
properly. A discussion of this subject can be found in 
Refs.~\cite{kievsky:g,kievsky:h}. Furthermore,
different benchmarks have been done in which the $nd$ and $pd$
phases calculated using the HH expansion are compared to those 
obtained using the FE technique~\cite{kievsky:i,kievsky:j}. As a
further application, the scattering wave functions obtained from the 
HH expansion can be used to describe electroweak few-nucleon reactions. 
It should be remarked
that at small energies, as those ones involved in astrophysical 
processes, the HH technique is the only one presently available that is
capable of yielding precise predictions for a few astrophysical 
reactions~\cite{viviani:f,marcucci:a,park:a,marcucci:b}.

The present paper is organized in the following way: 
in  Sec.~\ref{sec:jaco} the 
Jacobi coordinate systems are introduced, together with the 
hyperspherical variables. In Sec.~\ref{sec:hh} the hyperspherical 
harmonic functions are defined and their properties are 
briefly reviewed. In Sec.~\ref{sec:chh}, the practical implementation 
of the HH expansion for the $A=3$ and 4 bound states is described. 
In particular, the two ``standard'' 
approaches known as HH and correlated-HH (CHH) expansions are discussed.
In Sec.~\ref{sec:scatt}, the HH method devised for the study of 
low-energy scattering 
states is briefly reviewed. In Sec.~\ref{sec:res} a large variety 
of results obtained applying the HH method to the $A=3$ and 4 
bound and scattering states is presented and compared with those
obtained by other techniques and the available experimental data.
Some concluding remarks, with an outlook on the future of this 
technique, are presented in Sec.~\ref{sec:concl}. 
Finally, a few details of the calculations are given 
in the Appendices.

\section{Jacobi Coordinates and Hyperspherical Variables}
\label{sec:jaco}

Let us consider a generic isolated system of $A$ particles 
with spatial coordinates
$\r_i$ and masses $m_i$, $i=1,\ldots ,A$. In the center-of-mass
reference frame, $N=A-1$ vectors are sufficient to specify the spatial
configuration of the system. They can be taken as linear combinations of the
$\r_i$ and there is considerable freedom in choosing the corresponding
coefficients. Let us introduce the Jacobi vectors
$\jacb_i$, $i=1,\ldots ,N$, which, by definition, are such that 
the total kinetic energy operator can be written in the form
\begin{equation}
    T = - \sum_{i=1}^A {\hbar^2\over 2m_i} \nabla^2_{i} =
        -{\hbar^2\over m} \sum_{i=1}^N \nabla^2_{\jacb_i}
        -{\hbar^2\over 2M}  \nabla^2_{\bf X}\ ,
    \label{eq:lapl}
\end{equation}
where $m$ is a reference mass, $M=\sum_{i=1}^A m_i$ is the total mass
of the system, and ${\bf X}= (1/M) \sum_{i=1}^A m_i\r_i$ is the
center-of-mass coordinate.  In general, various choices of the Jacobi
coordinates are possible. One commonly used is the following
\be\label{eq:jac}
    \jacb_{N-j+1} = \sqrt{2 m_{j+1} M_j \over (m_{j+1}+M_j) m}
    \Bigl [\r_{j+1} - {\bf X}_j \Bigr ]\ ,
     \qquad j=1,\ldots,N\ ,
\ee
where
\be\label{eq:cm}
   M_j = \sum_{i=1}^j m_i\ ,\quad
  {\bf X}_j = {1\over M_j} \sum_{i=1}^j m_i \r_i \ ,
\ee
i.e. ${\bf X}_j$ is the coordinate of the center-of-mass of particles
$1,\ldots ,j$. In this paper we will be interested in cases where
all the particles have equal mass $m$ and correspondingly
Eq.~(\ref{eq:jac}) takes the form
\be\label{eq:jac1}
    \jacb_{N-j+1} = \sqrt{2 j \over j+1 }
    \Bigl [\r_{j+1} - {\bf X}_j \Bigr ]\ ,
     \qquad j=1,\ldots,N\ .
\ee
For $A=3$, the above definition of the Jacobi coordinates,
once a permutation of $(1,2,3)$ 
has been fixed, is unique. For $A=4$ two different
choices of  Jacobi vectors exist, specifically
\begin{equation}
\begin{array}{ll@{\qquad}ll}
      {\rm set} \,A & &{\rm set} \,B & \nonumber \\
      \jacb_{1A}=  & \sqrt{3\over2} ({\bf r}_m-
  { {\textstyle {\bf r}_i+{\bf r}_j +{\bf r}_k}\over {\textstyle 3} } )\ , &
     \jacb_{1B}=  & {\bf r}_m- {\bf r}_k\ , \nonumber \\
\noalign{\medskip}
         \jacb_{2A} = & \sqrt{4\over3} ({\bf r}_k-
  { {\textstyle {\bf r}_i+{\bf r}_j} \over {\textstyle 2} } )\ ,  &
  \jacb_{2B} = & \sqrt{2} ({ {\textstyle {\bf r}_m+{\bf r}_k}
  \over {\textstyle 2} }-
   { {\textstyle {\bf r}_i+{\bf r}_j}
   \over {\textstyle 2} })\ ,\label{eq:JcbV}\\
   \noalign{\medskip}
         \jacb_{3A} = & {\bf r}_j-{\bf r}_i\ , &
           \jacb_{3B} = & {\bf r}_j-{\bf r}_i\ , \nonumber
\end{array}
\end{equation}
where $(i,j,k,m)$ is a generic permutation of  $(1,2,3,4)$.
In the following, the wave function of the system will be
expanded into a complete basis. The basis elements can be defined
using set $A$ or set $B$ of the Jacobi coordinates. The completeness
of the basis assures that the expansions using one set or the other
are completely equivalent. In numerical applications the expansion
is truncated and this equivalence does not hold anymore.
Therefore it could be convenient to include basis elements defined
both in set $A$ and set $B$.
As an important example, let us consider the ground state of the
 $\alpha$-particle. The set $A$ of coordinates in
Eq.~(\ref{eq:JcbV}) can be more adequate for constructing 
contributions to the wave function corresponding to a $\{3+1\}$
cluster structure, namely $\hel+n$ or $\tri+p$. Set $B$,
instead, is more suitable for contributions coming from the $\{2+2\}$
configuration, such as a $d+d$ cluster structure. Therefore
the use of an expansion basis constructed using both 
coordinate sets can speed up 
the convergence~\cite{kamika:b,viviani:a}. 

For a given choice of the Jacobi vectors, the hyperspherical coordinates 
are given by the so-called hyperradius $\rho$, 
which results to be independent of the permutation order 
of the particles and is defined as
\begin{equation}\label{eq:rho}
         \rho^2=\sum_{i=1}^N \jac_i^2=
                \frac{2}{A}\sum_{i<j}( {\bf r}_i-{\bf r}_j)^2
               =2\sum_{i=1}^A({\bf r}_i-{\bf X})^2 \ ,
\end{equation}
and by a set $\Omega_N$ of angular variables. In the Zernike
and Brinkman~\cite{fabre,zerni} representation, they  are $2N$ polar
angles $\hat \jac_i\equiv (\theta_i,\phi_i)$  of the Jacobi
vectors ${\bf x}_i$, $i=1,\ldots ,N$, 
and  $N-1$ hyperspherical angles $\hypfi_i$, 
with $0\le\hypfi_i\le\pi/2$, given by the relation
\begin{equation}
   \cos\hypfi_i = { \jac_i \over \sqrt{\jac_1^2+\ldots+\jac_{i}^2}}\ ,
    \qquad i=2,\ldots,N\ ,
     \label{eq:phi}
\end{equation}
where $\jac_i$ is the modulus of the Jacobi vector
$\jacb_i$. Therefore, we have
\be\label{eq:omegaN}
 \Omega_N\equiv\{\hat \jac_1,\ldots,\hat \jac_N,\hypfi_2,\ldots,\hypfi_N\}
 \ .
\ee

\section{Hyperspherical Harmonic Functions}
\label{sec:hh}

The spherical harmonic functions  $Y_{\ell,m}(\theta,\phi)$ are a
very familiar tool in the study of quantum mechanical problems in a
three-dimensional space. It is well known that $r^\ell
Y_{\ell,m}(\theta,\phi)$ is a  harmonic polynomial 
of order  $\ell$ in the Cartesian
components $x,y,z$ of the vector ${\bf r}$, a harmonic polynomial
being defined to be a homogeneous polynomial satisfying the 
three-dimensional Laplace
equation. The spherical harmonic functions form  an irreducible
representation of the SO(3) group.  The hyperspherical harmonic 
functions are the generalization to the case of
$D=3N$ dimensional space. To this end, it is convenient to 
consider a homogeneous polynomial $h_{[G]}$ of
degree $G$ in the Cartesian coordinates of $N$ Jacobi vectors and
then introduce
\begin{equation}\label{eq:hp}
  Y_{[G]} = h_{[G]}/\rho^G\ ,
\end{equation}
where $\rho$ is the hyperradius as given in Eq.~(\ref{eq:rho}). 
Since $h_{[G]}$ is a
homogeneous polynomial,  $h_{[G]}/\rho^G$ does not depend anymore on
$\rho$ and therefore is a function only of the variables $\Omega_N$
specified in Eq.~(\ref{eq:omegaN}). Moreover, the $3N$-dimensional
Laplace operator $\Delta$ can be written in the form
\begin{equation}\label{eq:thh}
  \Delta= \sum_{i=1}^N \nabla^2_{\jacb_i} =
   \biggl ( {\partial^2 \over \partial\rho^2}
  +{3N-1\over \rho} {\partial \over \partial\rho} +{\Lambda_N^2(\Omega_N)
  \over \rho^2}\biggr)\ ,
\end{equation}
where $\Lambda_N^2$ is the $3N$-dimensional generalized angular
momentum operator  depending only on the
hyperangles $\Omega_N$.
By definition $\Delta h_{[G]}=0$, so that
\begin{eqnarray}
  \Delta h_{[G]} &=& \biggl ( {\partial^2 \over \partial\rho^2}
  +{3N-1\over \rho} {\partial \over \partial\rho} +{\Lambda_N^2(\Omega_N)
  \over \rho^2}\biggr) \rho^G Y_{[G]}(\Omega_N)\nonumber \\
  \noalign{\medskip}
   &=& \biggl ( \Lambda_N^2(\Omega_N) + G(G+D-2) \biggr) \rho^{G-2}
       Y_{[G]}(\Omega_N) = 0\ ,  \label{eq:equ1}
\end{eqnarray}
and dividing by $\rho^{G-2}$ we get
\begin{equation}\label{eq:equ2}
   \biggl ( \Lambda_N^2(\Omega_N) + G(G+D-2) \biggr)
       Y_{[G]}(\Omega_{N}) = 0 \ .
\end{equation}
A function satisfying the latter equation is an
eigenfunction of the generalized angular momentum operator and is
known as a hyperspherical harmonic function. The operator
$\Lambda_N^2$ is sometimes called the grand angular momentum operator, and $G$
the grand angular momentum quantum number. $[G]$ stands for a set
of quantum numbers as it will be specified in the following. 
The grand angular momentum operator
can be written in the form ~\cite{fabre}
\begin{eqnarray}
  \Lambda_i^2(\Omega_i)&=&{\partial^2 \over \partial \hypfi^2_i } +
    \biggl[3(i-2) {\rm cotan}\,\hypfi_i +2
    ({\rm cotan}\,\hypfi_i-\tan\hypfi_i)\biggr]
    {\partial \over \partial \hypfi_i }\nonumber\\
    && +{{L_i}^2 \over \cos^2 \hypfi_i}+
       { \Lambda^2_{i-1}(\Omega_{i-1}) \over \sin^2 \hypfi_i}
       \ , \label{eq:thhA}
\end{eqnarray}
where $-L^2_i$ is the angular momentum operator associated with
the $i$-th Jacobi vector. In particular
\begin{equation}\label{eq:thh1}
  \Lambda^2_{1}(\Omega_{1})= L^2_1 \ .
\end{equation}
The solutions of Eq.~(\ref{eq:equ2}) can be constructed by
following a recursive procedure~\cite{zerni}. In the case
$N=1$, namely $D=3$, the eigenfunctions are the spherical
harmonic functions $Y_{\ell_1 m_1} (\hat x_1)$. 

For $N=2$,
the equation to be solved is
\begin{equation}\label{eq:aut1}
  \Lambda_2(\Omega_2) Y_{[G]}(\Omega_2)= - G(G+4) Y_{[G]}(\Omega_2)
  \ ,\quad \Omega_2\equiv \{\hat\jac_1,\hat\jac_2,\hypfi_2\}   \ ,
\end{equation}
with $\Lambda_2(\Omega_2)$ following from Eq.~(\ref{eq:thhA}).
Let us look for a solution of the previous equation of the form
\begin{equation}\label{eq:aut2}
  Y_{[G]}(\Omega_2)= 
F(\cos 2\hypfi_2) (\cos\hypfi_2)^{\ell_2} (\sin\hypfi_2)^{\ell_1}
     Y_{\ell_1 m_1} (\hat x_1) Y_{\ell_2 m_2} (\hat x_2)\ ,
\end{equation}
where $F$ is a function to be determined. In terms of the variable
$z=\cos 2\hypfi_2$, from Eq.~(\ref{eq:aut1}) one gets
\begin{equation}\label{eq:aut3}
  (1-z^2)F''+ (\alpha - \beta z)  F' + \gamma F = 0 \ ,
\end{equation}
where
\begin{eqnarray}
  \alpha&=& \ell_2-\ell_1\ ,\quad
  \beta=\ell_1+\ell_2+3\ ,\nonumber \\
  \gamma&=&\frac{1}{4}\left[G(G+4) -(\ell_1+\ell_2)(\ell_1+\ell_2+4)\right] 
\label{eq:aut3b}
\ .
\end{eqnarray}
Eq.~(\ref{eq:aut3}) is the one satisfied by the Jacobi polynomial
$P^{\ell_1+1/2,\ell_2+1/2}_n(z) $ provided that
$G=2n+\ell_1+\ell_2$~\cite{abra}. Therefore, a solution 
of Eq.~(\ref{eq:aut1}) is given by
\begin{eqnarray}
   Y_{[G]}(\Omega_2)&=& {\cal N}^{\ell_2,\nu_2}_{n}
   (\cos\hypfi_2)^{\ell_2} (\sin\hypfi_2)^{\ell_1}
     Y_{\ell_1 m_1} (\hat \jac_1) Y_{\ell_2 m_2} (\hat \jac_2) \nonumber \\
   &\times&  P^{\ell_1+1/2,\ell_2+1/2}_n(\cos 2\hypfi_2)\ ,
\label{eq:aut5}
\end{eqnarray}
where ${\cal N}^{\ell_2,\nu_2}_{n}$ is a normalization factor which
will be specified later  and $\nu_2=2n+\ell_1+\ell_2+2$. 
It is easy to verify that $\rho^G Y_{[G]}(\Omega_2)$ 
is a homogeneous polynomial of
order $G$ in the Cartesian components of $\jacb_1, \jacb_2$. In fact, 
for $N=2$, one has $\jac_1=\rho \sin\hypfi_2$ and $\jac_2=\rho \cos
\hypfi_2$, and therefore
\begin{eqnarray}\label{eq:hop}
  \rho^G Y_{[G]}(\Omega_2) &=& {\cal N}^{\ell_2,\nu_2}_{n}
  \rho^{2n}
    \jac_1^{\ell_1} Y_{\ell_1 m_1}(\hat \jac_1)\;
   \jac_2^{\ell_2} Y_{\ell_2 m_2} (\hat \jac_2)\; \nonumber \\
&\times&
   P^{\ell_1+1/2,\ell_2+1/2}_n(\cos 2\hypfi_2)\ .
\end{eqnarray}
Since $P^{\ell_1+1/2,\ell_2+1/2}_n$ is a polynomial of degree $n$ 
in the variable $\cos 2\hypfi_2$  and
\begin{equation}\label{eq:hop2}
  \rho^2 \cos 2\hypfi_2 = 2\jac_2^2-\rho^2= \jac_2^2-\jac_1^2\ ,
\end{equation}
one has
\begin{eqnarray}\label{eq:hop3}
& \rho^{2n}& P^{\ell_1+1/2,\ell_2+1/2}_n(\cos 2\hypfi_2)=
   \rho^{2n} \sum_{m=0}^{n} a_m (\cos 2\hypfi_2)^m  \nonumber\\
   &=&  \sum_{m=0}^{n} a_m (\rho^2 \cos 2\hypfi_2)^m \rho^{2(n-m)}
         = \sum_{m=0}^{n} a_m (\jac_2^2-\jac_1^2)^m
   (\jac_2^2+\jac_1^2)^{n-m} \ ,
\end{eqnarray}
which is a homogeneous polynomial of degree $2n$. 
Moreover the two terms
in Eq.~(\ref{eq:hop}),
$\jac_1^{\ell_1} Y_{\ell_1 m_1}(\hat \jac_1)$ and
$\jac_2^{\ell_2} Y_{\ell_2 m_2}(\hat \jac_2)$, are homogeneous
polynomials of degree $\ell_1$ and $\ell_2$ respectively, so
that in
conclusion,  $\rho^G Y_{[G]}(\Omega_2)$ is a homogeneous polynomial of degree
$G=2n+\ell_1+\ell_2$. In this case the symbol
 $[G]$ stands for the set of quantum
numbers $\ell_1,m_1,\ell_2,m_2$ and $n$.

From the expression of the grand angular momentum operator 
$\Lambda^2_i$ given in Eq.~(\ref{eq:thhA}), it is possible to determine
the HH functions for a given $N$ in terms of those found
for the case $N-1$. Let us consider the function
$Y_{[G_{N-1}]}(\Omega_{N-1})$, satisfying the equation
\begin{equation}\label{eq:aut6}
  \Lambda^2_{N-1}(\Omega_{N-1}) Y_{[G_{N-1}]}(\Omega_{N-1}) =
  -G_{N-1}(G_{N-1}+D-5) Y_{[G_{N-1}]}(\Omega_{N-1})\ ,
\end{equation}
namely the eigenfunction of $\Lambda^2_{N-1}$ whose  grand angular momentum 
quantum number is
$G_{N-1}$, and look for the eigenfunction of $\Lambda^2_N$ of the
form
\begin{eqnarray}
\label{eq:aut7}
  Y_{[G]}(\Omega_N)&=& (\cos\hypfi_N)^{\ell_N} (\sin\hypfi_N)^{G_{N-1}}
\nonumber \\
    &\times& Y_{[G_{N-1}]}(\Omega_{N-1}) Y_{\ell_N m_N} (\hat \jac_N)
     F(\cos 2\hypfi_N)\ .
\end{eqnarray}
By inserting this expression for $Y_{[G]}$ in the corresponding 
eigenvalue equation and taking into account Eq.~(\ref{eq:aut6}), 
the following solution is found
\begin{equation}\label{eq:aut8}
  F(\cos 2\hypfi_N)= {\cal N}^{\ell_N,\nu_N}_{n_N}
  P^{\nu_{N-1},\ell_N+1/2}_{n_N}(\cos 2\hypfi_N)\ ,
\end{equation}
with
\begin{eqnarray}
  G&=&2n_N+\ell_N+G_{N-1}\ ,\quad
 \nu_{N-1}=G_{N-1}+{3(N-1)\over 2}-1\ ,\nonumber\\
 \nu_{N}&=&G+{3N\over 2}-1\ .
\label{eq:aut9}
\end{eqnarray}
The complete expression of the HH function can therefore be cast in
the form~\cite{fabre}
\begin{equation}
    Y_{[G]}(\Omega_N) = \left [
    \prod_{j=1}^N Y_{\ell_j,m_j}(\hat \jac_j) \right ]\times
      \left [ \prod_{j=2}^N
     {}^{(j)}{\cal P}^{G_{j-1},\ell_j}_{n_j}(\hypfi_j)
    \right ]
     \ ,
    \label{eq:hh}
\end{equation}
where
\begin{equation}\label{eq:poly}
    {}^{(j)}{\cal P}^{G_{j-1},\ell_j}_{n_j}(\hypfi_j)=
    {\cal N}^{\ell_j,\nu_j}_{n_j}
    (\cos\hypfi_j)^{\ell_j} (\sin\hypfi_j)^{G_{j-1}}
     P^{\nu_{j-1},\ell_j+{1\over2}}_{n_j}(\cos 2\hypfi_j)
     \ ,
\end{equation}
and the quantum numbers  $G_j$ and $\nu_j$ are defined to be
\begin{equation}
    G_j= \sum_{i=1}^j (\ell_i+2 n_i)\ , \quad n_1\equiv 0\ , \quad
    G\equiv G_N\ , \quad \nu_j=G_j+{3j\over 2}-1\ .
     \label{eq:go}
\end{equation}
The symbol $[G]$ distinguishes between different HH functions
having the same $G$ value. There are $3N-1$ quantum numbers which
specify a HH function: the $2N$ quantum numbers $\ell,m$
associated with  the spherical harmonic functions, and the $N-1$
quantum numbers $n$ associated with the hyperspherical polynomials.
In summary, $[G]$ stands for the following set of quantum numbers
\begin{equation}\label{eq:qn1}
   [G]\equiv \{ \ell_1,\ldots,\ell_N,\ m_1,\ldots,m_N, \
   n_2,\ldots,n_N\}\ .
\end{equation}
The normalization factors ${\cal N}$ are chosen so as to 
verify the orthonormality condition
\begin{equation}\label{eq:nohh}
  \int d\Omega_N \Bigl ( Y_{[G]}(\Omega_N)\Bigr)^*
  Y_{[G']}(\Omega_N)= \delta_{[G],[G']}\ ,
\end{equation}
where~\cite{fabre}
\begin{equation}\label{eq:domega}
  d\Omega_N=\sin\theta_1 d\theta_1d\phi_1 \prod_{j=2}^{N}
   \sin\theta_j d\theta_jd\phi_j (\cos\hypfi_j)^2
   (\sin\hypfi_j)^{3j-4} d\hypfi_j\ ,
\end{equation}
is the surface element on the hypersphere of unit hyperradius.
Their explicit expression is
\begin{equation}\label{eq:nhh}
  {\cal N}^{\ell_j,\nu_j}_{n_j}=
  \biggl[{2\nu_j\Gamma (\nu_j-n_j)n_j!\over
  \Gamma (\nu_j-n_j-\ell_j-{1\over 2})
   \Gamma (n_j+\ell_j+{3\over 2})}\biggr]^{1/2}\ .
\end{equation}
Again, by using the definition of the hyperspherical angles in
Eq.~(\ref{eq:phi}), it can be easily shown that the functions 
$\rho^G Y_{[G]}(\Omega_N)$
are homogeneous polynomials. In fact,
\begin{eqnarray}
   \rho^G Y_{[G]}(\Omega_N) &\propto& \left [
    \prod_{j=1}^N \jac_j^{\ell_j}
     Y_{\ell_j,m_j}(\hat \jac_j) \right ] \nonumber \\
&\times&
      \left [ \prod_{j=2}^N \rho^{2n_j}\;
     (\sin\hypfi_{j+1}\cdots\sin\hypfi_N)^{2n_j}{P}_{n_j}(\cos2\hypfi_j)
    \right ]
     \ .
\label{eq:hop4}
\end{eqnarray}
The first factor is a homogeneous polynomial of degree
$\ell_1+\cdots+\ell_N$. Moreover,
\begin{equation}\label{eq:hop5}
  \rho^2\cos2\hypfi_j =
   { \jac_j^2-(\jac_1^2+\cdots+\jac_{j-1}^2)
    \over (\sin\hypfi_{j+1}\cdots\sin\hypfi_N)^2}
   \ ,
\end{equation}
and
\begin{equation}\label{eq:hop6}
   ( \rho\sin\hypfi_{j+1}\cdots\sin\hypfi_N)^{2} =
    \jac_1^2+\cdots+\jac_{j}^2\ .
\end{equation}
Therefore
\begin{eqnarray}\label{eq:hop7}
 \lefteqn{(\rho\sin\hypfi_{j+1}\cdots\sin\hypfi_N)^{2n_j}
     {P}_{n_j}(\cos2\hypfi_j)}\quad\qquad&& \nonumber \\
   &=&  \rho^{2n_j}\;
     (\sin\hypfi_{j+1}\cdots\sin\hypfi_N)^{2n_j}
      \sum_{m=0}^{n_j} a_m (\cos 2\hypfi_j)^m  \nonumber\\
   &=&  \sum_{m=0}^{n_j} a_m
   \biggl(\jac_j^2-(\jac_1^2+\cdots+\jac_{j-1}^2)\biggr)^m
   \biggl(\jac_1^2+\cdots+\jac_{j}^2\biggr)^{n_j-m} \ ,
\end{eqnarray}
is a homogeneous polynomial of order $2n_j$.

The HH functions have several important properties. A few
of them will be reported here without proof. 
\begin{itemize}
\item 
The expansion
of a plane wave in the $3N$ dimensional space is 
given by ~\cite{avery}
\begin{equation}\label{eq:pw1}
  e^{i\sum_{i=1,N} \qb_i\cdot\jacb_i} =
  {  (2\pi)^{D/2} \over (Q\rho)^{D/2-1} }
   \sum_{[G]} i^G Y^*_{[G]}(\Omega^q_N)
  Y_{[G]}(\Omega_N) J_{G+{D\over 2}-1}(Q\rho)\ ,
\end{equation}
where $Q$ and $\Omega^q_N$ are the hyperspherical coordinates
associated with the $N$  Jacobi conjugate momenta $\qb_i$, and $J_\nu(Q\rho)$
are Bessel functions of the first kind. The sum is taken
over all the grand angular momentum quantum numbers $G$ and over the
corresponding 
degeneracy, namely the number of different HH functions for each $G$.

\item
By introducing the hypercoordinates  $\rho',\Omega'_N$  associated
with the $N$ vectors $\jacb_i'$ and using the plane wave expansion
given in Eq.~(\ref{eq:pw1}), the following relation is found
\begin{equation}\label{eq:pw5}
  \sum_{[G]}  Y^*_{[G]}(\Omega'_N)
  Y_{[G]}(\Omega_N) {\delta(\rho-\rho')\over \rho^{D-1}}
  = \prod_{i=1}^{N} \delta^3(\jacb_i-\jacb_i')\ .
\end{equation}
With the definition
\begin{equation}\label{eq:pw6}
  \delta^{D-1}(\Omega_N-\Omega'_N) = \prod_{i=1}^{N}
  \delta^2(\hat\jac_i-\hat\jac_i')
  \prod_{i=2}^{N} { \delta(\hypfi_i-\hypfi_i') \over
                (\cos\hypfi_i)^2   (\sin\hypfi_i)^{3i-4}}  \ ,
\end{equation}
we arrive at the relation
\begin{equation}\label{eq:pw7}
  \sum_{[G]}  Y^*_{[G]}(\Omega'_N)
  Y_{[G]}(\Omega_N)= \delta^{D-1}(\Omega_N-\Omega'_N)\ ,
\end{equation}
which assures the completeness of the HH basis. As a consequence, every
``regular'' function $f(\Omega_N)$ can be expanded in terms of the HH
functions. In fact
\begin{equation}\label{eq:pw8}
  f(\Omega_N) = \int d\Omega'_N\; \delta^{D-1}(\Omega_N-\Omega'_N) f(\Omega'_N)
  = \sum_{[G]}  a_{[G]}  Y_{[G]}(\Omega_N)\ ,
\end{equation}
where
\begin{equation}\label{eq:pw9}
  a_{[G]} = \int d\Omega'_N \; Y^*_{[G]}(\Omega'_N)f(\Omega'_N)
  \ .
\end{equation}

\item
The $3N$-dimensional Fourier transform of a function
$f(\jacb_1,\ldots,\jacb_N)$ is given by
\begin{eqnarray}
  \tilde f(\qb_1,\ldots,\qb_N) = {1\over (2\pi)^{D/2} }
  &\int& d^3\jacb_1\cdots d^3\jacb_N\; f(\jacb_1,\ldots,\jacb_N)\nonumber \\
  &\times& e^{i\sum_{i=1,N} \qb_i\cdot\jacb_i}\ ,
\label{eq:pw2}
\end{eqnarray}
and the inverse transform is
\begin{eqnarray}
  f(\jacb_1,\ldots,\jacb_N) = {1\over (2\pi)^{D/2} }
  &\int& d^3 \qb_1\cdots d^3 \qb_N\; \tilde f(\qb_1,\ldots,\qb_N) \nonumber \\
  &\times&e^{-i\sum_{i=1,N} \qb_i\cdot\jacb_i}\ .
\label{eq:pw3}
\end{eqnarray}
Note that
\begin{equation}\label{eq:pw4}
{1\over (2\pi)^{D} }
  \int d^3 \qb_1\cdots d^3 \qb_N\;
   e^{i\sum_{i=1}^{N} \qb_i\cdot(\jacb_i-\jacb_i')}
  = \prod_{i=1}^{N} \delta^3(\jacb_i-\jacb_i')\ .
\end{equation}

\item
From the previous discussion it follows that
the  spatial wave function of an $A$-body system
can be written in terms of the $N$ Jacobi vectors and in turn
expanded in the HH basis
\begin{equation}\label{eq:pw10}
  \Psi(\jacb_1,\ldots,\jacb_N) =
    \sum_{[G]}  u_{[G]}(\rho)
   Y_{[G]}(\Omega_N)\ ,
\end{equation}
where the expansion coefficients are now functions of the
hyperradius. The same expansion can be expressed as well in 
terms of the momentum space coordinates: 
\begin{equation}\label{eq:pw10m}
 \tilde{\Psi}(\qb_1,\ldots,\qb_N) =
    \sum_{[G]}  g_{[G]}(Q)
   Y_{[G]}(\Omega^q_N)\ .
\end{equation}
The two expansions are equivalent provided that
\begin{equation}
   g_{[G]}(Q)=(-i)^G\,\int_0^\infty d\rho\,
   \frac{\rho^{D-1}}{(Q\rho)^{D/2-1}}\,
   J_{G+ {D\over 2} -1}(Q\rho)\, u_{[G]}(\rho) \ .
\label{eq:vg}
\end{equation}
\end{itemize}

In this paper, HH functions with definite total angular momentum
$L$, $M$ are used. They are constructed 
(in configuration-space as an example) by the following coupling
scheme,
\begin{eqnarray}
  {}^{(A)}{\cal H}_{\{G\},LM}(\Omega_N) &=& \sum_{m_1,\ldots,m_N}
  (\ell_1 m_1 \ell_2 m_2 | L_2 M_2) (L_2 M_2 \ell_3 m_3 | L_3 M_3)
   \times \nonumber \\
  && \cdots \times (L_{N-1} M_{N-1} \ell_N m_N | LM)
    Y_{[G]}(\Omega_N)\ ,
   \label{eq:hh2}
\end{eqnarray}
where $Y_{[G]}(\Omega_N)$ is defined by Eq.~(\ref{eq:hh}), $(\ell_1
m_1 \ell_2 m_2 | L_2 M_2)$ and so on are Clebsch-Gordan coefficients
and $M_i=\sum_{j=1,i} m_j$. The symbol $\{G\}$ of the function
${}^{(A)}{\cal H}_{\{G\},LM}(\Omega_N)$ now stands for  the
following set of quantum numbers
\begin{equation}\label{eq:qn2}
   \{G\}\equiv \{ \ell_1,\ldots,\ell_N,\ L_2,\ldots,L_{N-1}, \
   n_2,\ldots,n_N\}\ .
\end{equation}
The explicit expression of a HH function for $A=3$ is
\begin{equation}\label{eq:hh3}
  {}^{(3)}{\cal H}_{\{\ell_1\ell_2 n_2\},LM}(\Omega_2)=
   \bigl[ Y_{\ell_1}(\hat \jac_1)
        Y_{\ell_2}(\hat \jac_2) \bigr  ]_{LM}
        {}^{(2)}{\cal P}^{\ell_1,\ell_2}_{n_2}(\hypfi_2)
        \ ,
\end{equation}
whereas for $A=4$,
\begin{eqnarray}\label{eq:hh4}
    {}^{(4)}{\cal H}_{\{\ell_1\ell_2 \ell_3 L_2 n_2n_3\},LM}(\Omega_3)&=&
  \Bigl [\bigl[ Y_{\ell_{1}}(\hat \jac_{1})
                Y_{\ell_{2}}(\hat \jac_{2})\bigr]_{L_{2}}
                Y_{\ell_{3}}(\hat \jac_{3}) \Bigr ]_{LM} \nonumber \\
    && \qquad \times
    {}^{(2)}{\cal P}^{\ell_1,\ell_2}_{n_2}(\hypfi_2)\;
     {}^{(3)}{\cal P}^{2n_2+\ell_1+\ell_2,\ell_3}_{n_3}(\hypfi_3)
     \ .
\end{eqnarray}

\section{HH and CHH Expansions}
\label{sec:chh}

In this section, a brief description of the HH expansion with or 
without the inclusion of correlation factors is presented. 
In the first subsection, we deal with the
``standard'' HH expansion for the systems of $A=3$ and $4$ nucleons.
The transformation coefficients relating HH functions given in different 
sets of Jacobi coordinates are introduced in Sec.~\ref{sec:tc}.
It is well known that for realistic NN interactions the HH expansion is
slowly convergent. An extensively used method to overcome such an
unpleasant behavior is presented in Sec.~\ref{sec:chhex},
where the correlated HH expansion is introduced. The calculation of
the correlation functions is discussed in~\ref{sec:corref}. In
Sec.~\ref{sec:deta}, some technical details of the practical
applications of these  expansions are reported.

\subsection{The HH Expansion}
\label{sec:hhex}
Let us consider first the  bound state of a trinucleon system
with total angular momentum $J,J_z$, and parity $\pi$.
The complete antisymmetrization  of the state is
guaranteed by writing the wave function as a sum of three Faddeev-like
amplitudes
\begin{equation}
     \Psi_3 = \sum_{p=1}^3 \psi(\jacb_1^{(p)},\jacb_2^{(p)})\ ,
      \label{eq:psit}
\end{equation}
where
\be\label{eq:jco3}
     \jacb_2^{(p)}= {\bf r}_i- {\bf r}_j\ ,\quad
      \jacb_1^{(p)}= \sqrt{4\over 3}   \left( {\bf r}_k-
      {{\bf r}_i+ {\bf r}_j\over 2}\right)\ ,
\ee
are the Jacobi vectors corresponding to the
$p$-th even permutation of the three particles
(corresponding hereafter to the ordering $i$, $j$, $k$ of the particles).
The generic amplitude $\psi(\jacb_1^{(p)},\jacb_2^{(p)})$ is, by
construction, antisymmetric with respect to the exchange of 
particles $i$ and $j$.
Each $\psi(\jacb_1^{(p)},\jacb_2^{(p)})$ can be expanded in HH
functions as
\be\label{eq:psit1}
     \psi(\jacb_1^{(p)},\jacb_2^{(p)})=\sum_{\alpha=1}^{N_c}
     \sum_{n=N^0_\a}^{N_\a}\sum_{l=0}^M\; c_{\a,n,l}\; f_{l}(\rho)\;
      B^{\rm HH}_{n\a}(p) \ ,
\ee
where
$f_l(\rho)$ is a complete set of functions for the variable
$\rho$ whose choice will be discussed in Sec.~\ref{sec:deta}, and
\be\label{eq:B3hh}
     B^{\rm HH}_{n\a}(p)= \Bigl\{
     {}^{(3)}{\cal H}_{\{\ell_{1\a}\ell_{2\a} n\},L_\a}(\Omega_{2}^{(p)})
      \; \bigl [ S_{a\alpha} s_k \bigr ]_{S_\alpha}
      \Bigr \}_{J J_z} \; \bigl [ T_{a\alpha} t_k \bigr ]_{T_\alpha
      T_z}\ .
\end{equation}
Here ${}^{(3)}{\cal H}$ is the HH function defined in
Eq.~(\ref{eq:hh3}) with 
\be\label{eq:omega3}
  \Omega_2^{(p)}\equiv \{ \hat{\jac}_1^{(p)},\hat{\jac}_2^{(p)},
    \hypfi_2^{(p)}\}\ , \qquad \cos\hypfi_2^{(p)}={\jac_2^{(p)}
  \over \rho}\ ,
\ee
and $s_k$ ($t_k$) denotes the spin (isospin)
state of particle $k$, whereas $S_{a\alpha}$ ($T_{a\alpha}$)
the spin (isospin) of the $i,j$ pair. Moreover,  the
total spin $S_\alpha$ is coupled with $L_\alpha$ to give the total
 angular momentum $JJ_z$. The set of quantum numbers
\be\label{eq:cha3}
   \a\equiv\{ \ell_{1\a},
   \ell_{2\a}, L_\alpha, S_{a\alpha}, S_\alpha,
   T_{a\alpha},T_\alpha\} \ ,
\ee
is usually denoted as ``channel". The sum over
$\alpha$ is truncated after the inclusion of the $N_c$ most
important channels. Note that we allow for the possibility to have
states of different total isospin $T_\alpha$ ($=1/2$ or $3/2$) in our expansion.
The quantum numbers of any channel 
must satisfy two requirements: in order to ensure the antisymmetry of the 
wave function, 
the  amplitudes $\psi$ have to change sign under the exchange of
the particles  $i$ and $j$, and therefore  the number
$\ell_{2\alpha}+S_{a\alpha}+ T_{a\alpha}$ must be  odd. In
addition, $\ell_{1\alpha}+\ell_{2\alpha}$ must be an even or odd number in
correspondence to the positive or negative parity of the considered
state. The quantum numbers of the most important channels are
given in Table~\ref{tab:chan3}.

\begin{table}[t]
\caption[Table]{\label{tab:chan3}
Quantum numbers and $N^0_\alpha$ values  for the first $14$
channels considered in the expansion of the 
even parity and $J=1/2$ wave function of a
three-nucleon system. }
\begin{indented}
\lineup
\item[]
\begin{tabular}{ccccccccc}
\br
    $\alpha$ &$\ell_{1\alpha}$&$\ell_{2\alpha}$&$L_\alpha$&
      $S_{a\alpha}$ & $S_\alpha$ &$T_{a\alpha}$& $T_\alpha$&
      $N^0_\alpha$  \\
\mr
  1  & 0 & 0 & 0 &   1 &  {1/ 2} & 0 & {1/2} &  0     \\
  2  & 0 & 0 & 0 &   0 &  {1/ 2} & 1 & {1/2} &  2     \\
  3  & 0 & 2 & 2 &   1 &  {3/ 2} & 0 & {1/2} &  0     \\
  4  & 2 & 0 & 2 &   1 &  {3/ 2} & 0 & {1/2} &  1     \\
  5  & 1 & 1 & 0 &   1 &  {1/ 2} & 1 & {1/2} &  2     \\
  6  & 1 & 1 & 1 &   1 &  {1/ 2} & 1 & {1/2} &  0     \\
  7  & 1 & 1 & 1 &   1 &  {3/ 2} & 1 & {1/2} &  1     \\
  8  & 1 & 1 & 2 &   1 &  {3/ 2} & 1 & {1/2} &  2     \\
  9  & 1 & 1 & 0 &   0 &  {1/ 2} & 0 & {1/2} &  4     \\
  10 & 1 & 1 & 1 &   0 &  {1/ 2} & 0 & {1/2} &  2     \\
  11 & 2 & 2 & 0 &   1 &  {1/ 2} & 0 & {1/2} &  4     \\
  12 & 2 & 2 & 2 &   1 &  {3/ 2} & 0 & {1/2} &  2     \\
  13 & 2 & 2 & 1 &   1 &  {1/ 2} & 0 & {1/2} &  2     \\
  14 & 2 & 2 & 1 &   1 &  {3/ 2} & 0 & {1/2} &  3     \\
\br
\end{tabular}
\end{indented}
\end{table}

An important property of the basis elements is the following: due
to the sum over the permutations in Eq.~(\ref{eq:psit}), some states
\begin{equation}
{\cal B}^{\rm HH}_{n\a}=\sum_{p=1}^3 B^{\rm HH}_{n\a}(p)
\end{equation}
are linearly dependent on others.  Such states can be removed from
the expansion basis by choosing appropriate values for the
quantities $N^0_\alpha$ in Eq.~(\ref{eq:psit1}). As a matter of
fact, the minimum $N^0_\alpha$ to be considered depends on the set of
channels and the HH functions included in the expansion of the 
wave function. 
For example, the states ${\cal B}^{\rm HH}_{n\a}$  of the
first two channels reported in Table~\ref{tab:chan3} with $n=0,1$
are identical. This means that, if the first two states of the
channel $\alpha=1$ are included in the calculation, we must use
$N^0_2=2$. A similar analysis has been made for all the channels
and the minimum values $N^0_\alpha$ needed to avoid linearly
dependent elements have been reported in Table~\ref{tab:chan3}. The
number of HH functions included in each channel is simply given by
$M_\a=N_\alpha-N^0_\alpha+1$.

As a result of a number of investigations on the
problem~\cite{kievsky:b,efros,demin}, it is well known
that the components of $\psi(\jacb_1^{(p)},\jacb_2^{(p)})$ with definite
values of the angular
momenta $\ell_{1\a}$, $\ell_{2\a}$ give contributions rapidly
decreasing when these values increase. As a matter of fact, the
components with $\ell_{1\a} + \ell_{2\a} > 6 $ give nearly
negligible contributions in the case of all the available two-body
NN interactions. On the contrary, very high values of the order $n$
of the Jacobi polynomials must be considered in
Eq.~(\ref{eq:psit1}).

We now generalize the HH approach to the four-body system.
The wave function of a $A=4$ system having total angular
momentum $J$, $J_z$, and parity $\pi$ can
be written as
\be\label{eq:psiq}
      { \Psi}_4 =\sum_{p=1}^{12} \Bigl[
          \psi_A(\jacb_{1A}^{(p)}, \jacb_{2A}^{(p)},
           \jacb_{3A}^{(p)})+
          \psi_B(\jacb_{1B}^{(p)}, \jacb_{2B}^{(p)},
          \jacb_{3B}^{(p)})\Bigr]\ ,
\ee
where the sum is taken over the twelve even permutations $p$ of the
particles. In the remainder,  we suppose that the permutation $p$
corresponds to the order  $i$, $j$, $k$, $m$ of the particles.
The vectors $\jacb_{1A}^{(p)}$, $\jacb_{2A}^{(p)}$,
$\jacb_{3A}^{(p)}$ and  $\jacb_{1B}^{(p)}$, $\jacb_{2B}^{(p)}$,
$\jacb_{3B}^{(p)}$  are the two
possible sets of Jacobi vectors defined in Eq.~(\ref{eq:JcbV}). Each
amplitude $\psi$ is then expanded in terms of the HH functions defined in
Eq.(\ref{eq:hh4}):
\be\label{eq:psiq1}
     \psi_X(\jacb_{1X}^{(p)}, \jacb_{2X}^{(p)},
           \jacb_{3X}^{(p)}) =  \sum_{\a=1}^{N_c^X} \;\;
     \sum_{n_2,n_3=0}^{N_\a}\; \sum_{l=0}^M \; c_{X,\a,n_2n_3,l}\;  f_l(\rho)\;
       B^{\rm HH}_{X,n_2n_3\a}(p)\ ,
\ee
where  $X\equiv A$ or $B$ and
\bea
     B_{A,n_2n_3\a}^{\rm HH} (p)
        & =&  \biggl\{
     {}^{(4)}{\cal H}_{\{\ell_{1\a}\ell_{2\a} \ell_{3\a} L_{2\a} 
        n_2n_3\},L_\a}(\Omega^{(A,p)}_3) \times \nonumber \\
 \noalign{\medskip}
      && \times\biggl [\Bigl[\bigl[ s_i s_j \bigr]_{S_{a\alpha}}
                s_k\Bigr]_{S_{b\alpha}} s_m  \biggr]_{S_\alpha}
         \biggr \}_{JJ_z}
           \biggl [\Bigl[\bigl[ t_i t_j \bigr]_{T_{a\alpha}}
              t_k\Bigr]_{T_{b\alpha}}
              t_m  \biggr]_{T_\alpha T_z}\ ,\label{eq:ampli1} \\
     B_{B,n_2n_3\a}^{\rm HH} (p)
        & =&  \biggl\{
     {}^{(4)}{\cal H}_{\{\ell_{1\a}\ell_{2\a} \ell_{3\a} L_{2\a} 
        n_2n_3\},L_\a}(\Omega^{(B,p)}_3) \times \nonumber \\
 \noalign{\medskip}
      && \times\Bigl[ \bigl[ s_i s_j \bigr]_{S_{a\alpha}}
                \bigl[ s_k  s_m \bigr]_{S_{b\alpha}}  \Bigr]_{S_\alpha}
         \biggr \}_{JJ_z}
            \Bigl[ \bigl[ t_i t_j \bigr]_{T_{a\alpha}}
             \bigl[ t_k  t_m \bigr]_{T_{b\alpha}}
               \Bigr]_{T_\alpha T_z}\ .\label{eq:ampli2}
\eea
Here, $\Omega^{(X,p)}_3$ denotes the hyperangular variables constructed
with the set $X$ of the Jacobi vectors relative to the permutation $p$
of the particles. Moreover,  $s_i$ ($t_i$) denotes the spin  (isospin)
state of particle $i$ and
\be\label{eq:cha4}
   \a\equiv\{ \ell_{1\alpha},
   \ell_{2\alpha}, \ell_{3\alpha}, L_{2\alpha}, L_\alpha,
   S_{a\alpha}, S_{b\alpha}, S_\alpha, T_{a\alpha}, T_{b\alpha}, T_\alpha
    \}\ ,
\ee
specifies the generic  four-body channel.
In Eqs.~(\ref{eq:ampli1}) and~(\ref{eq:ampli2}) $L_\a$ and $S_\a$ are
coupled to give $J$, $J_z$. In order to ensure the antisymmetry of the 
wave function,  
the  amplitudes $\psi_A$ and $\psi_B$  have to change sign under the
exchange of the particles  $i$  and $j$. Therefore, the number
$\ell_{3\alpha}+S_{a\alpha}+ T_{a\alpha}$ must be  odd.  In
addition, $\ell_{1\alpha}+\ell_{2\alpha}+ \ell_{3\alpha}$ must be an
even or odd number in correspondence to the positive or negative
parity of the  considered state. A few of the channels taken into account
in the calculations are given in Table~\ref{tab:chan4}.

\begin{table}[t]
\caption[Table]{\label{tab:chan4}
Quantum numbers  for the first $23$ channels
considered in the expansion of the even-parity and $J=0$ wave function of
a four-nucleon system for the set $A$ of Jacobi coordinates.
}
\begin{indented}
\lineup
\item[]
\begin{tabular}{ c   c c c c c  c c c  c c c  }
\br
$\alpha$  &$\ell_{1\a}$ & $\ell_{2\a}$ & $\ell_{3\a}$ & $L_{2\a}$
 & $L_\a$ & $S_{a\a}$ & $S_{b\a}$ & $S_\a$ & $T_{a\a}$ & $T_{b\a}$ & $T_\a$  \\ 
\mr
 1&0&0&0&0&0& 1&1/2& 0& 0&1/2&0 \\
 2&0&0&0&0&0& 0&1/2& 0& 1&1/2&0 \\
 3&0&0&2&0&2& 1&3/2& 2& 0&1/2&0 \\
 4&1&1&0&0&0& 1&1/2& 0& 0&1/2&0 \\
 5&1&1&0&0&0& 0&1/2& 0& 1&1/2&0 \\
 6&1&1&0&1&1& 1&1/2& 1& 0&1/2&0 \\
 7&1&1&0&1&1& 1&3/2& 1& 0&1/2&0 \\
 8&1&1&0&1&1& 0&1/2& 1& 1&1/2&0 \\
 9&0&2&0&2&2& 1&3/2& 2& 0&1/2&0 \\
10&2&0&0&2&2& 1&3/2& 2& 0&1/2&0 \\
11&1&1&0&2&2& 1&3/2& 2& 0&1/2&0 \\
12&1&0&1&1&0& 1&1/2& 0& 1&1/2&0 \\
13&1&0&1&1&0& 0&1/2& 0& 0&1/2&0 \\
14&0&1&1&1&0& 1&1/2& 0& 1&1/2&0 \\
15&0&1&1&1&0& 0&1/2& 0& 0&1/2&0 \\
16&1&0&1&1&1& 1&1/2& 1& 1&1/2&0 \\
17&1&0&1&1&1& 1&3/2& 1& 1&1/2&0 \\
18&1&0&1&1&1& 0&1/2& 1& 0&1/2&0 \\
19&0&1&1&1&1& 1&1/2& 1& 1&1/2&0 \\
20&0&1&1&1&1& 1&3/2& 1& 1&1/2&0 \\
21&0&1&1&1&1& 0&1/2& 1& 0&1/2&0 \\
22&1&0&1&1&2& 1&3/2& 2& 1&1/2&0 \\
23&0&1&1&1&2& 1&3/2& 2& 1&1/2&0 \\
 \br
\end{tabular}
\end{indented}
\end{table}

As in the three-body case, the contribution of channels with
$\ell_{1\alpha}+\ell_{2\alpha}+ \ell_{3\alpha}>6$ can be
disregarded. However, the sum over $n_2$ and $n_3$ usually
includes a very large number of terms. It has been found convenient
to limit the sum over $n_2$ and $n_3$ in Eq.~(\ref{eq:psiq1})
to include HH functions with $0\leq n_2+n_3 \leq N_\alpha$,
$N_\alpha$ being a non-negative integer.
For a given $N_\a$, the number $M_\alpha$ of
functions included in the expansion of the channel $\a$  is given by
\begin{equation}
  M_\alpha=(N_\alpha+1) (N_\alpha+2)/2\ .\label{eq:M}
\end{equation}
After antisymmetrization of the state, some of the functions
\be\label{eq:psiq2}
   {\cal B}^{\rm HH}_{X,n_2n_3\a}= \sum_{p=1}^{12}
   B^{\rm HH}_{X,n_2n_3\a}(p)
\ee
are linearly dependent on others and they must be
removed from the expansion. As an example, it can be verified that
${\cal B}^{\rm HH}_{A,0\, n_3\, \a=1}={\cal B}^{\rm
HH}_{A,0\, n_3\, \a=2}$ for all $n_3$ values (the channels $\a=1,2$ are specified
in Table~\ref{tab:chan4}). In this respect, it is
useful to know how many linearly independent states ${\cal B}$ exist
for a given set of values of $G$, $L$, $S$ and $T$, 
with $G=\ell_{1\alpha}+ \ell_{2\alpha}+\ell_{3\alpha}+2 n_2+2 n_3$. 
Such a number has been
calculated in Ref.~\cite{viviani:g}, where it has been shown that it is
significantly less than the degeneracy of the basis. As a result, a
large number of states has to be removed from the expansion and, after
that, the number $M_\a$ results to be appreciably lower than that given in
Eq.~(\ref{eq:M}).

\subsection{The Transformation Coefficients}
\label{sec:tc}

As discussed previously, in the case of $A$ identical fermions, 
antisymmetrical states can be obtained by appropriate sums over the
particle permutations. The calculation of the matrix elements of the potential
between  such states presents noticeable difficulties.  In general,  the
integrals to be evaluated involve the potential and two HH functions
constructed in terms of Jacobi vectors corresponding to different  
permutations of the particles. This task is 
simplified by the knowledge of  the following transformation coefficients
(TC),
\begin{equation}   
   {}^{(A)}{\cal H}_{\{G\},LM}(\Omega_N^{(p)}) = \sum_{\{G^\prime\}}
   a^{(p),G,L}_{\{G\},\{G^\prime\}} \ 
   {}^{(A)}{\cal H}_{\{G^\prime\},LM}(\Omega_{N})\ ,
   \label{eq:traco}
\end{equation}
where the sum is over all the quantum numbers $\{G^\prime\}$ 
specified in Eq.~(\ref{eq:qn2}), with the condition that the grand 
angular momentum $G$ defined in Eq.~(\ref{eq:go}) is conserved, 
$G^\prime=G$. As the number of functions with a given $G$ value is finite,
also the sum in Eq.~(\ref{eq:traco}) is over a finite number of terms. In
Eq.~(\ref{eq:traco}), $\Omega_{N}^{(p)}$ ($\Omega_{N}$) specifies the
hyperangular variables constructed with the 
permutation $p$ of the
particles.  To simplify the notation for $p=1$, in this section the
subscript ``$1$" will be omitted.

For the three-body system, these coefficients can be
easily calculated using recurrence relations~\cite{raynal}. 
However, the TC can also be calculated by taking into account the
orthonormality of the HH basis, namely
\begin{equation}   
     a^{(p),G,L}_{\{G\},\{G^\prime\}} =\int d\Omega_{N}\;
    [{}^{(A)}{\cal H}_{\{G^\prime\},LM}(\Omega_{N})]^\dag\;\;
     {}^{(A)}{\cal H}_{\{G\},LM}(\Omega_N^{(p)})
    \ .
    \label{eq:traco2}
\end{equation}
For $A=3$, the above integrals reduce to bi-dimensional integrals of
polynomial functions which can be calculated exactly using Gauss
quadrature as discussed in~\ref{sec:rare3}.
For $A=4$, Eq.~(\ref{eq:traco2}) results in a
5-dimensional integration. In that case it is more useful to use 
a recurrence formula, derived in Ref.~\cite{viviani:g}, and briefly outlined 
in~\ref{sec:arare}.

The spin-isospin states appearing in
Eqs.~(\ref{eq:B3hh}), (\ref{eq:ampli1}) and~(\ref{eq:ampli2}) can be
transformed in terms of spin-isospin states constructed using the
reference order $1,2,3,\ldots$ of the particles. The resulting
states have the same total spin $S_\a$ and total isospin $T_\a$ as the
initial states. Therefore, the complete 
functions $B^{\rm HH}_{n_2\a}(p)$ and 
$B^{\rm HH}_{X,n_2n_3\a}(p)$ can be transformed to the reference
permutation $p=1$ and a given choice of Jacobi vectors. 
For example, for $A=4$,
\be\label{eq:rrevai}
   B^{\rm HH}_{X,n_2n_3\a}(p)=\sum_{\mu'}a^{X,n_2n_3\a}_{n'_2n'_3\a'}(p)
   B^{\rm HH}_{A,n'_2n'_3\a'}(p=1) \ ,\qquad
   \mu'\equiv \{n_2', n_3', \a'\}\ ,
\ee
where the
sum is restricted to the indices $n'_2$ and $n'_3$ and channels
$\alpha'$ such that $G=\ell_{1\alpha}+ \ell_{2\alpha}+\ell_{3\alpha}+2 n_2+2 n_3= \ell_{1\alpha'}+
   \ell_{2\alpha'}+\ell_{3\alpha'}+2 n'_2+2 n'_3$, $L_\a=L_{\a'}$, $S_\a=S_{\a'}$ and
$T_\a=T_{\a'}$. 
Hence, summing over all the even permutations, it is possible to
define a completely antisymmetric
basis state given in the reference system $p=1$:
\be\label{eq:sym1}
    \sum_{p=1}^{12} B^{\rm HH}_{X,n_2n_3\a}(p)
   \equiv{\cal B}^{G L_\a S_\a T_\a,  J\pi}_{X,\mu}=
    \sum_{\mu'} A^{G L_\a S_\a T_\a, J\pi}_{X,\mu,\mu'}
   B^{\rm HH}_{A,n'_2n'_3\a'}(p=1) \ ,
\ee
with $ \mu\equiv\{ n_2,n_3,\ell_{1\alpha},
   \ell_{2\alpha}, \ell_{3\alpha}, L_{2\alpha},
   S_{a\alpha}, S_{b\alpha}, T_{a\alpha}, T_{b\alpha} \} $ and
\be\label{eq:sym2}
  A^{G L_\a S_\a T_\a, J\pi}_{X,\mu,\mu'}=
\sum_{p=1}^{12} a^{X,n_2n_3\a}_{n'_2n'_3\a'}(p) \ .
\ee
The same analysis holds for the $A=3$ system, where now $\mu\equiv
\{n_2, \ell_{1\alpha}, \ell_{2\alpha}, S_{a\alpha}, T_{a\alpha} \} $
and $\mu'\equiv \{n_2',\a'\}$.
Finally, the wave function of an $A=3,4$ system can be expanded as
\be\label{eq:expr}
   \Psi_A^{J\pi}=\sum_{GLSTX}\sum_{\mu,l}C^{GLST}_{X,\mu,l} 
   {\cal B}^{GLST,J\pi}_{X,\mu} \; f_l(\rho)\ ,
\ee
where the coefficients $C^{GLST}_{X,\mu,l}$ are linear parameters
(for $A=3$, the sum over $X$ is restricted to the single 
choice of Jacobi vectors specified in Eq.~(\ref{eq:jco3})).
The above expansion can also be given in momentum space:
\be\label{eq:expm}
   \Psi_A^{J\pi}=\sum_{GLSTX}\sum_{\mu,l} C^{GLST}_{X,\mu,l} 
   \tilde{\cal B}^{GLST,J\pi}_{X,\mu} \; g_{G,l}(Q) \;\ ,
\ee
where $\tilde{\cal B}^{GLST,J\pi}_{X,\mu}$ is now the hyperspherical-spin-isospin
amplitude in momentum space and, according to Eq.~(\ref{eq:vg}),
\begin{equation}
   g_{G,l}(Q)=(-i)^G\,\int_0^\infty d\rho\,
   \frac{\rho^{3N-1}}{(Q\rho)^{3N/2-1}}\,
   J_{G + 3 N/2 -1}(Q\rho)\, f_{l}(\rho) \ .
\label{eq:vg1}
\end{equation}

Sometimes it will be convenient to consider HH functions having a 
specific value of $j$, the total angular momentum of particles
$1,2$. To this aim, it is useful to write the basis 
in the $jj$-coupling scheme
\begin{equation}
  {\cal B}_{X,\mu}^{GLST,J\pi} =
  \sum_\nu {\cal A}^{GLST,J\pi}_{X,\mu,\nu} \; \Xi^{GTJ\pi}_{\nu} \  .
  \label{eq:PSI3jj}
\end{equation}
The basis functions $\Xi^{GTJ\pi}_{\nu}$ are HH functions in which the
angular-spin part for $A=3$ is coupled as
\be
  \left [ \Bigl ( Y_{\ell_2}(\hat x_2)
          {S_a} \Bigr)_{j_a}
           \Bigl ( Y_{\ell_1}(\hat x_1) s_3 \Bigr)_{j_b}
   \right]_{JJ_Z} \;\;\ ,
     \label{eq:PHIjj3}
\ee
and, for $A=4$, as 
\be
  \biggl \{ \left [ \Bigl ( Y_{\ell_3}(\hat x_3)
          {S_a} \Bigr)_{j_3}
           \Bigl ( Y_{\ell_2}(\hat x_2) s_3 \Bigr)_{j_2}
   \right]_{J_2}
          \Bigl ( Y_{\ell_1}(\hat x_1) s_4
   \Bigr)_{j_1}\biggr\}_{JJ_z} \;\;\ .
     \label{eq:PHIjj4}
\ee
The coefficients ${\cal A}^{GLSTJ\pi}_{X,\mu,\nu}$ are related to
$A^{GLSTJ\pi}_{X,\mu,\mu'}$ via Wigner coefficients.
Now, the integer index $\nu$ labels all possible choices of
\bea
   \nu\equiv\{n_2,\ell_2,S_a,j_a,\ell_1,j_b,T_a\}\ ,
\quad {\rm for}\,A=3 \ , \\
   \nu\equiv\{n_3,\ell_3,S_a,j_3,n_2,\ell_2,j_2,J_2,\ell_1,j_1,T_a,T_b\} \ ,
\quad {\rm for}\,A=4 
   \ ,
\label{eq:nu}
\eea
compatible with the given values of $G$, $T$, $J$ and $\pi$.

\subsection{The PHH and CHH Expansion}
\label{sec:chhex}

The rate of convergence of the  HH
expansion results to be rather slow when the
particle interaction contains large repulsion at small distances. For
example, in the calculation of the trinucleon binding energy with the 
AV18 NN potential, HH functions with $G$ up to $180$ were
considered~\cite{kievsky:a}. The resulting expansion included a 
large number of terms (approximately $600$). Only in that
way, the calculated  binding energy  was found to be in
agreement with the corresponding estimates obtained by other
techniques~\cite{kameyama,chen:a,friar:b,ishikawa,sasakawa}. 
The calculation rapidly
becomes more and more involved when larger systems are considered. For the
$\a$-particle, for example, approximately $4000$ states have to be
included in Eq.~(\ref{eq:psiq}). 

In order to overcome this difficulty, in Ref.~\cite{rosati} the
radial dependence of each amplitude has been modified by the
inclusion of a suitably chosen correlation factor 
(correlated-HH or, briefly, CHH expansion).
The role of the correlation factors is to speed up the convergence of
the expansion by improving the description of the system when two
particles are close to each other. In such configurations, there
are large cancellations between the contributions from kinetic and
potential energy terms and therefore the wave function must be very
precisely constructed. Thus, it can be convenient to include from the
very beginning  in the wave function suitable terms for describing those
configurations. In this way, the number of basis functions
necessary to get convergence is strongly reduced.

For the three-body system, the functions $B_{n\a}^{\rm HH}(p)$ in
Eq.~(\ref{eq:psit1}) are replaced by the following ones,
\be\label{eq:B3chh}
     B^{\rm CHH}_{n\a}(p)= F^{(3)}_{\a p}\;
     B^{\rm HH}_{n\a}(p)\ ,
\end{equation}
where $ F^{(3)}_{\a p}$ is a correlation factor which, in general,
will depend  on the
channel $\a$ and the permutation $p$ of the particles. In order to
maintain the antisymmetry of the wave function, $ F^{(3)}_{\a p}$ is taken to
be symmetric under the exchange of particles $i$ and $j$. 
A choice of $ F^{(3)}_{\a p}$, which 
has been extensively investigated, is 
\begin{equation}
      F^{(3)}_{\a p}
      = f_{\alpha}(r_{ij}) g_{\alpha}(r_{jk}) g_{\alpha}(r_{ik})\ ,
     \label{eq:corre3}
\end{equation}
where $f_\a$ and $g_\a$ are unidimensional functions determined as discussed
in~\ref{sec:corref}.

For a four-body system, the functions
\be\label{eq:B4chh}
     B^{\rm CHH}_{X,n_2n_3\a}(p)= F^{(4)}_{\a p}\;
     B^{\rm HH}_{X,n_2n_3\a}(p)
\end{equation}
are used to replace $B^{\rm HH}_{X,n_2n_3\a}(p)$ in
Eq.~(\ref{eq:psiq1}). In this case, the correlation factors have
been taken of the form~\cite{viviani:a}
\begin{equation}
    F^{(4)}_{\a p}
      = f_{\alpha}(r_{ij}) g_{\alpha}(r_{ik}) g_{\alpha}(r_{jk})
         g_{\alpha}(r_{im}) g_{\alpha}(r_{jm}) h_{\alpha}(r_{km}) \ ,
     \label{eq:corre4}
\end{equation}
where $f_{\alpha}$, $g_{\alpha}$ and $h_{\alpha}$ are
unidimensional functions of the interparticle distances. The choice of
these functions is also discussed in~\ref{sec:corref}.

The correlation factors so far introduced  are of the
Jastrow type, namely they are given by products over all the particle
pairs. A simplified choice consists in correlating only one pair. The
expansion basis obtained is called the pair-correlated hyperspherical
harmonic (PHH) expansion, which, for a three-body system, is given
by~\cite{kievsky:e}
\be\label{eq:B3phh}
     B^{\rm PHH}_{n\a}(p)= f_\a(r_{ij})     B^{\rm HH}_{n\a}(p)\ .
\end{equation}
The use of this (simplified) correlation factor has been shown to
provide very accurate results.

It should be noticed that, with the inclusion of the correlation
factors, the linear dependence between the states discussed in the
previous subsections has to be reconsidered. In fact, 
for the $A=3$ case, 
the minimum values $N^0_\a$ of the index $n$ entering in 
Eq.~(\ref{eq:psit1}) can be taken to be $N^0_\a=0$
for all the considered channels. Also for $A=4$, the linear dependent
states reduce to a small number (the actual number depends on the
particular choice of the correlation factors).

As for the HH expansion, it is possible to define a completely antisymmetric 
correlated state for $A=3$
\be\label{eq:cstate3}
     {\cal B}^\lambda_{n\a}= \sum_{p=1}^3 B^\lambda_{n\a}(p)\ ,
\end{equation}
and for $A=4$ 
\be\label{eq:cstate4}
     {\cal B}^\lambda_{X,n_2n_3\a}= \sum_{p=1}^{12}
     B^\lambda_{X,n_2n_3\a}(p) \, ,
\end{equation}
where $\lambda\equiv$ CHH or PHH. Accordingly it is possible to 
expand the $A=3,4$ wave function in such a basis
\bea\label{eq:cexp}
     \Psi_3^{J\pi}&=\sum_{ln\a}C^\a_{nl}f_l(\rho){\cal B}^\lambda_{n\a} \ ,\cr 
     \Psi_4^{J\pi}&=\sum_{lXn_2n_3\a}C^\a_{Xn_2n_3l}f_l(\rho)
     {\cal B}^\lambda_{X,n_2n_3\a} \;\;\ .
\eea
A difference with the HH expansion is that the basis states are not labelled
with the grand angular momentum $G$, due to the presence of the correlation
factors. Furthermore, the presence of correlation factors complicates  the 
transformation of the states given in permutation $p$ to the
reference system $p=1$. As a consequence, multidimensional integrals
have to be calculated to obtain the matrix elements when using the
correlated basis states. This can be done very efficiently for $A=3$
whereas, for $A=4$, a stochastic integration method should be used.

\subsection{Details of the Calculation}
\label{sec:deta}

In general, the $A=3$ and 4 bound state wave functions 
can be cast in the form
\begin{equation}
  |\Psi^{J\pi}_A\rangle=\sum_{\zeta}\, c_\zeta\, 
  |\Psi_\zeta \rangle \ ,
  \label{eq:psi2}
\end{equation}
where $|\Psi_\zeta\rangle$ is one of the complete set of antisymmetric states
defined in the preceding sections, and $\zeta$ is an index denoting all the
quantum numbers necessary to completely specify  
the basis elements.
The coefficients of the expansion can be calculated using the 
Rayleigh-Ritz variational principle, which states that
\begin{equation}
  \langle\delta_c \Psi^{J\pi}_A\,|\,H-E\,|\Psi^{J\pi}_A\rangle
   =0 \ ,
   \label{eq:rrvar}
\end{equation}
where $\delta_c \Psi^{J\pi}_A$ indicates the variation of 
the wave function for arbitrary infinitesimal 
changes of the linear coefficients $c_\zeta$. 
The problem of determining $c_\zeta$ and the energy $E$ 
is then reduced to a generalized eigenvalue problem, 
\begin{equation}
  \sum_{ \zeta'}\,\langle\Psi_\zeta\,|\,H-E\,|\, \Psi_{\zeta'}\,\rangle 
\,c_{\zeta'}=0
  \ .
  \label{eq:gepb}
\end{equation}
The main difficulty of the method is to compute the 
matrix elements of the Hamiltonian $H$ with respect to the basis states
$|\Psi_\zeta\rangle$. Usually $H$ is given as a sum of terms (kinetic energy,
two-body potential, etc.). The calculation of the matrix elements of
some parts of $H$ can be more conveniently performed in coordinate 
space, while for other parts it could be easier to work in momentum
space. 

In the case of the HH expansion, the basis states are given in
coordinate (momentum) space in Eq.~(\ref{eq:expr}) (Eq.~(\ref{eq:expm})).
The criteria to choose 
hyperradial functions $f_l(\rho)$ have been the following: 
(i) $f_l(\rho)\rightarrow 0$ for $\rho\rightarrow\infty$;
(ii) $f_l(\rho)$ should constitute an orthonormal basis;
(iii) $f_l(\rho)$ should be easy to handle, when the Fourier 
transform of Eq.~(\ref{eq:vg}) is performed. A possible choice used here  
is given by 
\begin{equation}
 f_l(\rho)=\gamma^{D/2} \sqrt{\frac{l!}{(l+D-1)!}}\,\,\, 
 L^{(D-1)}_l(\gamma\rho)\,\,e^{-\frac{\gamma}{2}\rho} \ ,
 \label{eq:fllag}
\end{equation}
where $L^{(D-1)}_l(\gamma\rho)$ are Laguerre polynomials~\cite{abra}. 
Here, there is only one non-linear parameter, $\gamma$,  
to be variationally optimized. In particular, 
$\gamma$ can be chosen in the interval 
2.5--4.5 fm$^{-1}$ for the AV18 and CDBonn potentials and 
4--8 fm$^{-1}$ for the N3LO-like potential, for both $A=3$ and 4.
The corresponding functions $g_{G,l}(Q)$ can be written as
\bea
  g_{G,l}(Q)&=&\frac{(-i)^G}{\gamma^{D/2}}
  \sqrt{\frac{l!}{(l+D-1)!}}\, \sum_{k=0}^l b^l_k\,2^{k+D}\,\,
  \Gamma(G+k+D) \nonumber\\
   && \qquad \times \,\,\frac{u^{k+D}}{(1-u^2)^{\frac{D}{4}-\frac{1}{2}}}\,\,
  P^{1-G-D/2}_{k+D/2}(u) \ ,
\label{eq:gllag}
\eea
where $u=\frac{1}{\sqrt{1+(2Q/\gamma)^2}}$, $P_n^m$ is an 
associated Legendre function and $b^l_k$ is given by
\begin{equation}
   b^l_k=\frac{(-1)^k}{k!}\, \left( \begin{array}{c}
                                l+D-1 \\ l-k 
                               \end{array} \right)
  \ ,
  \label{eq:ak}
\end{equation}
so that $L^{(D-1)}_l(x)=\sum_{k=0}^l b^l_k\,x^k$~\cite{abra}.

In such a case, the normalization ($N$) and kinetic
energy ($T$) operator matrix elements can be easily computed, both
in coordinate or in momentum space. Explicitly they are
\begin{eqnarray}
 N^{GLSTJ\pi}_{X\mu l,X'\mu' l'}&=& 
  \sum_{\mu''}(A^{GLSTJ\pi}_{X,\mu\mu''})^*
 A^{GLSTJ\pi}_{X'\mu'\mu''} \delta_{l,l'} \ ,
  \label{eq:norm} \\
  T^{GLSTJ\pi}_{X\mu l,X'\mu' l'}&=& 
 \sum_{\mu''}(A^{GLSTJ\pi}_{X,\mu\mu''})^*
 A^{GLSTJ\pi}_{X',\mu'\mu''} T_{l,l'} \ ,
  \label{eq:kin}
\end{eqnarray}
with 
\begin{eqnarray}
  T_{l,l'}&=& 
-\frac{\hbar^2}{2m} \int\,d\rho\,\rho^{3A-4}\,f_{l'}(\rho)\,
  \bigg[ \frac{\partial^2}{\partial\rho^2}
      +\frac{3A-4}{\rho}\frac{\partial}{\partial\rho}
      -\frac{G(G+3A-4)}{\rho^2} \bigg]\,f_{l}(\rho)\nonumber \\
  &=& \frac{\hbar^2}{2m}\,\int\,dQ\,Q^{3A-2}\,g_{G,l'}(Q)\,g_{G,l}(Q) \ ,
  \label{eq:rkin}
\end{eqnarray}
where use has been made of the fact that the HH functions are 
eigenfunctions of the operator $\Lambda^2_N(\Omega_N)$ defined in
Eq.~(\ref{eq:equ2}) corresponding to the eigenvalues $G(G+3A-4)$, and
that the  functions $f_l(\rho)$ form an orthonormal set with respect to
the weight $\rho^{3 A -4}$. 
The calculation of the two-body potential energy matrix elements is more
conveniently performed using the hyperspherical-angular-spin-isospin 
basis elements $\Xi^{GTJ\pi}_\nu$, which have a well defined angular momentum
$j$ between particles $1,2$, and  are totally antisymmetric. Therefore, 
the following relation holds
\begin{equation}\label{eq:v12}
<\Xi^{G'T'J\pi}_{\mu'}|V|\Xi^{GTJ\pi}_{\mu}>=
\frac{A(A-1)}{2}<\Xi^{G'T'J\pi}_{\mu'}|v(1,2)|\Xi^{GTJ\pi}_{\mu}> \;\;\ .
\end{equation}
The potential $v(1,2)$ acts on the particle pair $1$,$2$, accordingly
to the quantum number $j$ and all other quantum numbers in $\mu$,
with the exception of $S_{a\a},T_{a\a}$ and $\ell_{1,\a}$ 
($\ell_{1,\a}$,$\ell_{2,\a}$) for $A=3$ ($A=4$), which are conserved. 
This reduces
considerably the total number of matrix elements needed in the
calculation. Moreover, using the TC introduced in 
Sec.~\ref{sec:tc}, the above equation reduces to a two-dimensional
integral in one hyperangle ($\cos\phi_3$ for $A=4$ or $\cos\phi$
for $A=3$) and $\rho$. 

In the case of non-local potentials, one has
\begin{equation}
v(1,2)=V(\bfx'_N,\bfx_N)
\label{eq:v12cs}
\end{equation}
in coordinate space, and
\begin{equation}
v(1,2)=\widetilde V(\bfk'_N,\bfk_N) 
\label{eq:v12ms}
\end{equation}
in momentum space, where the spin-isospin-dependence is understood. 
The integrals of Eq.~(\ref{eq:v12}) then  
are three-dimensional, and 
can be calculated using Tchebyshev and Laguerre weights and
points for the hyperangular and the hyperradial variables, 
respectively~\cite{abra}. 
A sufficiently
dense grid can be used to obtain relative errors $< 10^{-6}$ for these
integrals. More details about the procedure to
compute matrix elements of the NN and 3N interactions are given 
in~\ref{sec:mehh}.

In the case of the CHH or PHH expansion, the matrix elements of the
norm, kinetic energy and potential energy are all calculated
numerically. These expansions have been used to treat the case of
local potentials. The corresponding integrals reduce to three dimensions
for $A=3$ and to six dimensions for $A=4$. In the first case
a Gauss, Tchebyshev, Laguerre integration has been used for the 
variables $\mu_{12}=\hat x_1\cdot\hat x_2$, $z=\cos 2\phi$, and $\rho$,
whereas a quasirandom technique has been used for 
$A=4$.

\section{The HH Technique for Scattering States}
\label{sec:scatt}

We consider in this section the application of the 
HH expansion to a scattering problem. In particular, we focus our 
attention to elastic processes of the type
$N+Y\rightarrow N+Y$, where $N$ is a nucleon and $Y$ is a bound nuclear 
system ($A_Y+1=A=3$, $4$), in the low
energy region, where the nucleus $Y$ cannot be broken.

The wave function $\Psi_{N-Y}^{L S J J_z}$ describing a $N-Y$
scattering state with incoming orbital angular momentum $L$ and channel spin
$S$ ($\vec{S}=\vec{\frac{1}{2}}+\vec{S_Y}$), parity $\pi=(-)^L$, 
and total angular momentum $J, J_z$,
can be written as 
\begin{equation}
    \Psi_{N-Y}^{LSJJ_z}=\Psi_C^{LSJJ_z}+\Psi_A^{LSJJ_z} \ ,
    \label{eq:psica}
\end{equation}
where $\Psi_C^{LSJJ_z}$ describes the system in the region where the particles
are close to each other and their mutual interactions are strong, 
while $\Psi_A^{LSJJ_z}$ describes the relative motion between the nucleon $N$
and the nucleus $Y$ in the asymptotic region, where the $N-Y$ nuclear 
interaction 
is negligible. The function $\Psi_C^{LSJJ_z}$, which has to 
vanish in the limit of large intercluster
separations, can
be expanded either in the HH or CHH basis as it has been done 
in the case of bound states (see Eqs.~(\ref{eq:psit}) and~(\ref{eq:psiq}) 
for the HH and Eq.~(\ref{eq:cexp}) for the CHH expansion). Therefore, 
applying Eq.~(\ref{eq:psi2}), 
the function $\Psi_C^{LSJJ_z}$ can be casted in the form 
\begin{equation}
  |\Psi^{LSJJ_z}_C\rangle=\sum_{\zeta}\, c_\zeta\,
  |\Psi_\zeta \rangle \ ,
  \label{eq:psis2}
\end{equation}
where $|\Psi_\zeta\rangle$ is defined in Sec.~\ref{sec:chh}.

The function $\Psi_A^{LSJJ_z}$ is the appropriate 
asymptotic solution of the relative $N-Y$ Schr\"odinger equation, 
and can be written
as a linear combination of the following functions
\begin{equation}
  \Omega_{LSJJ_z}^\lambda= \frac{\cal C}{\sqrt{N_p}} \sum_{p=1}^{N_p}
  \Bigl [ [\chi_N\otimes \phi_Y]_{S} \otimes 
  Y_{L}(\hat{y}_p) \Bigr ]_{JJ_z} {\cal R}^\lambda_L(y_p) \ ,
  \label{eq:psiom}
\end{equation}
where ${\cal C}$ is a normalization factor (see below).
Here the sum over $p$ has to be done over the even $N_p$ permutations of the
$A$ nucleons necessary to antisymmetrize the functions 
$\Omega_{LSJJ_z}^\lambda$, 
$\phi_Y$ ($\chi_N$ ) is the $Y$ ($N$) wave function, 
${\bf y}_p$ is the distance 
between $N$ and the center of mass of $Y$, 
$Y_{L}(\hat{y}_p)$ is the standard spherical harmonic function, 
and the functions ${\cal R}^\lambda_L(y_p)$ are the regular ($\lambda\equiv R$)
and irregular ($\lambda\equiv I$) radial solutions of the relative two-body 
$N-Y$ Schr\"odinger equation without the nuclear interaction. 
These regular and irregular functions, denoted as 
${\cal F}_L(y_p)$ and ${\cal G}_L(y_p)$ respectively, have the form
\begin{eqnarray}
{\cal F}_L(y_p)&=&{F_L(\eta,\xi)\over (2L+1)!!q^L\xi C_L(\eta)} \ , 
\nonumber \\
{\cal G}_L(y_p)&=&(2L+1)!! q^{L+1}C_L(\eta)f_R(y_p){G_L(\eta,\xi)\over \xi}
\ ,
\label{eq:risol}
\end{eqnarray}
where $q$ is the
modulus of the $N-Y$ relative momentum 
(related to the total kinetic energy in the center of mass (c.m.) system by
$T_{c.m.}={q^2\over 2\mu}$, $\mu$ being the $N-Y$ reduced mass), 
$\eta=2\mu e^2/q$ and $\xi=qy_p$ are the usual Coulomb parameters, 
and the regular (irregular) Coulomb function $F_L(\eta,\xi)$ 
($G_L(\eta,\xi)$) and the 
factor $C_L(\eta)$ is defined in the standard 
way~\cite{chen:b}. The factors $(2L+1)!! q^L C_L(\eta)$
have been introduced so that ${\cal F}$ and ${\cal G}$
have a well defined limit for $q\rightarrow 0$.
The function $f_R(y_p)=[1-\exp(-b y_p)]^{2L+1}$ 
has been introduced to regularize $G_L$ at small values of $y_p$. 
The trial parameter $b$ is
determined by requiring that $f_R(y)\rightarrow 1$ for 
large values of $y_p$,
thus not modifying the asymptotic behaviour of the 
scattering wave function. A value of $b\simeq R_Y^{-1}$, 
$R_Y$ being the dimension of nucleus $Y$, has been found appropriate.
The non-Coulomb case of Eq.~(\ref{eq:risol}) is
obtained in the limit $e^2\rightarrow 0$. In this case $F_L(\xi)/\xi$ and
$G_L(\xi)/\xi$ reduce to the regular and irregular Riccati-Bessel 
functions and
the factor $(2L+1)!!C_L(\eta)\rightarrow 1$ for $\eta\rightarrow 0$.
With the above definitions, $\Psi_A^{LSJJ_z}$  can be written in the form
\begin{equation}
  \Psi_A^{LSJJ_z}= \sum_{L^\prime S^\prime}
 \bigg[\delta_{L L^\prime} \delta_{S S^\prime} 
\Omega_{L^\prime S^\prime JJ_z}^R
  + {\cal R}^J_{LS,L^\prime S^\prime}(q)
     \Omega_{L^\prime S^\prime JJ_z}^I \bigg] \ ,
  \label{eq:psia}
\end{equation}
where the parameters ${\cal R}^J_{LS,L^\prime S^\prime}(q)$ give the
relative weight between the regular and the irregular components 
of the wave function. They
are closely related to the reactance matrix (${\cal K}$-matrix)
elements which can be written as
\begin{equation}
 {\cal K}^J_{LS,L^\prime S^\prime}=
 (2L+1)!!(2L'+1)!!q^{L+L'+1}C_L(\eta)C_{L^\prime}(\eta)
 {\cal R}^J_{LS,L^\prime S^\prime} \;\;\ .
\end{equation}
By definition of the ${\cal K}$-matrix, its eigenvalues are
$\tan\delta_{LSJ}$, $\delta_{LSJ}$ being the phase shifts.
The sum over $L^\prime$ and $S^\prime$ in Eq.~(\ref{eq:psia}) is over all 
values compatible with a given $J$ and parity $\pi$. In particular, the sum 
over $L^\prime$ is limited to include either even or odd values since
$(-1)^{L^\prime}=\pi$.

The matrix elements ${\cal R}^J_{LS,L^\prime S^\prime}(q)$ and 
the linear coefficients occurring in the expansion of $\Psi^{LSJJ_z\pi}_C$ 
are determined applying the Kohn variational principle~\cite{kohn}, 
which states that the functional
\begin{equation}
   [{\cal R}^J_{LS,L^\prime S^\prime}(q)]=
    {\cal R}^J_{LS,L^\prime S^\prime}(q)
     - \left \langle \Psi^{L^\prime S^\prime JJ_z }_{N-Y} \left |
         H-E \right |
        \Psi^{LSJJ_z}_{N-Y}\right \rangle
\label{eq:kohn}
\end{equation}
has to be stationary with respect to variations of the trial parameters 
in $\Psi^{LSJJ_z}_{N-Y}$. 
Here $E$ is the total energy of the system, and the normalization of
the asymptotic states (factor ${\cal C}$ in Eq.~(\ref{eq:psiom}) )
has been fixed by the condition:
\be
   \langle \Omega^R_{LSJJ_z}| H-E | \Omega^I_{LSJJ_z} \rangle
  -\langle \Omega^I_{LSJJ_z}| H-E | \Omega^R_{LSJJ_z} \rangle =1 \ .
\ee
Using Eqs.~(\ref{eq:psis2}) and~(\ref{eq:psia}), 
the variation of the diagonal functionals of Eq.~(\ref{eq:kohn}) with
respect to the linear parameters $c_\zeta$ leads to the following 
system of linear equations:
\be
  \sum_{\zeta'} c_\zeta < \Psi_\zeta|H-E|\Psi_{\zeta'}>= 
     -D^\lambda_{LSJJ_z}(\zeta) \ .
  \label{eq:set1}
\ee
Two different inhomogeneous terms $D^\lambda$ corresponding to
$\lambda\equiv R,I$ are introduced and are defined as 
\be
  D^\lambda_{LSJJ_z}(\zeta)= < \Psi_\zeta|H-E|\Omega^\lambda_{LSJJ_z}> \ .
\ee
The matrix elements ${\cal R}^J_{LS,L'S'}$ are obtained 
varying the diagonal functionals of Eq.~(\ref{eq:kohn}) with respect to them. 
This leads to the following set of algebraic equations
\be
  \sum_{L'' S''} {\cal R}^J_{LS,L''S''} X_{L'S',L''S''}= Y_{LS,L'S'} \ ,
\label{eq:set2}
\ee
with the coefficients $X$ and $Y$ defined as
\bea
X_{LS,L'S'}&= <\Omega^I_{LSJJ_z}+\Psi^{LSJJ_z,I}_C|H-E|\Omega^I_{L'S'JJ_z}> \ ,
\nonumber \\
Y_{LS,L'S'}&=-<\Omega^R_{LSJJ_z}+\Psi^{LSJJ_z,R}_C|H-E|\Omega^I_{L'S'JJ_z}> \ ,
\eea
where $\Psi^{LSJJ_z,\lambda}_C$ is the solution of the set of 
Eq.~(\ref{eq:set1}) with the corresponding inhomogeneous term. A 
second order estimate of ${\cal R}^J_{LS,L'S'}$ is 
given by the quantities $[{\cal R}^J_{LS,L'S'}]$, obtained by 
substituting in Eq.~(\ref{eq:kohn}) the
first order results.

In the particular case of $q=0$ (zero-energy scattering),
the scattering can occur only in the channel $L=0$ and the observables 
of interest are the scattering lengths. Within the 
present approach, they are defined as
\be
  ^{(2J+1)}a_{NY}=-\lim_{q\rightarrow 0}{\cal R}^J_{0J,0J}\ ,
\ee
and the corresponding asymptotic states are then given by
\begin{equation}
  \Psi_A^{0JJJ_z}= \bigg[\Omega_{JJ_z}^R
  - \,^{(2J+1)}\!a_{NY} \Omega_{JJ_z}^I \bigg] \ .
  \label{eq:psia0}
\end{equation}

\section{Results}
\label{sec:res}

In this section results are presented for bound and 
zero-energy scattering states obtained within the HH approach,
with and without the inclusion of correlation factors, 
using several interaction models. 
The section is divided into the following parts: 
in Sec.~\ref{subsec:cent}, the triton and $^4$He bound states are
studied with 5 different central potentials, 
which have been used by several groups to produce 
benchmark calculations. 
For some of them, different versions appear in
the literature causing problems at the moment of comparisons.
For this reason,
the potential parameters are specified and the HH results 
are compared to those ones obtained with other techniques. 
In Sec.~\ref{subsec:gs} the 
$A=3$ and 4 bound states are studied using the most recent 
models for the nuclear interaction, consisting of two- and 
three-nucleon potentials. Their predictions for binding energies and other 
ground state properties are presented. When possible, comparisons with
other methods are also given.
Finally, in Sec.~\ref{subsec:ss}, the results for the 
$nd$, $pd$, $n^3$H, and $p^3$He scattering lengths are presented and 
discussed.

\subsection{Bound States with Central Potentials}
\label{subsec:cent}

Let us firstly consider the $A=3,4$ bound states using 5 
central two-nucleon potential models, i.e., the Volkov~\cite{volkov}, 
Afnan-Tang S3 (ATS3)~\cite{afnan}, Minnesota~\cite{thompson}, and 
Malfliet-Tjon version V (MT-V) and I/III (MT-I/III)~\cite{malfliet}. 
These potentials have been used by several groups to describe bound 
states of light nuclei. We review here briefly their main characteristics. 
Each potential $V(r)$ can be written as 
\begin{equation}
V(r)=\sum_i v_i f(\mu_i, r) \ , 
\label{eq:vr}
\end{equation}
where the function
$f(\mu_i,r)$ is either $\exp(-\mu_i r^2)$ for the
Gaussian-type  potentials or $\exp(-\mu_i r)/r$ for the 
Yukawa-type potentials.
The operators $v_i$ act on the spin-isospin degrees of freedom
and are written as 
\begin{equation}
v_i=V_i \times
(W_i+M_i P_r+B_iP_{\sigma}-H_iP_{\tau}) \ ,
\label{eq:vi}
\end{equation}
where $P_r$, $P_{\sigma}$ and $P_{\tau}$ are the space-, spin- and 
isospin-exchange operators.  
The values of parameters $V_i$, $\mu_i$,  
$W_i$, $M_i$, $B_i$ and $H_i$ 
of the potentials used in the present work 
are listed in Table~\ref{tb:nnpar}. 
Few remarks are here in order: 
(i) the Volkov and MT-V are spin-independent models,
while the other 3 potentials are spin dependent; 
(ii) the MT-I/III version acts
only on s-waves; (iii) it is 
customary to include the point-Coulomb interaction ($e^2=1.44$ MeV
fm) with the Minnesota potential; 
(iv) for all the considered potentials, the total
orbital angular momentum is a good quantum number and therefore we
have included in the wave functions only the channels with $L=0$.
Finally, note that the Volkov, ATS3, Minnesota and MT-V 
potentials are equal to those used in Ref.~\cite{varga}, while 
the version of the MT-I/III here adopted is
the same as the one reported in Table~I of
Ref.~\cite{schelling}.

\begin{table}
\caption{\label{tb:nnpar}
List of the parameters of the central 
NN potentials used in this paper introduced in Eqs.~(\protect\ref{eq:vr})
and~(\protect\ref{eq:vi}). 
The potential
strengths $V_i$ are in units of MeV for Gaussian- (G)
and MeV~fm for Yukawa-type (Y) potentials, respectively.
The parameters $W_i$, $M_i$, $B_i$ and $H_i$ are dimensionless and
the ranges $\mu_i$ are in units of fm$^{-2}$ for G or fm$^{-1}$ for 
Y potentials, respectively. The Majorana mixture parameter 
$M$ of the Volkov potential and the 
parameter $u$ in the Minnesota potential are $0$ and 1, 
respectively.}
\begin{indented}
\lineup
\item[]
\begin{tabular}{@{}lcccccccc}
\br
Potential & Type & $i$ & $V_i$ & 
$\mu_i$ & $W_i$ & $M_i$ & $B_i$ & $H_i$ \\
\mr
MT-V   &Y  & 1  & 1458.047  & 3.11 & 1.0  & 0.0  & 0.0 & 0.0       \\
\protect\cite{malfliet}    &  & 2  & $-$578.089  & 1.55 & 1.0  & 0.0  & 0.0 & 0.0       \\
\mr
MT-I/III  & Y   & 1 & 1438.72   & 3.11  & 1.0  & 0.0 & 0.0 & 0.0       \\
\protect\cite{malfliet} &  & 2 & $-570.4255$& 1.55 & 1.0  & 0.0 & 0.0 & 0.0       \\
            &  & 3 & $-56.4585$ & 1.55 & 0.0  & 0.0 & 1.0 &  0.0      \\
\mr
Volkov   & G & 1    & 144.86   &  0.82$^{-2}$ & $1.0-M$  & $M$  &
 0.0     &  0.0      \\
 \protect\cite{volkov}  & & 2    & $-$83.34   & 1.60$^{-2}$ & $1.0-M$ & $M$ &
 0.0      & 0.0  \\
\mr
ATS3  & G   & 1 & 1000.0   & 3.0  & 1.0  & 0.0 & 0.0 & 0.0       \\
\protect\cite{afnan} &  & 2 & $-$326.7   & 1.05 & 0.5  & 0.0 & 0.5 & 0.0       \\
        &  & 3    & $-166.0$ & 0.80  & 0.5  & 0.0  & $-0.5$ &  0.0      \\
           &  & 4    & $-43.0$  & 0.60  & 0.5  & 0.0  & 0.5  & 0.0       \\
           &  & 5    & $-23.0$  & 0.40 & 0.5  & 0.0  & $-0.5$ & 0.0       \\
\mr
Minnesota & G  & 1& 200.0 &1.487 &$0.5u$&$1.0-0.5u$&0.0& 0.0       \\
\protect\cite{thompson}&  &2& $-178.0$& 0.639 & $0.25u$& $0.5-0.25u$ & $0.25u$ &
$0.5-0.25u$    \\
         &   & 3 & $-91.85$& 0.465& $0.25u$& $0.5-0.25u$ &$-0.25u$ &
$-0.5+0.25u$
\\ 
\br
\end{tabular}
\end{indented}
\end{table}

The triton and $^4$He binding energies have been calculated for these 
5 central interaction models using the HH or the CHH approaches. The results
are compared to those ones obtained by other techniques in 
Table~\ref{tb:centr}. In all of these calculation, 
$\hbar^2/m=41.47$ MeV fm$^2$. Some of the techniques considered 
in the comparison have already been described in Sec.~\ref{sec:intro}, 
and they are the FY equations method~\cite{nogga:a}, 
the SVM~\cite{suzuki:a,usukura:a} and the 
CRCG expansion technique~\cite{kamika:b}. 
Furthermore, the results of the 
effective-interaction hyperspherical harmonics (EIHH) 
method of Ref.~\cite{barnea:b} and the ATMS (Amalgamation of Two-body
correlation into the Multiple Scattering process) method of 
Ref.~\cite{akaishi} 
have been reported. 

\begin{table} [t]
\caption[table]{\label{tb:centr}
The triton and $^4$He binding energies in MeV, calculated 
for various central interaction models, are compared with
the results obtained by other methods.
}
\begin{indented}
\lineup
\item[]
\begin{tabular}{llcc}
\br
{\rm Potential} & {\rm Method} & $B(\tri)$ & $B(\heq)$ \\
\mr
{\rm Volkov}    & {\rm HH}   & 8.465\n   & 30.420    \\ 
                & {\rm SVM}  & 8.46\n\n  & 30.424    \\
\mr
{\rm ATS3}      & {\rm HH}   & 8.758\n   & 31.618    \\
                & {\rm CHH}  & 8.758\n   & 31.61\n   \\
                & {\rm SVM}  & 8.753\n   & 31.616    \\
\mr
{\rm Minnesota} & {\rm HH}   & 8.3858    & 29.947    \\
                & {\rm CHH}  & 8.3858    & 29.95\n   \\
                & {\rm SVM}  & 8.380\n   & 29.937    \\
                & {\rm EIHH} & 8.3856    & 29.96\n   \\
\mr
{\rm MT-V}      & {\rm HH}   & 8.2527    & 31.347    \\
                & {\rm CHH}  & 8.2527    & 31.357    \\
                & {\rm SVM}  & 8.2527    & 31.360    \\
                & {\rm EIHH} & 8.244\n   & 31.358    \\
                & {\rm CRGC} &           & 31.357    \\
                & {\rm FY}   & 8.2527    & 31.364    \\
                & {\rm ATMS} &           & 31.364    \\
\mr
{\rm MT-I/III}  & {\rm HH}   & 8.5357    & 30.310    \\
                & {\rm CHH}  & 8.5357    & 30.31\n   \\
                & {\rm FY}   & 8.5357    & 30.312    \\
\br
\end{tabular}
\end{indented}
\end{table}

From inspection of the table, we can conclude that: 
(i) for the Volkov potential with Majorana parameter
$M=0$, our results agree very well with the estimates 
of the SVM~\cite{varga} method. Note that the Volkov potential, given
as a sum of Gaussians,  has a very soft core and therefore the
induced two-body correlations in the ground state wave function
are weaker than in the other cases. In fact, we have found that
the convergence of the HH expansion in this case is quite fast.  
Since only the inclusion of HH states with fairly low  values of the grand 
angular momentum quantum number is sufficient to obtain convergence, a 
successful HH calculation for this potential was already possible 
more than 30 years ago for $A=3$~\cite{wage} and more than
20 years ago for $A=4$~\cite{ballot}.  
(ii) Both for the ATS3 and Minnesota (with exchange parameter 
$u$=1) potential models, 
there is a good
agreement between the different theoretical estimates.
Note that both these potentials are given as a sum of Gaussians
but have a rather strong repulsion at short interparticle
distances. This induces important two-body correlations in the
wave function and consequently the convergence is slower than 
before. A grand angular momentum quantum number of $40$ is sufficient for an 
accuracy of $1$ keV in the binding energy.
(iii) The MT-V potential is given 
as a superposition of Yukawians having a strong
repulsive core with a $1/r$ divergence. This model, as well as the MT-I/III,
represents the most challenging  problem for the HH expansion, due
to the difficulty of constructing accurate two-body correlations at
short interparticle distances, where the cancellation between the 
kinetic and potential energy terms is critical. The introduction of
correlation factors in the HH expansion allows for very accurate results 
in both the $A=3$ and $4$ systems. The HH without correlation factors
requires a large number of basis elements to get convergence. For
$A=3$, the maximum value for the 
grand angular momentum quantum number ($G_{\rm max}$) considered has been $200$.
For $A=4$, it is possible to extrapolate the HH result to consider
infinite number of basis elements,  
using the procedure of Ref.~\cite{viviani:b}. For the MT-V potential,
the extrapolated HH result is 31.358 MeV, 
very close to other accurate estimates.
For the (s-wave) MT-I/III we observe that
our estimate is already close to the very precise calculation of
Ref.~\cite{schelling}. The ``missing'' binding energy in this case is estimated
to be $21$ keV, bringing our estimated binding energy to be $30.331$ MeV.

\subsection{Bound State with Realistic Interactions}
\label{subsec:gs}

In this section, the $A=3$ and 4 bound states are studied 
using the most recent 
models for the nuclear interaction, consisting of two- and 
three-nucleon potentials. 
We first summarize the main characteristics of these interaction models.  

Among the realistic models for the two-nucleon interaction,  
we have considered 
the Argonne AV8$'$~\cite{pudliner:b}, $v_{14}$
(AV14)~\cite{wiringa:f}  and AV18~\cite{wiringa:a} models, 
the Nijmegen 2 (NJ2)~\cite{stoks}, 
and the CDBonn~\cite{machleidt:b} models. 
We have also considered one of the models developed by Doleschall and
collaborators~\cite{doleschall}, labelled as ISuj model~\cite{isuj}.
Note that the Argonne and the NJ2 models are local 
and expressed in coordinate-space, the ISuj model is also 
expressed in coordinate-space but is non-local, and the CDBonn model 
is non-local and expressed in momentum-space. The AV8$'$ 
potential is a reduction
of the AV18 potential including only central, tensor and spin-orbit components,
with modified parameters in order to reproduce the deuteron binding energy.
We have included it, since it has been used before to produce a benchmark for
the $\a$-particle binding energy~\cite{kamada:a}.
Among the potential models derived within an 
effective field theory approach, we have considered those ones 
calculated up to next-to-next-to-next-to-leading order 
by the Idaho~\cite{entem} (N3LO-Idaho), 
and by the J\"ulich~\cite{epelbaum:b,epelbaum:c}
(N3LO-J\"ulich) groups. Note that the N3LO-Idaho model is obtained 
within a dimensional regularization scheme 
with cutoff parameter $\Lambda$=500 MeV. The 
N3LO-J\"ulich model, instead, uses the spectral function 
regularization scheme
with cutoff parameter $\tilde\Lambda$, 
and an exponential regulator 
function in order to remove the divergences in the Lippmann-Schwinger 
equation with cutoff parameter  $\Lambda$. The cutoff combination
$[ \Lambda,\tilde{\Lambda} ]=[ 550,600 ]$ MeV has been chosen in the 
present study. 
Finally, we have considered also 
two effective low-momentum interactions 
which have been derived from the AV18 model 
using 
a renormalization group approach. The 
smooth  cutoff $\Lambda_{low-k}$ of the exponential regulator function 
has been chosen equal to 2 or 3 fm$^{-1}$ 
($V_{low-k,2.0}$ and $V_{low-k,3.0}$)~\cite{bogner:d}. Note that 
these two $V_{low-k}$ models are different than the ones 
adopted in some of the earlier works on 
$V_{low-k}$~\cite{bogner:a,bogner:b,bogner:c,schwenk,nogga:vlowk}, 
where a sharp cutoff formulation was used.

Two remarks are here in order: (i) in the application of the HH expansion
using the NJ2, ISuj, CDBonn, 
N3LO-Idaho, N3LO-J\"ulich, and $V_{low-k}$ potential models, 
a truncation of the model space has been performed at $j_{max}$=6, 
$j_{max}$ being the maximum two-body total angular momentum considered. 
The chosen 
value allows for an accuracy of 1 keV for the triton binding energy. (ii) 
The point-like Coulomb interaction has been added when the  
$^3$He and $^4$He nuclei are considered. 
However, in the case of the AV18 and NJ2 calculations, 
the full electromagnetic interaction has been 
added~\cite{stoks,wiringa:c}.

As already discussed in Sec.~\ref{sec:intro}, realistic models for the 
nuclear Hamiltonian include two-nucleon and three-nucleon interaction
terms. 
Among the available two- and three-nucleon interaction models, 
we have used in the present work the AV18 together with the 
Urbana IX~\cite{pudliner:a}
(AV18/UIX) model and the CDBonn together 
with the Tucson-Melbourne~\cite{tucson} model (CDBonn/TM). 
In the TM model, the regularization cutoff 
parameter $\Lambda_{\rm{TM}}$ has been fixed to
4.795 $m_\pi$, in order to reproduce the triton
binding energy when used in conjunction with the CDBonn potential. 
The parameters $a$, $b$, $c$ and $d$ of the potentials 
are the ones of Table~XI of Ref.~\cite{nogga:c}. However, 
in the $\heq$ calculation, for the sake of
comparison, the CDBonn/TM model has been used with
$\Lambda_{\rm{TM}}=4.784\, m_\pi$, the same as in Ref.~\cite{nogga:b}.

Within an effective field theory approach, three-nucleon 
interactions arise when next-to-next-to leading  order (N2LO) contributions 
are considered. Therefore, the $A=3$ and 4 nuclei have been 
studied also using the N3LO-Idaho two-nucleon together with the 
local N2LO three-nucleon interaction as developed by Navrat\`il in 
Ref.~\cite{navratil:e} (N3LO-Idaho/N2LO). 
The two free parameters in the
N2LO three-nucleon interaction 
model have been chosen from the combination that reproduces
the $A=3,4$ binding energies~\cite{navratil:e}.

\begin{table}[t]
\caption{The triton and the $^4$He binding energies $B$ (MeV), 
calculated with the AV18, CDBonn, N3LO-Idaho, 
two-nucleon 
interaction models, and with the AV18/UIX, CDBonn/TM, 
and N3LO-Idaho/N2LO two- and three-nucleon interactions.  
Note that in the CDBonn/TM case, the parameter $\Lambda_{\rm{TM}}$
has been chosen to be 4.795 $m_\pi$ for $\tri$ and
4.784 $m_\pi$ for $\heq$.
The HH results are compared with the ones 
obtained with different approaches. See 
text for explanations. Note that a recently updated FE/FY 
result~\protect\cite{nogga:private}
for the $^4$He binding energy calculated with the CDBonn potential 
is given in parentheses.
}
\label{tb:comp}
\begin{indented}
\lineup
\item[]
\begin{tabular}{@{}llll}
\br                              
Potential & Method & $B$($^3$H) & $B$($^4$He) \cr
\mr
AV18   & PHH                                & 7.624\e & --    \cr
       & HH                                 & 7.624\e & 24.22\e \cr
       & FE/FY~\protect\cite{nogga:b,nogga:c}  & 7.621\e & 24.23\e \cr
       & FE/FY~\protect\cite{deltuva:c}        & 7.621\e & 24.24\e \cr
       & FE/FY~\protect\cite{lazauskas}        & 7.616\e & 24.22\e \cr
\mr
CDBonn & HH                                 & 7.998\e & 26.13\e \cr
       & FE/FY~\protect\cite{nogga:b,nogga:c,nogga:private}  & 8.005\e 
& 26.23 (26.16)\e \cr
       & FE/FY~\protect\cite{deltuva:d}        & 7.998\e & 26.11\e \cr
       & NCSM~\protect\cite{navratil:d}     & 7.99(1)\n &  \cr
\mr
N3LO-Idaho & HH                                  & 7.854\e & 25.38\e \cr
           & FE/FY~\protect\cite{nogga:private}   & 7.854\e & 25.37\e \cr
           & FE/FY~\protect\cite{deltuva:d}         & 7.854\e & 25.38\e \cr
           & NCSM~\protect\cite{navratil:e}      & 7.852(5)& 25.39(1) \cr
\mr
AV18/UIX    & PHH                           & 8.479\e  & $-$   \cr
            & HH                            & 8.479\e  & 28.47\e \cr
       & FE/FY~\protect\cite{nogga:b,nogga:c}  & 8.476\e  & 28.53\e \cr
            & FE/FY~\protect\cite{lazauskas}   & 8.473\e  &  \cr
\mr
CDBonn/TM   & HH                            & 8.474\e  &  29.00\e \cr
       & FE/FY~\protect\cite{nogga:b,nogga:c}  & 8.482\e  & 29.09\e \cr
\mr
N3LO-Idaho/N2LO & HH                             & 8.474\e  & 28.37\e   \cr
                & NCSM~\protect\cite{navratil:d} & 8.473(5) & 28.34(2) \cr
\br
\end{tabular}
\end{indented}
\end{table}

The $A=3$ and 4 binding energies and ground-state properties are listed in 
Tables~\ref{tb:comp}, \ref{tb:h3}, \ref{tb:he3}, and~\ref{tb:he4}.
In Table~\ref{tb:comp}, using some representative models among 
the ones quoted above, i.e. the 
AV18, CDBonn, N3LO-Idaho, AV18/UIX, CDBonn/TM and 
N3LO-Idaho/N2LO, 
the PHH and HH results for the triton and 
$^4$He binding energies are compared with those
ones obtained with other approaches.
In particular, we have referred to the
FE (for $A=3$) and FY (for $A=4$)
approach of Refs.~\cite{nogga:b,deltuva:c,deltuva:d,lazauskas},
and the NCSM approach of 
Ref.~\cite{navratil:c,navratil:d,navratil:e}. 

From inspection of Table~\ref{tb:comp}, we can conclude that for 
any of the considered realistic potential models, the agreement 
between the HH or PHH results and those from the other techniques
here considered is excellent. Such an agreement 
is present for both local and non-local potentials. 
To be noticed that the AV18 result of 
Ref.~\cite{lazauskas} does not include the $T=3/2$ contribution. Moreover,
the CDBonn and CDBonn/TM results of Ref.~\cite{nogga:b}
include the $n-p$ mass difference. Once this contribution is subtracted,
the agreement with the other methods is quite good. 
In the same row of the table
a recently updated result for $^4$He is given in 
parentheses~\cite{nogga:private}.

\begin{table}[t]
\caption{The triton binding energies $B$ (MeV), 
the proton $r_p$ and neutron $r_n$ radii (fm),
the expectation values of the kinetic energy operator 
$\langle T\rangle$ (MeV), the 
mixed-symmetry $S'$, $P$, $D$ and $T=3/2$ probabilities (all in \%)
calculated with the AV8$'$, AV14, AV18, NJ2, ISuj, CDBonn, N3LO-Idaho, 
N3LO-J\"ulich, $V_{low-k,2.0}$ and $V_{low-k,3.0}$ two-nucleon 
interaction models, and with the AV18/UIX, CDBonn/TM (with
$\Lambda_{\rm{TM}}=4.795\, m_\pi$), 
and N3LO-Idaho/N2LO two- and three-nucleon interactions.  
}
\label{tb:h3}
\begin{indented}
\lineup
\item[]
\begin{tabular}{@{}lcccccccc}
\br                              
Potential & $B$ & $\langle T\rangle$ & $r_p$ & $r_n$  
& $P_{S'}$ & $P_{P}$ & $P_{D}$ & $P_{T=3/2}$ \cr
\mr
AV8$'$          & 7.767 & 47.605 & 1.642 & 1.807 & 1.273 & 0.067 & 8.579 & 0.0     \cr
AV14            & 7.684 & 45.678 & 1.667 & 1.828 & 1.126 & 0.076 & 8.967 & 0.0     \cr
AV18            & 7.624 & 46.727 & 1.653 & 1.824 & 1.293 & 0.066 & 8.510 & 0.0025  \cr
NJ2             & 7.651 & 47.520 & 1.647 & 1.816 & 1.292 & 0.064 & 8.329 & 0.0032  \cr
ISuj            & 8.475 & 32.950 & 1.568 & 1.715 & 1.433 & 0.026 & 4.776 & 0.0028  \cr
CDBonn          & 7.998 & 37.630 & 1.618 & 1.771 & 1.310 & 0.047 & 7.018 & 0.0049  \cr
N3LO-Idaho      & 7.854 & 34.555 & 1.655 & 1.808 & 1.365 & 0.037 & 6.312 & 0.0009  \cr
N3LO-J\"ulich   & 7.292 & 43.467 & 1.726 & 1.892 & 1.690 & 0.018 & 4.316 & 0.0017 \cr
V$_{low-k,2.0}$ & 8.597 & 29.509 & 1.574 & 1.713  & 1.330 & 0.018 & 4.062 & 0.0016  \cr
V$_{low-k,3.0}$ & 8.085 & 32.790 & 1.614 & 1.765  & 1.317 & 0.042 & 6.749 & 0.0020  \cr
\mr
AV18/UIX        & 8.479 & 51.275 & 1.582 & 1.732 & 1.054 & 0.135 & 9.301 & 0.0025  \cr
CDBonn/TM       & 8.474 & 39.364 & 1.580 & 1.722 & 1.202 & 0.101 & 6.971 & 0.0049  \cr
N3LO-Idaho/N2LO & 8.474 & 36.482 & 1.611 & 1.752 & 1.242 & 0.121 & 6.815 & 0.0009  \cr
\mr
Exp.    & 8.482 &        & 1.60  &       &       &       &       &         \cr
\br
\end{tabular}
\end{indented}
\end{table}

In Tables~\ref{tb:h3}, \ref{tb:he3}, and~\ref{tb:he4} we present our
results for the $A=3,4$ bound states properties, among which the 
expectation value of the kinetic energy, 
the proton and neutron radii and different occupation
probabilities. In the case of inclusion of NN
forces only, we notice that all the potentials, except for the ISuj
model, do not reproduce
the experimental trinucleon binding energy. The ISuj model has been
constructed modifying the off-energy shell part of AV18 in order to 
reproduce the NN scattering data and also some $A=3$ observables, as the
trinucleon binding energy. This model predicts a quite low 
$D$-state probability as the models derived from $\chi$PT. The binding
energy and the other quantities predicted by the $V_{low-k}$ models depend 
sizably on
$\Lambda_{low-k}$. Here, we have chosen two representative values of
such a parameter.

In the cases in which a 3N interaction 
is included, the experimental binding  energy is well reproduced, since one of
the parameter of such a term is chosen accordingly. 
In all the calculations the 
$n-p$ mass difference contribution has not been included. This contribution 
can be estimated perturbatively by taking the mean value of the operator
\begin{equation}
  K_{\Delta } =\sum_{i=1,A}
  {1\over 2} \left( {{1\over 2m_p}- {1\over 2 m_n} }\right )
  \nabla_i^2 \;\; \tau_z(i) \ .
  \label{eq:dima}
\end{equation}
This effect has been found to be quite tiny. For example, 
using the AV18 interaction model, 
the change in binding energy of
$\tri$, $\het$, and 
$\heq$ has been found to be about $+6$ keV, $-6$ keV, and  $-0.15$ keV,
respectively~\cite{nogga:b,viviani:b}. 
By inspecting the tables, it can be observed the presence of 
quite large differences
in the occupation probabilities and the expectation value of the
kinetic energy for the different models. For example, the mean value
of the kinetic energy for CDBonn, ISuj,  N3LO-Idaho and $V_{low-k}$-type potentials
is noticeably smaller than that found with the AV18 or NJ2 potentials. 
This is due to the fact that
the repulsion at short interparticle distances is softer for the former 
potentials than for the latter ones. Also the
percentages of the $P$- and $D$-waves are significantly smaller for the former 
potentials. This fact has an important consequence in the application  of the
HH method, since for potentials with a softer repulsion at short interparticle
distances the convergence is usually faster. 
For example, for $A=4$, the convergence has been reached
including HH states up to $G_{\rm max}=80, 52, 32, 22$ for AV18, CDBonn,
N3LO-Idaho, and $V_{low-k}$ potential models, respectively.

\begin{table}[t]
\caption{Same as Table~\protect\ref{tb:h3}, but for $^3$He.}
\label{tb:he3}
\begin{indented}
\lineup
\item[]
\begin{tabular}{@{}lcccccccl}
\br                              
Potential & $B$ & $\langle T\rangle$ & $r_p$ & $r_n$ 
& $P_{S'}$ & $P_{P}$ & $P_{D}$ & $P_{T=3/2}$ \cr
\mr
AV8$'$          & 7.108 & 46.696 & 1.846 & 1.666 & 1.470 & 0.066 & 8.542 & 0.0\n\n\n    \cr
AV14            & 7.033 & 44.813 & 1.867 & 1.691 & 1.315 & 0.076 & 8.967 & 0.0\n\n\n    \cr
AV18            & 6.925 & 45.685 & 1.872 & 1.678 & 1.530 & 0.065 & 8.467 & 0.0080 \cr
NJ2             & 6.994 & 46.607 & 1.857 & 1.668 & 1.491 & 0.063 & 8.282 & 0.0083 \cr
ISuj            & 7.712 & 32.194 & 1.757 & 1.590 & 1.643 & 0.026 & 4.795 & 0.0059 \cr
CDBonn          & 7.263 & 36.767 & 1.819 & 1.637 & 1.542 & 0.046 & 7.000 & 0.0109 \cr
N3LO-Idaho      & 7.128 & 33.789 & 1.855 & 1.675 & 1.607 & 0.037 & 6.313 & 0.0062\n \cr
N3LO-J\"ulich   & 6.590 & 42.471 & 1.948 & 1.751 & 1.985 & 0.018 & 4.327 & 0.0095\n \cr
V$_{low-k,2.0}$ & 7.847 & 28.861 & 1.755 & 1.591 & 1.539 & 0.018 & 4.086 & 0.0063\n \cr
V$_{low-k,3.0}$ & 7.358 & 32.087 & 1.810 & 1.632 & 1.537 & 0.042 & 6.736 & 0.0073\n \cr
\mr
AV18/UIX        & 7.750 & 50.211 & 1.771 & 1.602 & 1.242 & 0.132 & 9.248 & 0.0075 \cr
CDBonn/TM       & 7.720 & 38.495 & 1.767 & 1.597 & 1.409 & 0.099 & 6.966 & 0.0106 \cr
N3LO-Idaho/N2LO & 7.733 & 35.745 & 1.794 & 1.628 & 1.450 & 0.119 & 6.818 & 0.0057 \cr
\mr
Exp.    & 7.718 &        & 1.77  &       &       &       &       &         \cr
\br
\end{tabular}
\end{indented}
\end{table}

The protonic radii are well reproduced by the models including a 3N
interaction. To be noticed that the ISuj and $V_{low-k,2.0}$ models, which 
reproduce quite well the $A=3,4$ binding energies with only two-body forces, 
tend to predict too small protonic radius. This could be related to the 
corresponding small $D$-wave percentage.

A few comments on the estimated percentages of the $T=3/2$ and $T> 0$
components in the $A=3$ and 4 systems are in order.
For the percentage of the $T=1$ component 
in the $\heq$ ground state wave function, 
50\%  of it is due to the effect of the Coulomb interaction between
the protons, the remaining 50\% is due to the 
CSB terms in the nuclear interaction. This can be seen from
Table~\ref{tb:he4}, since the models where the CSB is absent (AV8$'$ and AV14)
predict a percentage approximately half of those calculated with AV18, CDBonn,
N3LO-Idaho, etc., which were fitted to both $pp$ and $np$ data including CSB
by construction. The difference in $P_{T=1}$ for 
these latter models is however small, due to the important role of the Coulomb
interaction.
On the contrary, the $T=3/2$ component in triton and the 
$T=2$ component in $\heq$ are largely dominated by CSB of nuclear
origin (different pion masses, etc.). The values reported in
Table~\ref{tb:h3} and~\ref{tb:he4} show that, depending on the interaction, 
rather  different values for the triton $P_{T=3/2}$ and 
the $\heq$ $P_{T=2}$ 
are obtained (note that the standard models of 3N have
little effect on the isospin admixtures~\cite{viviani:b}).
The origin of the rather
large differences found for the triton $P_{T=3/2}$ and the $\heq$ 
$P_{T=2}$ (a factor 5 between CDBonn and
N3LO-Idaho) must be related to quite different off-shell behavior of
the CSB terms of the interactions. The 
$T=3/2$ component in $^3$He is also affected by the 
Coulomb interaction, which strongly reduces 
the difference between the $P_{T=3/2}$ results.
The knowledge of the $T=1$ and $2$ 
percentages is important for parity-violating electron scattering experiments
on $\heq$, aimed at studying admixture of strange quark $s\bar s$ pairs
in nucleons and nuclei~\cite{BmK01,RHD94,Acha07}. A preliminary study of this
important aspect has been published in Ref.~\cite{viviani07}.
It could play an important role also in the study of the
reaction $d+d\rightarrow \alpha+\pi^0$. This reaction is possible only if 
isospin symmetry is violated, namely it probes directly the CSB terms in the
nuclear Hamiltonian~\cite{IUCF2,Gea04}. 

Finally, we would like to comment on the capability of these potential models
in reproducing simultaneously the $A=3,4$ binding energies. It is well known
that there is an almost linear relation between the $\tri$ and $\heq$ binding 
energies. By inspection of Tables~\ref{tb:h3}, \ref{tb:he3}, 
and~\ref{tb:he4} we observe that the models which reproduce the $\tri$ binding
energy (ISuj, AV18/UIX, CDBonn/TM, and N3LO-Idaho/N2LO) slightly
overpredict the $\heq$ binding energy. 
This overprediction is of about $0.6$ MeV
for ISuj and CDBonn/TM and reduces to $0.16$ MeV for AV18/UIX or to $0.06$ MeV
for N3LO-Idaho/N2LO. Another important aspect is the
correct description of the mass difference between the
$^3$H and $^3$He, directly related to the CSB components in the Hamiltonian.
As already mentioned, the present calculations do not include the $n-p$ mass 
difference. Taking it into account together with other small contributions, as
explained in Ref.~\cite{nogga:b}, the AV18/UIX theoretical estimate for the mass
difference $D$ is around $750$ keV, slightly smaller than the experimental one
$D_{\rm exp}=764$ keV. 
The $\tri$-$\het$ mass difference values for the CDBonn/TM and
N3LO-Idaho/N2LO models are $D=754$ keV and $741$ keV,
respectively, as can be seen from Tables~\ref{tb:h3} and \ref{tb:he3}. 
Taking into accout the $n-p$ mass difference contribution ($\approx 12$ keV), it
can be seen that the CDBonn/TM well reproduces $D_{\rm exp}$, 
whereas the N3LO-Idaho/N2LO model gives a slightly smaller value. 
For the ISuj
model, the value $D=763$ keV coincides with $D_{\rm exp}$.
To be noticed that additional contributions could come from CSB terms in the
3N interaction. However, the 3N models considered in the present analysis do not
contain terms which violate isospin symmetry.

\begin{table}[t]
\caption{
The $\heq$ binding energies $B$ (MeV), 
the proton $r_p$ and neutron $r_n$ radii (fm),
the expectation values of the kinetic energy operator 
$\langle T\rangle$ (MeV), the 
$P$, $D$, $T=1$, and $T=2$ probabilities (all in \%)
calculated with the AV8$'$, AV14, AV18, NJ2, ISuj, CDBonn, N3LO-Idaho, 
N3LO-J\"ulich, $V_{low-k,2.0}$ and $V_{low-k,3.0}$ two-nucleon 
interaction models, and with the AV18/UIX, CDBonn/TM (with
$\Lambda_{\rm{TM}}=4.784\, m_\pi$), 
and N3LO-Idaho/N2LO two- and three-nucleon interactions. 
}
\label{tb:he4}
\begin{indented}
\lineup
\item[]
\begin{tabular}{@{}lcccccccc}
\br                              
Potential & $B$ & $\langle T\rangle$ & $r_p$ & $r_n$  
          & $P_{P}$ & $P_{D}$ & $P_{T=1}$ & $P_{T=2}$ \cr
\mr
AV8$'$     & 24.90 & \n98.80 & 1.507 & 1.502 & 0.408 & 14.50 & 0.0013 & 0.0001 \cr
AV14       & 24.23 & \n95.76 & 1.526 & 1.521 & 0.393 & 14.41 & 0.0013 & 0.0001 \cr
AV18       & 24.21 & \n97.84 & 1.514 & 1.509 & 0.347 & 13.74 & 0.0028 & 0.0052 \cr
NJ2        & 24.42 & 100.27  & 1.506 & 1.501 & 0.334 & 13.37 & 0.0016 & 0.0074 \cr
ISuj       & 28.97 & \n68.66 & 1.383 & 1.378 & 0.097 & \n6.10 & 0.0020 & 0.0072 \cr
CDBonn     & 26.13 & \n77.58 & 1.458 & 1.453 & 0.223 & 10.74 & 0.0029 & 0.0108 \cr
N3LO-Idaho & 25.38 & \n69.24 & 1.518 & 1.513 & 0.172 & \n9.29 & 0.0035 & 0.0024 \cr
N3LO-J\"ulich   & 22.20 & \n86.85 & 1.636 & 1.631 & 0.081 & \n6.22 & 0.0065 & 0.0063 \cr
V$_{low-k,2.0}$ & 29.66 & \n60.77 & 1.389 & 1.384 & 0.064 & \n4.69 & 0.0020 & 0.0023 \cr
V$_{low-k,3.0}$ & 26.57 & \n65.68 & 1.455 & 1.450 & 0.201 & 10.04  & 0.0025 & 0.0038 \cr
\mr
AV18/UIX        & 28.46 & 113.30 & 1.430 & 1.425 & 0.732 & 16.03 & 0.0025 & 0.0050 \cr
CDBonn/TM       & 29.00 & \n84.56 & 1.396 & 1.391 & 0.454 & \n9.94 & 0.0021 & 0.0105 \cr
N3LO-Idaho/N2LO & 28.36 & \n74.93 & 1.476 & 1.471 & 0.608 & 10.79  & 0.0028 & 0.0020 \cr
\mr
Exp. & 28.30 & & 1.47 & & & & & \cr
\br
\end{tabular}
\end{indented}
\end{table}

\subsection{$A=3$ and 4 Scattering States}
\label{subsec:ss}

The $A=3$ and 4 zero-energy scattering states have been studied 
with the following interaction models: 
the two-nucleon AV14, AV18 and 
N3LO-Idaho, and the two- and three-nucleon  
AV18/UIX and N3LO-Idaho/N2LO models. In a consistent
calculation of the $A=3,4$ scattering lengths, the electromagnetic
interaction cannot be disregarded. 
In particular, the magnetic moment interaction
gives contribution in processes in which the incident nucleon is either
a proton or a neutron. In the first case, besides the Coulomb and the
magnetic moment interactions,
other contributions of long range as vacuum polarization and 
two-photon-exchange are present.
The treatment of these terms remains outside the aims
of the present work. Accordingly, in the description of
$pd$ and $p^3$He, we consider only the point Coulomb interaction
and, in some cases, the magnetic moment interaction too.

The results for the $nd$ and $pd$ doublet and quartet scattering lengths 
are given in Tables~\ref{tb:ndsl} and~\ref{tb:pdsl} and are compared with 
those obtained by other techniques~\cite{av14nd,av18nd} and the 
available experimental data~\cite{ndexp71,ndexp03}.
The $pd$ calculation has been performed with and without the inclusion of the 
$T=3/2$ components of 
the wave function (see Eq.~(\ref{eq:psica})). Furthermore, we performed
two different calculations using the AV18 potential since, by construction,
it includes the long range electromagnetic terms.
In the $nd$ case, the results labelled as AV18$^*$ do not include 
these terms. In the $pd$ case the AV18$^*$ results include
only the point Coulomb interaction, whereas those labelled as AV18
include the complete electromagnetic potential with the exception
of the two-photon exchange and the vacuum polarization terms.
The calculations have been performed with the PHH expansion in the
$A=3$ case with local potential models, and with the HH
expansion in all other cases. The dimension of the basis and the 
number of the integration 
points have been increased in order to reach an accuracy of $0.002$ fm
in the calculation of the scattering lengths.

From inspection of the tables we can conclude that:
(i) both the $nd$ and $pd$ quartet scattering lengths are almost 
model-independent. 
(ii) The $nd$ doublet scattering length $^2a_{nd}$ is very
sensitive to the choice of the two-nucleon potential model, when no 
3N interaction is included. Once the 3N interaction is included, 
and therefore 
the triton binding energy is well reproduced, $^2a_{nd}$ 
becomes little model-dependent.
This is a well-known
feature, related to the fact that $^2a_{nd}$ and the triton binding energy
are linearly correlated (the so-called
Phillips line~\cite{phillips}). 
(iii) The $pd$ doublet scattering length
$^2a_{pd}$ is positive and quite model-dependent, if only the two-nucleon 
interaction is included. Once the three-nucleon interaction is added, 
$^2a_{pd}$ becomes small. 
Some model-dependence remains, but the problem of extrapolating
to zero energy the experimental results makes very difficult any
meaningful comparison between theory and experiment.
(iv) The $nd$ quartet scattering length $^4a_{nd}$ is in very good 
agreement with the experimental data. Conversely, some disagreement 
is observed in the doublet scattering length $^2a_{nd}$ 
for most of the NN plus 3N interactions considered. The results
presented have been calculated without including the neutron-proton
mass difference. This contribution will further decrease the value of
$^2a_{nd}$. To be noticed that a recent measurement~\cite{ndexp03}
of the coherent $nd$ scattering length predicts 
$^2a_{nd}=[0.645\pm0.003({\rm expt})\pm0.007({\rm theory})]$ fm.
This very precise measurement is not well described by any of the
potential models analyzed in this section, with the exception of the 
N3LO-Idaho/N2LO.

\begin{table}[h]
\caption{The $nd$ doublet and quartet 
scattering lengths $^2a_{nd}$, $^4a_{nd}$, in fm, are 
calculated with the AV14, AV18, AV18$^*$, and N3LO-Idaho two-nucleon 
interaction models, and with the AV14/TM (with $\Lambda_{\rm TM}=5.13\,m_\pi$), 
AV18/UIX, AV18$^*$/UIX, and N3LO-Idaho/N2LO 
two- and three-nucleon interactions, 
and are compared with the results obtained with the FE 
technique~\protect\cite{av14nd,av18nd}
and the available experimental 
data~\protect\cite{ndexp71,ndexp03}.}
\label{tb:ndsl}
\begin{indented}
\lineup
\item[]
\begin{tabular}{@{}llcc}
\br                              
Potential & Method & $^2a_{nd}$ & $^4a_{nd}$ \cr
\mr
AV14 & PHH                            & 1.189 & 6.379 \cr
AV14 & FE~\protect\cite{av14nd}      & 1.204 & 6.380 \cr
AV18 & PHH                            & 1.258 & 6.345 \cr
AV18 & FE~\protect\cite{av18nd}      & 1.248 & 6.346 \cr
AV18$^*$ & PHH                        & 1.275 & 6.325 \cr
AV18$^*$ & FE~\protect\cite{av18nd}  & 1.263 & 6.326 \cr
N3LO-Idaho & HH                      & 1.100 & 6.342 \cr
\mr
AV14/TM & PHH                         & 0.586 & 6.371 \cr
AV18/UIX & PHH                        & 0.590 & 6.343 \cr
AV18/UIX & FE~\protect\cite{av18nd}  & 0.578 & 6.347 \cr
AV18$^*$/UIX & PHH                    & 0.610 & 6.323 \cr
AV18$^*$/UIX & FE~\protect\cite{av18nd}& 0.597 & 6.326 \cr
N3LO-Idaho/N2LO & HH                 & 0.675 & 6.342 \cr
\mr 
Exp.~\protect\cite{ndexp71} &     & 0.65$\pm$0.04 & 6.35$\pm$0.02 \cr
Exp.~\protect\cite{ndexp03} &     & 0.645$\pm$0.003$\pm$0.007  &     \cr
\br
\end{tabular}
\end{indented}
\end{table}

\begin{table}[h]
\caption{Same as Table~\protect\ref{tb:ndsl} but for 
$pd$. Also listed are the results obtained without the 
inclusion of the $T=3/2$ component in the 
wave functions ($^2a_{nd}(T=1/2)$ and $^4a_{nd}(T=1/2)$).} 
\label{tb:pdsl}
\begin{indented}
\lineup
\item[]
\begin{tabular}{@{}llcccc}
\br                              
Potential & Method
& $^2a_{pd}$ & $^4a_{pd}$ 
& $^2a_{pd}(T=1/2)$ & $^4a_{pd}(T=1/2)$ \cr
\mr
AV14 & PHH        & 0.937  & 13.773  & 0.941 & 13.773 \cr
AV14 & FE~\protect\cite{av14nd}& & & 0.965 & 13.764 \cr
AV18 & PHH        & 1.134 & 13.662 & 1.150 & 13.662 \cr
AV18$^*$ & PHH     & 1.185 & 13.588 & 1.198 & 13.589 \cr
N3LO-Idaho & HH  & 0.876 & 13.646 & 0.866 & 13.646 \cr
\mr
AV18/UIX & PHH    &-0.089 & 13.662 &-0.074 & 13.663 \cr
AV18$^*$/UIX & PHH &-0.035 & 13.588 &-0.019 & 13.590 \cr
N3LO-Idaho/N2LO & HH & 0.072 & 13.647 & 0.082 & 13.647 \cr
\br
\end{tabular}
\end{indented}
\end{table}

The results for the $n^3$H and $p^3$He singlet and triplet 
scattering lengths are given in Table~\ref{tb:a4sl}
and are compared with other theoretical calculations, performed using the FY
equations method~\cite{lazauskas,deltuva:d}, and the available experimental  
data~\cite{Rauch85,Hale90,alley,tegner}. 
For $n^3$H, the scattering lengths can be
obtained from the experimental values of the total cross section $\sigma_T$
and the coherent scattering length $a_c$,
\begin{equation} 
   \sigma_T= \pi (|{}^1a|^2+3|{}^3a|^2)\ , \quad 
    a_c={1\over4} {}^1a + {3\over4}{}^3a\ . \label{eq:def}
\end{equation}
The  $n\tri$ cross section has been accurately measured over a wide energy range
and the extrapolation to zero energy does not present any problem. The 
value obtained is $\sigma_T=[1.70\pm 0.03]$ b~\cite{Phill80}. The coherent
scattering  length has been measured by neutron-interferometry
techniques. The most recent values reported in the literature 
are  $a_c=[3.59\pm0.02]$ fm~\cite{Rauch85} and $a_c=[3.607\pm0.017]$ fm, the 
latter value being  
obtained from $p\het$ data using an approximate
Coulomb-corrected R-matrix theory~\cite{Hale90}.
However, in the ${}^1a$-${}^3a$ plane, the ellipse defined 
by the total cross section and corresponding to the experimental value of
$\sigma_T=1.70$ b and  the straight line corresponding
to the coherent scattering length  $a_c\approx 3.6$ fm
are almost tangent. Therefore, a small change in the $a_c$ value
produces a large variation of ${}^1a$ and ${}^3a$. This is also the
reason for the large uncertainty in the values reported
in Table~\ref{tb:a4sl}.  

Note that there is a fairly good agreement
between the results of the various theoretical calculations.
The calculated scattering lengths show a scaling behaviour
with respect to the $\tri$ binding energy similar to the Phillips line, 
as discussed in Ref.~\cite{viviani:e}. The total cross sections obtained with
the AV18/UIX and N3LO-Idaho/N2LO models are 1.73 b and 1.68 b, in agreement
with the experimental value. The theoretical coherent scattering lengths
are $3.73$ fm and $3.65$ fm, respectively, at variance with respect to the
experimental value. The origin of this discrepancy is still unclear.

For $p^3$He, scattering length measurements were reported in 
Refs.~\cite{alley,tegner}. However, 
these have rather large errors. In fact, the 
$p^3$He data have been extrapolated to zero energy 
from measurements taken above 1 MeV, 
and therefore suffer of large systematic uncertainties.

\begin{table}[h]
\caption{ 
The $n^3$H, $p^3$He singlet and triplet scattering lengths
$^1a_{n^3{\rm H}}$, $^3a_{n^3{\rm H}}$, $^1a_{p^3{\rm He}}$, 
$^3a_{p^3{\rm He}}$, 
in fm, are 
calculated with the AV18, and N3LO-Idaho two-nucleon 
interaction models, and with the AV18/UIX
and N3LO-Idaho/N2LO two- and three-nucleon interactions. 
The available results 
from the FY equations method~\protect\cite{lazauskas,deltuva:d} and the 
available experimental data~\protect\cite{Rauch85,Hale90,alley,tegner}
are also reported.  
}
\label{tb:a4sl}
\begin{indented}
\lineup
\item[]
\begin{tabular}{@{}llcccc}
\br                              
Potential & Method & $^1a_{n^3{\rm H}}$ & $^3a_{n^3{\rm H}}$ 
& $^1a_{p^3{\rm He}}$ & $^3a_{p^3{\rm He}}$\cr
\mr
AV18 & HH & 4.29 & 3.73 & 12.9 & 10.0 \cr
AV18 & FY~\protect\cite{lazauskas} & 4.27 & 3.71 &  & \cr
AV18 & FY~\protect\cite{deltuva:d} & 4.28 & 3.71 &  & \cr
N3LO-Idaho & HH & 4.20 & 3.67 & 11.5 & 9.2 \cr
N3LO-Idaho & FY~\protect\cite{deltuva:d} & 4.23 & 3.67 &  & \cr
\mr
AV18/UIX & HH & 4.10 & 3.61 & 11.5 & 9.1 \cr
AV18/UIX & FY~\protect\cite{lazauskas} & 4.04 & 3.60 &  & \cr
N3LO-Idaho/N2LO & HH & 3.99 & 3.54 & 11.0 & 8.6 \cr
\mr
Exp.~\protect\cite{Rauch85}      &    & $4.98\pm 0.29$
                                      & $3.13\pm 0.11$ & & \cr
Exp.~\protect\cite{Hale90}       &    & $4.45\pm 0.10$
                                      & $3.32\pm 0.02$ & & \cr
Exp.~\protect\cite{alley}        &    &        
                                      &                & $10.8\pm 2.6$
                                                       & $\n8.1\pm 0.5$ \cr
Exp.~\protect\cite{tegner}       &    &
                                      &                & & $10.2\pm1.5$ \cr
\br
\end{tabular}
\end{indented}
\end{table}

\section{Summary and Conclusions}
\label{sec:concl}

In this article we have presented a theoretical approach suitable
to describe with high precision bound states and zero-energy 
scattering states for
the $A=3,4$ nucleon systems.
A particular attention has been given to the presentation
of the various models available 
for the two- and three-body nuclear interaction. 
In fact, in recent times, different NN and 3N interactions
have been proposed and applied by many authors
to the study of few-nucleon systems. The capability
of these interactions to describe all the complexity of the
nuclear dynamics is at present an intense subject of research.
Bound states and low-energy scattering states are the first
structures to be examined using the interaction models 
under consideration. To this aim, any reliable technique employed 
should allow for meaningful comparison with the available
experimental data. In other words, the solution of the
Schr\"odinger equation should be accurate enough in order
to eliminate uncertainties in the theoretical predictions.

The method discussed in this paper is based on the expansion of the wave function 
over the HH functions and the solution of the Schr\"odinger equation
is obtained by using a variational principle.
Nowadays computing facilities make it possible to consider very 
large sets of HH functions, so that the problem of convergence
can be carefully studied for specific observables as binding
energies or scattering lengths.
Several papers~\cite{kievsky:a,kievsky:e,viviani:b,viviani:c,viviani:d} 
have been devoted
to applying and verifying this approach for different systems, 
and the obtained results have been successfully compared to
those of other accurate techniques.

There are two main motivations behind the present work. The first one is
to highlight the details of the HH method and its capability 
and accuracy in studying $A=3$ and 4 nuclear systems. The extension to larger
$A$-values is an important problem and studies in this direction are 
currently in progress. The second 
motivation is that the results  presented here for 
different types of interaction models should turn out 
to provide a useful comparison for other
available or future techniques. 
The main merit of the HH method is that it can be implemented to treat
the $A=3,4$ systems with interactions given in configuration or
in momentum space including the electromagnetic potential. In particular,
the description of low-energy scattering states with charged
particles does not arise particular difficulties and, in fact,
the HH technique has been applied successfully for the description
of the $pd$ capture reaction~\cite{viviani:f,marcucci:b}
and the $p^3$He weak capture reaction~\cite{marcucci:a,park:a}
in the keV region.

From the results here presented we can conclude that there are some
problems, not yet completely solved, in the theoretical description 
of the $A=3,4$
systems. The simultaneous reproduction of the $^3$H, $^3$He and 
$^4$He binding energies is not totally satisfactory. Several
combinations of modern NN plus 3N interaction models tend to overestimate the
$\alpha$-particle binding energy, once the 3N interaction 
strength has been fixed to
reproduce the triton binding energy. The nucleons 
in the $\alpha$-particle are in average very close to each other, so that
a greater sensitivity to 3N interaction 
terms should be manifested in this system.
The N2LO 3N interaction model in conjunction with the 
N3LO-Idaho NN interaction gives
the better description of the three binding energies mentioned above.
This model contains a number of operators very similar to 
those present in the TM and UIX model, but having different 
relative strengths. The overestimation given by the 
CDBonn/TM potential is appreciable, suggesting that a further analysis
of the TM model could be useful.

Another problem is the underestimation of the $nd$
doublet scattering length by all the models considered except
the N3LO-Idaho/N2LO.
A high precision measurement of the coherent $nd$ scattering length
is available and, from this quantity, the $nd$ doublet scattering
length can be evaluated. This has been done using the 
theoretical estimates for the $nd$ quartet scattering length since 
this quantity results to be almost model independent. In any case,
after considering the theoretical uncertainty introduced, the theoretical
estimates given in Table~\ref{tb:ndsl} are lower with respect to the
mean experimental value by $5\;$\% to $10\;$\%. This is not the case 
for the N3LO-Idaho/N2LO potential. It seems that this model is the 
only one among those studied here able to reproduce simultaneously the 
experimental values of the $A=3$ and 4 binding energies 
and the $A=3$ scattering lengths.
 
Although many efforts have been done, and are still in progress, for 
a more accurate determination of the nuclear interaction, a long way 
seems to be in front of us. To this aim, the extension of the HH method 
to treat systems with $A>4$ and scattering states at medium and high 
energies will be very useful. Intense efforts are pursued 
in this directions by the authors.

\appendix


\section{The Transformation Coefficients for $A=3$}\label{sec:rare3}

Let us discuss how to compute the transformation coefficients 
(TC) for $A=3$. It is convenient to start
from  Eq.~(\ref{eq:traco2}), which can be written as
\begin{equation}   
    a_{\ell_1\ell_2n_2,\ell_1^\prime\ell_2^\prime n_2^\prime}^{(p),G,L} =
   \int d\Omega_2
    \left[{}^{(3)}{\cal H}_{\{\ell_1'\ell_2' n_2'\},LM}(\Omega_2)\right]^\dag
    \;\;{}^{(3)}{\cal H}_{\{\ell_1\ell_2 n_2\},LM}(\Omega_2^{(p)})
    \ ,
    \label{eq:traco2b}
\end{equation}
where the functions ${}^{(3)}{\cal H}_{\{\ell_1\ell_2 n_2\},LM}(\Omega_2)$
are defined in Eq.~(\ref{eq:hh3}). First of all, let us recall the following
identity satisfied by the spherical harmonics
\begin{equation}
  c^\ell Y_{\ell,m}(\hat c)=\sum_{\ell_a+\ell_b=\ell} a^{\ell_a}
     b^{\ell_b} \sqrt{4\pi} D_{\ell,\ell_a,\ell_b}
     \left[ Y_{\ell_a}(\hat a) Y_{\ell_b}(\hat b)\right]_{\ell,m}\ ,
  \label{eq:abc}
\end{equation}
where the vector variables are related by ${\bf c}={\bf a}+{\bf b}$ and
\be
  D_{\ell,\ell_a,\ell_b}=
    \sqrt{(2\ell+1)! \over (2\ell_a+1)! (2\ell_b+1)!}\ .
\ee
The Jacobi vectors constructed with permutation $p$ are linearly related
to the Jacobi vectors $\jacb_{1},\jacb_2$ for the permutation 1
(corresponding to the order $1,2,3$ of the particles), namely
\begin{equation}
  \jacb_{i}^{(p)}=\sum_{j=1,2} \alpha_{ij}^{(p)} \jacb_j\ ,\qquad i=1,2\ ,
  \label{eq:jacjac}
\end{equation}
where $\alpha_{ij}^{(p)}$ are numerical coefficients. It can be
shown that
\bea
  \lefteqn{(\sin\hypfi_2^{(p)})^{\ell_1} (\cos\hypfi_2^{(p)})^{\ell_2}
   \left[ Y_{\ell_1}(\hat \jac_1^{(p)}) Y_{\ell_2}(\hat
   \jac_2^{(p)})\right]_{L,M}=}  &&    \nonumber\\
   \qquad\qquad\qquad\sum_{\ell_1'\ell_2'}
   C^{(p)}_{\ell_1\ell_2,\ell_1'\ell_2'}(\sin\hypfi_2,\cos\hypfi_2)
   \left[ Y_{\ell_1}(\hat \jac_1) Y_{\ell_2}(\hat
   \jac_2)\right]_{L,M} \ , \label{eq:appa1}
\eea
where 
\bea
C^{(p)}_{\ell_1\ell_2,\ell_1'\ell_2'}(x,y) &=&
  \sum_{\lambda_1+\lambda_2=\ell_1} \sum_{\lambda_1'+\lambda_2'=\ell_2}
     x^{\lambda_1+\lambda_1'} y^{\lambda_2+\lambda_2'} \nonumber \\
    &\times&  (\a_{11}^{(p)})^{\lambda_1} (\a_{12}^{(p)})^{\lambda_2}
     (\a_{21}^{(p)})^{\lambda_1'} (\a_{22}^{(p)})^{\lambda_2'}\nonumber \\       
    &\times&  D_{\ell_1,\lambda_1,\lambda_2}
       D_{\ell_2,\lambda_1',\lambda_2'}
     (-)^{\lambda_1+\lambda_1'+\lambda_2+\lambda_2'}
      \hat\ell_1\hat\ell_2\hat\ell_1'\hat\ell_2'
      \hat\lambda_1\hat\lambda_2\hat\lambda_1'\hat\lambda_2' \nonumber \\
     &\times&   \left( \begin{array}{ccc}
        \lambda_1 & \lambda_1' & \ell_1' \\
        0 & 0 & 0 
       \end{array}   \right)
        \left( \begin{array}{ccc}
        \lambda_2 & \lambda_2' & \ell_2' \\
        0 & 0 & 0 
        \end{array}   \right)
           \left\{ \begin{array}{ccc}
            \lambda_1 & \lambda_2 & \ell_1 \\
            \lambda_1'  & \lambda_2'  & \ell_2 \\
            \ell_1'  & \ell_2' & L
           \end{array} \right\}\ , \label{eq:appa2}
\eea
and $\hat\ell=\sqrt{2\ell+1}$. 

The integrand in Eq.~(\ref{eq:traco2b}) depends on the angular variables
$\hat\jac_1$ and $\hat\jac_2$ through the spherical harmonics and the
argument $\cos2\hypfi_2^{(p)}$ of the Jacobi polynomial, which is
given by
\bea
  \cos2\hypfi_2^{(p)}= 2&&\Bigl[ (\alpha^{(p)}_{21})^2 (\sin\hypfi_2)^2    
   +(\alpha^{(p)}_{22})^2 (\cos\hypfi_2)^2\nonumber\\
  && \qquad\qquad
    +2 \alpha^{(p)}_{21} \alpha^{(p)}_{22} \mu \sin\hypfi_2 \cos\hypfi_2 
   \Bigr] -1 \ ,
\label{eq:appa3}
\eea
where $\mu=\hat\jac_1\cdot\hat\jac_2$. It is possible to perform 
analytically part of
the angular integration (keeping $\mu$ fixed) with the result
\bea
  \lefteqn{\int d\hat\jac_1\; d\hat\jac_2 \; 
  \left[ Y_{\ell_1'}(\hat \jac_1) Y_{\ell_2'}(\hat
   \jac_2)\right]^\dag_{L,M} 
   \left[ Y_{\ell_1}(\hat \jac_1) Y_{\ell_2}(\hat
   \jac_2)\right]_{L,M} =} && \nonumber \\
   && \qquad\qquad\qquad\qquad\qquad\qquad\qquad\qquad\sum_\lambda
   A_\lambda^{\ell_1'\ell_2'\ell_1\ell_2,L} {1\over 2}
   \int d\mu P_\lambda(\mu) \ ,
   \nonumber 
\eea
where
\bea
   A_\lambda^{\ell_1'\ell_2'\ell_1\ell_2,L}&=&
    (-)^{L+\ell_2+\ell_2'} \widehat{\ell_1}
     \widehat{\ell_2}\widehat{\ell_1'}\widehat{\ell_2'}
     (2\lambda+1) \nonumber\\
    &&\times \left\{ \begin{array}{ccc}
            \ell_1' & \ell_2' & L \\
            \ell_2  & \ell_1  & \lambda 
       \end{array}   \right\}
     \left( \begin{array}{ccc}
        \ell_1' & \ell_1 & \lambda \\
        0 & 0 & 0 
       \end{array}   \right)
     \left( \begin{array}{ccc}
        \ell_2' & \ell_2 & \lambda \\
        0 & 0 & 0 
       \end{array}   \right)\ .
    \label{eq:amu}
\eea
Finally, using Eqs.~(\ref{eq:appa2}) and~(\ref{eq:amu}), we arrive  at
the expression  for the TC
\begin{eqnarray}   
    a_{\ell_1\ell_2n_2,\ell_1^\prime\ell_2^\prime n_2^\prime}^{(p),G,L} &=&
   {\cal N}_{n_2'}^{\ell_2',\nu_2}
   {\cal N}_{n_2}^{\ell_2,\nu_2}
   {1\over 2}\int_0^{\pi\over 2} d\hypfi_2 \int_{-1}^{+1}d\mu\,
    (\cos\hypfi_2)^{2+\ell_2'}
    (\sin\hypfi_2)^{2+\ell_1'} \nonumber\\
   && \times  P_{n_2'}^{\ell_1'+1/2,\ell_2'+1/2}(\cos2\hypfi_2)
    P_{n_2}^{\ell_1+1/2,\ell_2+1/2}(\cos2\hypfi_2^{(p)}) \nonumber\\
   && \times 
    \sum_{\lambda_1,\lambda_2,\lambda}
     C^{(p)}_{\ell_1\ell_2,\lambda_1\lambda_2}(\sin\hypfi_2,\cos\hypfi_2)
     A_\lambda^{\ell_1'\ell_2'\lambda_1\lambda_2,L}
     P_\lambda(\mu)
    \ ,
    \label{eq:traco3}
\end{eqnarray}
where ${\cal N}_{n_2}^{\ell_2,\nu_2}$ is given in Eq.~(\ref{eq:nhh})
and $\nu_2=G+2$.
This integral can be calculated easily using Gauss quadrature, since it is a
polynomial in $\mu$, $\cos\hypfi_2$ and $\sin\hypfi_2$.


\section{The Transformation Coefficients for $A=4$}\label{sec:arare}

Let us now consider the calculation of the 
transformation coefficients (TC) for $A=4$ ($N=3$) for a generic
permutation $p$ of the particles.
In the following, the hypervariables constructed starting from
a generic choice of the Jacobi vectors 
$\jacb_1^{(p)}$, $\jacb_2^{(p)}$, $\jacb_3^{(p)}$, 
corresponding to the permutation $p$ of the particles
and either set $A$ or $B$ (see Eq.~(\ref{eq:JcbV})),  
will be denoted by $\Omega_3^{(p)}$.
We will also consider the hyperangular variables 
$\Omega_3$ constructed in terms of a reference set of Jacobi vectors
$\jacb_1$, $\jacb_2$, $\jacb_3$
corresponding for example to those defined in set $A$ 
and to the order $1$, $2$, $3$, $4$ of the particles.
Let us define
\begin{equation}
  y_p=\cos 2\phi_{2}^{(p)}\ , \quad z_p=\cos 2\phi_{3}^{(p)}\ ,
  \label{eq:xz}
\end{equation}
where the hyperangles $\phi_{2}^{(p)}$, $\phi_{3}^{(p)}$ are defined in
Eq.~(\ref{eq:phi}) in terms of the moduli of the Jacobi vectors. 
In terms of these variables, the
expression of a generic $A=4$ HH function is   
\begin{eqnarray}
  {}^{(4)}{\cal H}^G_{\{\ell_1,\ell_2,\ell_3, L_2 ,n_2,
    n_3\},LM}(\Omega_3^{(p)}) & = & 
   \left [ \Bigl ( Y_{\ell_1}(\hat \jac_{1}^{(p)}) 
    Y_{\ell_2}(\hat \jac_{2}^{(p)}) \Bigr )_{L_2}  Y_{\ell_3}(\hat
    \jac_{3}^{(p)}) \right     ]_{LM}
    \nonumber \\
 &\times&  {\cal N}^{\ell_1,\ell_2,\ell_3}_{ n_2, n_3} 
   (1-y_p)^{\ell_1 \over 2}    (1+y_p)^{\ell_2 \over 2} 
    \nonumber \\
   &\times& (1-z_p)^{\ell_1+\ell_2+2n_2 \over 2}    (1+z_p)^{\ell_3\over 2}    
    \nonumber \\
   &\times& P^{\ell_1+{1\over 2}, \ell_2+{1\over 2}}_{n_2}(y_p)
      P^{\ell_1+\ell_2+2n_2+2, \ell_3+{1\over 2}}_{n_3}(z_p)\ ,
      \label{eq:hh4b}
\end{eqnarray}
where the normalization factor is
\begin{eqnarray}
  {\cal N}^{\ell_1,\ell_2,\ell_3}_{ n_2, n_3} &=&
   \left( {1\over 2}\right)^{\ell_1+\ell_2+n_2+{\ell_3\over2}}\nonumber \\
  &\times& \prod_{j=2}^3 \left [ { 2\nu_j \Gamma(\nu_j-n_j) n_j! \over
   \Gamma(\nu_j-n_j-\ell_j-1/2) \Gamma(n_j+\ell_j+3/2) } \right
   ]^{1\over 2}\ ,
   \label{eq:norma4}
\end{eqnarray}
with $\nu_j=G_j+(3j-5)/2$ and $G_j$ defined in Eq.~(\ref{eq:go}).
The superscript $G$ in Eq.~(\ref{eq:hh4b}) has been inserted to remember that
the possible choices of 
the quantum numbers $\{\ell_1,\ell_2,\ell_3, L_2 ,n_2, n_3\}$ are restricted to
the case $\ell_1+\ell_2+\ell_3+2(n_2+n_3)=G$.

Let us start with  the state having
quantum numbers $n_2=n_3=0$  ($G= G_m=\ell_1+\ell_2+\ell_3$).
Since the values of $G_m$ to be considered are small,  the TC for this state
can be easily calculated by means of any of the methods proposed in the
literature. For example, one can employ the technique of
Ref.~\cite{E95}. This method is based on the fact that
Eq.~(\ref{eq:traco}) has to be verified for any spatial configuration of the
Jacobi vectors set $\{\jacb_i\}$. For given values of $G$ and $L$, the sum
in the right-hand side of Eq.~(\ref{eq:traco}) is over $N_{GL}$ terms. If
the equation is required to be satisfied   for $N_{GL}$ values of the set
$\{\jacb_i\}$, one gets a system of $N_{GL}$ linear equation for the
required TC. The HH functions we are interested in have
$G=G_m=\ell_1+\ell_2+\ell_3\leq 6$, therefore the number  $N_{GL}$ of
linear equations is small and a standard numerical technique can be
employed.

The Jacobi vectors 
$\jacb_1^{(p)}$, $\jacb_2^{(p)}$, $\jacb_3^{(p)}$ 
are linearly related
to the Jacobi vectors $\jacb_{1},\jacb_2,\jacb_3$, hence
the following relations hold
\begin{equation}
   (\jac_{3}^{(p)})^2= \sum_{i,j=1}^3 \Gamma^{(p)}_{ij} \jacb_i
   \cdot \jacb_j\ , \qquad
   (\jac_{2}^{(p)})^2 = \sum_{i,j=1}^3 \Delta^{(p)}_{ij} \jacb_i\cdot \jacb_j\ ,
   \label{eq:rij}
\end{equation}
where the numerical coefficients $\Gamma^{(p)}_{ij}$  and $\Delta^{(p)}_{ij}$
can be easily computed by expressing the vectors $\jacb_{3}^{(p)}$  and
$\jacb_{2}^{(p)}$ in terms of $\jacb_1$,  $\jacb_2$ and  $\jacb_3$.

Let us now assume to know the TC for the HH functions ${}^{(4)}{\cal
H}^G_{\{\ell_1,\ell_2,\ell_3, L_2 ,n_2, n_3\},LM}(\Omega_3^{(p)})$ in terms of the
HH functions constructed with $\Omega_3$, namely
\bea
  {}^{(4)}{\cal H}^G_{\{\ell_1,\ell_2,\ell_3, L_2 ,n_2,
    n_3\},LM}(\Omega_3^{(p)}) &=&
    \sum_{\ell_1^\prime,\ell_2^\prime,\ell_3^\prime, L_2^\prime,
    n_2^\prime,  n_3^\prime}
     a^{(p),G,L}_{\ell_1,\ell_2,\ell_3, L_2 ,n_2, n_3;
    \ell_1^\prime,\ell_2^\prime,\ell_3^\prime, L_2^\prime,
    n_2^\prime,  n_3^\prime} 
    \nonumber \\
  &&\qquad  \times {}^{(4)}{\cal H}^G_{\{\ell_1',\ell_2',\ell_3', L_2' ,n_2',
    n_3'\},LM}(\Omega_3) \ .
    \label{eq:tracoK}
\eea
It is important to note that the sum over $\{\ell_1',\ell_2',\ell_3', L_2'
,n_2',n_3'\}$ is restricted by the condition
$\ell_1'+\ell_2'+\ell_3'+2(n_2'+n_3')=G$. This can be understood from the fact
that $\rho^G\times {}^{(4)}{\cal H}^G$ 
is a harmonic polynomial of degree $G$. The
transformation given in Eq.~(\ref{eq:rij}) clearly cannot change the degree of
such a polynomial and therefore ${}^{(4)}{\cal H}^G(\Omega_3^{(p)})$ can be
expressed only in terms of HH functions with the same $G$. Also the 
transformation~(\ref{eq:rij}) is equivalent to a rotation, and therefore 
$L,M$ cannot change and the TC cannot depend on $M$.

Let us consider first the function ${}^{(4)}{\cal
H}^{G+2}_{\{\ell_1,\ell_2,\ell_3, L_2 ,n_2, n_3+1\},LM}(\Omega_3^{(p)})$.
Using the expression of the Jacobi polynomials $P_{n_3+1}(z_p)$ in terms of
$P_{n_3}(z_p)$ and $P_{n_3-1}(z_p)$, one obtains
\bea
   \lefteqn{a^{(p),G+2,L}_{\ell_1,\ell_2,\ell_3, L_2 ,n_2, n_3+1; 
    \ell_1^\prime,\ell_2^\prime,\ell_3^\prime, L_2^\prime,
    n_2^\prime,  n_3^\prime} =} && \nonumber \\
   && \qquad\quad 2 b_{n_3} 
    \sum_{\ell_1^{\prime\prime},\ell_2^{\prime\prime}, 
    \ell_3^{\prime\prime} , L_2^{\prime\prime},
    n_2^{\prime\prime},  n_3^{\prime\prime} } 
   a^{(p),G,L}_{\ell_1,\ell_2,\ell_3, L_2 ,n_2, n_3;
    \ell_1^{\prime\prime},\ell_2^{\prime\prime},
    \ell_3^{\prime\prime}, L_2^{\prime\prime},
    n_2^{\prime\prime},  n_3^{\prime\prime}} 
     \sum_{i,j=1}^3 
       \Gamma^{(p)}_{i,j} I_{i,j} \ ,
    \label{eq:tracoK2P}
\eea
where
\begin{equation}
  b_{n_3} ={ {\cal N}^{\ell_1,\ell_2,\ell_3}_{ n_2, n_3+1}
                   \over {\cal N}^{\ell_1,\ell_2,\ell_3}_{ n_2, n_3} }
  \; { (2n_3+\alpha_3+\beta_3+1)(2n_3+\alpha_3+\beta_3+2)
                     \over
       2(n_3+1)(n_3+\alpha_3+\beta_3+1) } 
                 \ , \label{eq:ri2b} 
\end{equation}
and $\alpha_3=\ell_1+\ell_2+2 n_2 +2$, $\beta_3=\ell_3+1/2$. The
terms $I_{i,j}$, $i,j=1,3$ are defined by
\begin{eqnarray}
     I_{i,j}=\int d\Omega_3
   &&\left [
    {}^{(4)}{\cal H}^{G+2}_{\{\ell_1',\ell_2',\ell_3', L_2' ,n_2',
    n_3'\},LM}(\Omega_3)
    \right]^\dag
     { \jacb_i \cdot \jacb_j \over \rho^2 }\;\nonumber \\
   && \qquad\qquad \times{}^{(4)}{\cal H}^G_{\{\ell_1,\ell_2,\ell_3, L_2 ,n_2,
     n_3\},LM}(\Omega_3)
       \ ,
      \label{eq:Iij}
\end{eqnarray}
where $d\Omega_3$ is given by
\begin{equation}
  d\Omega_3= {1\over 128\sqrt{2}}d\hat \jac_1 d\hat \jac_2 d\hat \jac_3 dy d z
  (1-y)^{1\over 2} (1+y)^{1\over 2} (1-z)^{2} (1+z)^{1\over 2}
    \ , \label{eq:domega1}
\end{equation}
and $y=\cos 2\phi_2$, $z=\cos 2\phi_3$.
The integrals $I_{i,j}$ involve only functions constructed within the same set
of Jacobi vectors $\jacb_i$, $i=1$, $3$. Moreover, the factors
$(\jacb_i \cdot \jacb_j)/\rho^2$, have the  following
expressions  in terms of the hyperangular variables
$\Omega_3\equiv(\hat \jac_1, \hat \jac_2, \hat \jac_3, y,z)$
(remember that $x_3=\rho\cos\hypfi_3$, etc),
\begin{eqnarray} 
i,j&=&1,1\  , \qquad {\jac_1^2\over \rho^2}={ (1-y)(1-z)\over 4}
     \ ,\label{eq:11} \\
i,j&=&2,2\  , \qquad {\jac_2^2\over\rho^2}={(1+y)(1-z) \over 4} 
     \ ,\label{eq:22} \\
i,j&=&3,3\  , \qquad {\jac_3^2\over\rho^2}={ 1+z \over 2}
     \ ,\label{eq:33} \\
i,j&=&1,2\  , \qquad {\jacb_1 \cdot \jacb_2\over\rho^2}
          =-{4\pi\over\sqrt{3}} { (1-z) \sqrt{1-y^2} \over 4} 
           \left [ Y_1(\hat \jac_1) Y_1(\hat \jac_2) \right ]_{0,0}
          \ ,\label{eq:12} \\
i,j&=&1,3\  , \qquad {\jacb_1 \cdot \jacb_3\over\rho^2}
          =-{4 \pi\over\sqrt{3}}{ \sqrt{1-z^2} \sqrt{1-y} \over 2\sqrt{2} } 
          \left [ Y_1(\hat \jac_1) Y_1(\hat \jac_3) \right ]_{0,0}
          \ ,\label{eq:13} \\
i,j&=&2,3\  , \qquad {\jacb_2 \cdot \jacb_3\over\rho^2}
          =-{4\pi\over\sqrt{3}} { \sqrt{1-z^2} \sqrt{1+y} \over 2\sqrt{2} } 
          \left [ Y_1(\hat \jac_2) Y_1(\hat \jac_3) \right ]_{0,0}
          \ .\label{eq:23} 
\end{eqnarray}
Therefore, the integrals $I_{i,j}$ reduces to products of simple
integrals of the kind 
\begin{equation} 
  I_1[\ell',m',\ell,m,\ell'',m'']=\int d\hat x \left [
   Y_{\ell^\prime m^\prime}(\hat x) \right]^*
   Y_{\ell m}(\hat x)  Y_{\ell'' m''}(\hat x) 
   \ , \label{eq:YYY}
\end{equation}  
and
\begin{equation}
  I_2[a,b,c,d,e,f,n,m]=\int_{-1}^{+1} dx (1-x)^a (1+x)^b P^{c,d}_n(x)
     P^{e,f}_m(x) \ , \label{eq:PP}
\end{equation}
where the non-negative numerical coefficients $a$,$\ldots$,$f$ are
integers or half-integers and $ P^{c,d}_n(x)$, $P^{e,f}_m(x)$ are Jacobi
polynomials. Such integrals can be calculated analytically, or by using simple
quadrature formulas (for more details, see the appendix of Ref.~\cite{viviani:g}).

Eq.~(\ref{eq:tracoK2P}) can be used to evaluate the transformation
coefficients  of  the HH functions with grand angular momentum 
$G+2$ and $n_3>0$, once those corresponding to $G$ are known. For the
HH function with $n_3=0$ one needs another recurrence formula,
obtained from that one applied to the HH function
with $n_3=0$, $n_2+1$. By proceeding as in the previous case, one 
obtains
\begin{eqnarray}
   \lefteqn{ a^{(p),G+2,L}_{\ell_1,\ell_2,\ell_3, L_2 ,n_2+1, 0; 
    \ell_1^\prime,\ell_2^\prime,\ell_3^\prime, L_2^\prime,
    n_2^\prime,  n_3^\prime} = c^\prime_{n_2} A_{n_2}  
     a^{(p),G+2,L}_{\ell_1,\ell_2,\ell_3, L_2 ,n_2-1, 2; 
     \ell_1^\prime,\ell_2^\prime,\ell_3^\prime, L_2^\prime,
     n_2^\prime,  n_3^\prime}      } \nonumber \\
   & &
     \qquad\qquad
     +\sum_{\ell_1^{\prime\prime},\ell_2^{\prime\prime}, 
     \ell_3^{\prime\prime} , L_2^{\prime\prime},
      n_2^{\prime\prime},  n_3^{\prime\prime} } 
      a^{(p),G,L}_{\ell_1,\ell_2,\ell_3, L_2 ,n_2, 0;
      \ell_1^{\prime\prime},\ell_2^{\prime\prime},
      \ell_3^{\prime\prime}, L_2^{\prime\prime},
       n_2^{\prime\prime},  n_3^{\prime\prime}}
     \nonumber \\
  & &
     \qquad\qquad \qquad\qquad \qquad\times
       \left [ 4 b^\prime_{n_2} \Delta^{(p)}_{ij} +
        2(b^\prime_{n_2}-a^\prime_{n_2}) \Gamma^{(p)}_{ij}\right]
         I_{i,j} \ ,
    \label{eq:tracoK3P}
\end{eqnarray}
where
\begin{eqnarray}
  A_{n_2} &=&  { {\cal N}^{\ell_1,\ell_2,\ell_3}_{ n_2+1, 0}
                     \over {\cal N}^{\ell_1,\ell_2,\ell_3}_{ n_2-1, 2} }
                \;   { 8 \over (1+\alpha_3+\beta_3)(2+\alpha_3+\beta_3) }
                 \ , \label{eq:AAA}  \\
  a^\prime_{n_2} &=&{ {\cal N}^{\ell_1,\ell_2,\ell_3}_{ n_2+1, 0 }
                     \over {\cal N}^{\ell_1,\ell_2,\ell_3}_{ n_2, 0} }
    \;   { (2n_2+\alpha_2 +\beta_2 + 1)(\alpha_2^2-\beta_2^2)
                     \over
       2(n_2+1)(n_2+\alpha_2+\beta_2+1)(2n_2+\alpha_2 +\beta_2) } 
                 \ , \label{eq:ri4a}  \\
  b^\prime_{n_2} &=&{ {\cal N}^{\ell_1,\ell_2,\ell_3}_{ n_2+1, 0}
                   \over {\cal N}^{\ell_1,\ell_2,\ell_3}_{ n_2,0} }
  \;  { (2n_2+\alpha_2+\beta_2+1)(2n_2+\alpha_2+\beta_2+2)
                     \over
       2(n_2+1)(n_2+\alpha_2+\beta_2+1) } 
                 \ , \label{eq:ri4b}  \\
  c^\prime_{n_2} &=&- { {\cal N}^{\ell_1,\ell_2,\ell_3}_{ n_2+1, 0}
                     \over {\cal N}^{\ell_1,\ell_2,\ell_3}_{ n_2-1, 0} }
    \;   { (n_2+\alpha_2)(n_2+\beta_2)(2n_2+\alpha_2 +\beta_2 +2)
                     \over
       (n_2+1)(n_2+\alpha_2+\beta_2+1)(2n_2+\alpha_2 +\beta_2) } \;
       \nonumber\\
    &&\qquad\qquad\qquad\qquad
      \times  (1-\delta_{n_2,0})
                 \ , \label{eq:ri4c}  
\end{eqnarray}
with $\alpha_2=\ell_1+1/2$ and $\beta_2=\ell_2+1/2$.

In practice, the recurrence relations are used as follows. After the
calculation of the TC for the function with $(n_2= 0,n_3= 0)$,  those of the
state $(0,1)$ are obtained by Eq.~(\ref{eq:tracoK2P}) and those of the state
$(1,0)$ by Eq.~(\ref{eq:tracoK3P}). It can be noticed that, in the latter
case, the term proportional to $A_{n_2}$ does not  contribute,  as
$c^\prime_{n_2=0}=0$. Then, Eq.~(\ref{eq:tracoK2P}) is used again for
calculating the TC of the states $(0,2)$ and $(1,1)$. The coefficients of
the state $(2,0)$ can be now derived from Eq.~(\ref{eq:tracoK3P}), since
those of the state $(0,2)$, entering that expression, are already available.
The procedure can then be continued for larger  values of $(n_2,n_3)$.


\section{The Correlation Factors}\label{sec:corref}

The correlation factors in Eqs.~(\ref{eq:corre3})
and~(\ref{eq:corre4}) have the Jastrow form,  namely they are
products of one-dimensional functions. Therefore, for each channel
a few functions need to be chosen. One possibility would be to
determine them by some preliminary variational procedure, as in
Ref.~\cite{corto87} for a simplified problem (central interaction).
However, such an approach would be numerically very involved and
the following simpler procedure is preferred.

In a generic nuclear system, when all the remaining particles are far
from a given pair, the dependence of
the total wave function on the coordinates of the pair of particles is mainly
determined by their mutual interaction. Therefore, the radial 
wave function of the
relative motion of the $i$, $j$  pair in the angular-spin state
$\beta\equiv {}^{2S_\beta+1}(\ell_\beta)_{j_\beta}$, can be approximately
described by the solution of an equation of the form
\begin{equation}
       \sum_{\beta^\prime} \biggl \{
            -\htm\Bigl [ {d^2\over dr^2}+{2\over r}{d\over dr} -
                      {\ell_\beta(\ell_\beta+1) \over r^2} \Bigr ]
           \delta_{\beta\beta^\prime} +
          V_{\beta\beta^\prime}(r)+\lambda_{\beta\beta^\prime}(r)
          \biggr \}
          \phi_{\beta^\prime}(r)=0\ ,\label{eq:fr}
\end{equation}
where  $V_{\beta\beta^\prime}(r)=<\beta^\prime | V(i,j) |
\beta>$  and $V(i,j)$ is the interparticle potential.
Depending on the quantum numbers, the
state $\beta$  can be a single state or coupled to other ones. The
additional term $\lambda_{\beta\beta^\prime}(r)$ in Eq.~(\ref{eq:fr})
has the role of simulating the average effect on the pair from the other
particles.  There is a large arbitrariness
in choosing  $\lambda_{\beta\beta^\prime}(r)$, since the important
condition to be satisfied
is $|\lambda_{\beta\beta^\prime}(r)|\ll |V_{\beta\beta^\prime}(r)|$
when $r$ is small. In Ref.~\cite{rosati},  it was found that a
satisfactory choice is
\begin{equation}\label{eq:lchh}
  \lambda_{\beta\beta^\prime}(r)= \Lambda_\beta
     \exp(-c r)
      \delta_{\beta\beta^\prime} \ .
\end{equation}
The value of $1/c$ should be greater than the range of the
potential $V_{\beta\beta^\prime}(r)$, but its precise value has
been found to be unimportant (the value adopted is $1/c=2.0$ fm).
The depth $\Lambda_\beta$ is then fixed so that  $\phi_\beta(r)$
satisfy some appropriate healing condition. In this work, we have chosen
$\phi_\beta(r)\ra r^{\ell_\beta}$ when $r\ra\infty$. The functions
$\phi_\beta(r)/r^{\ell_\beta}$ have been used to construct the
appropriate correlation factor $F_\alpha$ for a given channel $\a$.

Let us consider first the $A=3$ case. The functions $f_{\a}(r)$ are
related to the reference pair $(i,j)$ characterized by definite
values of the angular momentum, spin and isospin for each channel;
therefore these functions can be taken as solutions of
Eq.~(\ref{eq:fr}). However, since the total wave function has been
constructed in the $LS$-coupling scheme, in general the total angular
momentum $j_p$ of the reference pair has not a definite value. For
the first three channels of Table~\ref{tab:chan3}, 
such a problem does not exist; in fact,  $\ell_1=0$ and, since
$J=1/2$, one has $j_p=S_{a\a}$. As a consequence, the
correlations $f_\a$, $\a=1$, $3$, correspond to the states ${}^3S_1$,
${}^1S_0$ and ${}^3D_1$, respectively.
The channels with $\a>3$ have a minor relevance to produce the
structure  of the system than the first three ones. Therefore, 
we have chosen $f_\a(r)=\phi_{^3S_1}(r)$ 
for the channels with $\ell_2=0$, $S_{a\a}=1$, $T_{a\a}=0$,
$f_\a(r)=\phi_{^1S_0}(r)$ for the channels with $\ell_2=0$,
$S_{a\a}=0$, $T_{a\a}=1$, and $f_\a(r)=\phi_{^3D_1}(r)/r^2$ for the
channels with $\ell_2=2$, $S_{a\a}=1$, $T_{a\a}=0$. Furthermore, 
for the channels with
$\ell_2=1$, we have used the solution of Eq.~(\ref{eq:fr}) by
taking into account only the central part of the pair potential in
the state ${}^3P$.
Finally, for the PHH expansion, $g_\a(r)=1$, while in the CHH case,
the functions $g_\a(r)$ have be taken equal to  $\bar\phi(r)$,
the latter function being chosen to be the solution of
Eq.~(\ref{eq:fr}) with $\ell_\beta=0$ and
\be\label{eq:ffch3b}
  V_{\beta,\beta'}(r)= {1\over 2} \Bigl( V^{(c)}_S(r)+V^{(c)}_T(r) \Bigr )
      \delta_{\beta\beta^\prime} \ .
\ee
Here $V^{(c)}_S(r)$ ($V^{(c)}_T(r)$) 
is the central part of the
nuclear potential projected on the state ${}^1S_0$ (${}^3S_1$). At large
interparticle distances,
the correlation factors go to $1$ in order to recover the HH
expansion, which is well suited for describing such configurations.

\begin{table}[t]
\caption[Table]{\label{tab:corre4} \small
Correlation functions for the 22 channels 
used for the CHH expansion of the ground state of the four-nucleon system.
The functions $f_1=\phi_{^{3}S_1}$, $f_2=\phi_{^{1}S_0}$,
$f_3=\phi_{^{3}D_1}/r^2$, and $f_4=\phi_{^{3}P_1}/r$
are the solutions of Eq.~(\ref{eq:fr}) for the
state $\beta\equiv\,^{2S+1}\!L_j$. The function $\bar\phi$ is calculated as
explained in the text.}
\begin{indented}
\lineup
\item[]
\begin{tabular}{ c  c  c c c c c  c c c  c c c  c c c c}
\br
$\alpha$  & set &$\ell_1$ & $\ell_2$ & $\ell_3$ & $\ell_{12}$ & $L$ &
  $S_a$ & $S_b$ & $S$ & $T_a$ & $T_b$ & $T$ & $f_a$ & $f_b$ 
& $f_c$ & $f_d$ \\ 
\mr
1 &$A$ &0&0&0&0&0 &1&1/2&0  &0&1/2&0 & $f_1$ & $\bar\phi$ & $\bar\phi$ & $\bar\phi$ \\
2 &$A$ &0&0&0&0&0 &0&1/2&0  &1&1/2&0 & $f_2$ & $\bar\phi$ & $\bar\phi$ & $\bar\phi$ \\
3 &$A$ &0&0&2&0&2 &1&3/2&2  &0&1/2&0 & $f_3$ & $\bar\phi$ & $\bar\phi$ & $\bar\phi$ \\
4 &$A$ &0&2&0&2&2 &1&3/2&2  &0&1/2&0 & $f_1$ & $\bar\phi$ & $\bar\phi$ & $\bar\phi$ \\
5 &$A$ &0&2&2&2&0 &1&1/2&0  &0&1/2&0 & $f_3$ & $\bar\phi$ & $\bar\phi$ & $\bar\phi$ \\
6 &$A$ &0&2&2&2&1 &1&1/2&1  &0&1/2&0 & $f_3$ & $\bar\phi$ & $\bar\phi$ & $\bar\phi$ \\
7 &$A$ &0&2&2&2&1 &1&3/2&1  &0&1/2&0 & $f_3$ & $\bar\phi$ & $\bar\phi$ & $\bar\phi$ \\
8 &$A$ &0&2&2&2&2 &1&3/2&2  &0&1/2&0 & $f_3$ & $\bar\phi$ & $\bar\phi$ & $\bar\phi$ \\
9 &$B$ &2&0&2&2&0 &1& 1 &0  &0& 0 &0 & $f_3$ & $\bar\phi$ & $\bar\phi$ & $f_3$ \\
10&$B$ &2&0&2&2&1 &1& 1 &1  &0& 0 &0 & $f_3$ & $\bar\phi$ & $\bar\phi$ & $f_3$ \\
11&$B$ &2&0&2&2&2 &1& 1 &2  &0& 0 &0 & $f_3$ & $\bar\phi$ & $\bar\phi$ & $f_3$ \\
12&$A$ &0&1&1&1&0 &1&1/2&0  &1&1/2&0 & $f_4$ & $\bar\phi$ & $\bar\phi$ & $\bar\phi$ \\
13&$A$ &0&1&1&1&1 &1&1/2&1  &1&1/2&0 & $f_4$ & $\bar\phi$ & $\bar\phi$ & $\bar\phi$ \\
14&$A$ &0&1&1&1&1 &1&3/2&1  &1&1/2&0 & $f_4$ & $\bar\phi$ & $\bar\phi$ & $\bar\phi$ \\
15&$A$ &0&1&1&1&2 &1&3/2&2  &1&1/2&0 & $f_4$ & $\bar\phi$ & $\bar\phi$ & $\bar\phi$ \\
16&$A$ &1&1&0&0&0 &1&1/2&0  &0&1/2&0 & $f_1$ & $\bar\phi$ & $\bar\phi$ & $\bar\phi$ \\
17&$A$ &1&1&0&1&1 &1&1/2&1  &0&1/2&0 & $f_1$ & $\bar\phi$ & $\bar\phi$ & $\bar\phi$ \\
18&$A$ &1&1&0&1&1 &1&3/2&1  &0&1/2&0 & $f_1$ & $\bar\phi$ & $\bar\phi$ & $\bar\phi$ \\
19&$A$ &1&1&0&2&2 &1&3/2&2  &0&1/2&0 & $f_1$ & $\bar\phi$ & $\bar\phi$ & $\bar\phi$ \\
20&$B$ &0&0&0&0&0 &1& 1 &0  &0& 0 &0 & $f_1$ & $\bar\phi$ & $\bar\phi$ & $f_1$ \\
21&$B$ &0&0&2&0&2 &1& 1 &2  &0& 0 &0 & $f_3$ & $\bar\phi$ & $\bar\phi$ & $f_1$ \\
22&$B$ &2&0&0&2&2 &1& 1 &2  &0& 0 &0 & $f_1$ & $\bar\phi$ & $\bar\phi$ & $f_3$ \\
\br
\end{tabular}
\end{indented}
\end{table}

Let us now consider the $A=4$ case. 
The four-body correlation factors are given in term of 
the functions $f_\a$, $g_\a$ and $h_\a$ in Eq.~(\ref{eq:corre4}).
The functions $f_{\a}(r_{ij})$ related to the reference pair $(i,j)$ have been
determined by using the same criteria as for the $A=3$ case.
The functions $g_{\a}(r_{ik})$ and
$h_{\a}(r_{km})$ with $k$, $m$ different from the reference pair
indices $i$, $j$ (see Eq.~(\ref{eq:corre4}))
correlate pairs which are not in a definite angular-spin-isospin
state. For simplicity reasons, 
for the channels constructed with the set $A$ of the Jacobi
vectors ($\a=1-8$ and $12-19$, as listed in
Table~\ref{tab:chan4}), we have chosen $g_{\a}=h_{\a}=\bar\phi$,
where  the function $\bar\phi$ has been calculated as in the $A=3$ case,
namely using the potential given in Eq.~(\ref{eq:ffch3b}).
Other choices for these functions have been proved 
not to influence appreciably the final result.  

The correlation functions of the channels constructed with the set $B$
of the Jacobi vectors ($\a=9-11$ and $20-22$) have been
chosen in a similar way. In this case, the functions correlating
pairs with definite angular-spin-isospin quantum numbers are
$f_{\a}$ and $h_{\a}$, and they have been chosen to be
$\phi_{^3S_1}$ or $\phi_{^3D_1}$ depending on the value of $\ell_3$
and $\ell_1$, respectively. The functions $g_{\a}$ have been chosen
to be $\bar\phi$, as in the other channels. A summary of the
correlation factors used in the various channels is given in
Table~\ref{tab:corre4}.


\section{The Matrix Elements of the Interaction in the HH Expansion}
\label{sec:mehh}

In this appendix, a more detailed discussion how to compute the matrix
elements of the interaction between the basis functions $\Xi^{GTJ\pi}_{\nu}
f_l(\rho)$ 
of Eq.~(\ref{eq:PSI3jj}) is reported. The states $\Xi^{GTJ\pi}_{\nu}$ are 
constructed in terms of the Jacobi vectors defined with permutation $p=1$
and the standard choice of Jacobi vectors given in Eq.~(\ref{eq:jac1}).
In this case, the Jacobi vectors are specified in Eq.~(\ref{eq:jac1}), namely
\be
   \jacb_N=\r_2-\r_1\ ,\quad \jacb_{N-1}=\sqrt{4\over 3}
   \left( \r_3-{\r_1+\r_2\over 2}\right)\ ,\ \ldots\ .\label{eq:appd1}
\ee
Let us remember the general expression of the $jj$-coupling
spin-isospin-HH states $\Xi^{GTJ\pi}_{\nu}$:
\bea
  \Xi^{GTJ\pi}_{\nu} &=& \biggl\{ \left [ \Bigl ( Y_{\ell_N}(\hat x_N) {S_2}
              \Bigr)_{j_N} 
           \Bigl ( Y_{\ell_{N-1}}(\hat x_{N-1}) s_3 \Bigr)_{j_{N-1}}
           \right]_{J_{N-1}} \cdots
           \Bigl ( Y_{\ell_1}(\hat x_1) s_A \Bigr)_{j_1}\biggr\}_{JJ_z} 
      \nonumber\\
  &&\times  \left\{\Bigl[ (T_2 t_3)_{T_3}t_4\Bigr]_{T_4}\cdots t_A\right\}_{TT_z}\ ,
   \label{eq:PHIjjA}
\eea
where $S_2$ ($T_2$) is the spin  (isospin) state of the particles 1 and 2,
etc. Moreover, let us use the short-cut
$\{T\}\equiv\{T_2,T_3,\ldots,T_{A-1},T_A\}$, where
$T_i$ is the total isospin of particles $1,\ldots,i$ and
$T_A\equiv T$.
Let us consider first the matrix elements of the NN interaction, which
in general is assumed to be non-local and non-isospin conserving. In
configuration space, the matrix element in Eq.~(\ref{eq:v12}) is given by 
\bea
  V^{J\pi}_{G'T'\nu',GT\nu}&=&
  <\Xi^{G'T'J\pi}_{\nu'} |v(1,2)|\Xi^{GTJ\pi}_{\nu} >
   \nonumber\\
  &=& \int d^3\jacb_N d^3\jacb_N' d^3\jacb_{N-1}\cdots
   d^3\jacb_1\;
  \left[\Xi^{G'T'J\pi}_{\nu'}(\Omega_N')\right]^\dag f_{l'}(\rho')  
   \nonumber\\
  &&\qquad  \times V(\jacb_N',\jacb_N)  
   \Xi^{GTJ\pi}_{\nu}(\Omega_N) f_l(\rho)\ , \label{eq:appd2}
\eea
where the dependence of $V$ on the spin and isospin operators of particles 1
and 2 is understood. Above, $\Omega_N$ denotes the hyperangular variables
constructed from $\{\jacb_1,\ldots,\jacb_{N-1},\jacb_N\}$ and
$\Omega_N'$ those
constructed from $\{\jacb_1,\ldots,\jacb_{N-1},\jacb_N'\}$.
Moreover $\rho=\sqrt{\jac_1+\cdots +\jac_N^2}$ and
$\rho'=\sqrt{\jac_1+\cdots +\jac_{N-1}^2+(\jac_N')^2}$.
Note that $V$ cannot change the total angular momentum $j_N$ of particles
$1,2$. Let us suppose to know
\bea
 V^{j_N,\{T'\},\{T\}}_{\ell_N'S_2',\ell_NS_2}(\jac_N',\jac_N)&=&
  \int d\hat\jac_N' d\hat\jac_N  
   \Bigl(Y_{\ell_N'}(\hat x_N') {S_2'} \Bigr)_{j_N}^\dag
   \left\{\Bigl[ (T_2' t_3)_{T_3'}t_4\Bigr]_{T_4'}\cdots t_A\right\}^\dag_{T'T_z}
   \nonumber \\     
   && \times V(\jacb_N',\jacb_N)  
   \nonumber \\     
   && \times \Bigl(Y_{\ell_N}(\hat x_N) {S_2} \Bigr)_{j_N}
   \left\{\Bigl[ (T_2 t_3)_{T_3}t_4\Bigr]_{T_4}\cdots t_A\right\}_{TT_z}  \ ,
   \label{eq:vnnpro}
\eea
which is usually quite easy to obtain.

The integration over $d^3\jacb_{N-1}\cdots d^3\jacb_1$ is
easily performed applying the properties of the HH function. First of all, the
hyperspherical angles $\hypfi_2,\ldots,\hypfi_{N-1}$ do not depend on 
$\jac_N$, and therefore are the same for both $\Omega_N$ and $\Omega_N'$.
Second, it results that 
\be
  d^3\jacb_{N-1}\cdots d^3\jacb_1 = 
   (\rho_{N-1})^{D-4} d\rho_{N-1} d\Omega_{N-1}\ ,
  \label{eq:rhon1}
\ee
where $\Omega_{N-1}\equiv\{\hat\jac_1,\ldots,\hat\jac_{N-1},\hypfi_2,\ldots,
\hypfi_{N-1}\}$ are the hyperspherical coordinates related
to the Jacobi vectors $\{\jacb_{1},\ldots,\jacb_{N-1}\}$ and
\be
    \rho_{N-1}=\sqrt{\jac_1^2+\cdots+\jac_{N-1}^2}\ .
\ee
Third, $\jacb_N$ ($\jacb_N'$) depends only on $\rho$ and $\hypfi_N$
($\rho'$ and $\hypfi_N'$) and hence $V$ does not depend on the $\Omega_{N-1}$
variables. Therefore, we can integrate over $d\Omega_{N-1}$ 
and use the orthonormality relation of Eq.~(\ref{eq:nohh}). Also the traces over the
spin variables of particles $3,\ldots,A$ can be easily computed,
and the final results is:
\bea
   V^{J\pi}_{G'T'\nu',GT\nu}&=&
   \int \jac_N^2 d\jac_N \; (\jac_N')^2d\jac_N'\; 
    (\rho_{N-1})^{D-4} d\rho_{N-1}\;
   \nonumber \\     
   &\times& {}^{(N)}{\cal P}^{G_{N-1},\ell_N'}_{n_N'}(\hypfi_N') f_{l'}(\rho')    
     V^{j_N,\{T'\},\{T\}}_{\ell_N'S_2',\ell_NS_2}(\jac_N',\jac_N)
   \nonumber \\     
   &\times&  {}^{(N)}{\cal P}^{G_{N-1},\ell_N}_{n_N}(\hypfi_N) f_l(\rho) 
    \nonumber\\
   &\times& \delta_{j_N,j_N'}\cdots\delta_{j_1,j_1'}
      \delta_{J_{N-1},J_{N-1}'}\cdots\delta_{J_2,J_2'}
   \nonumber\\
   &\times&   \delta_{\ell_{N-1},\ell_{N-1}'}\cdots\delta_{\ell_1,\ell_1'}
        \delta_{n_{N-1},n_{N-1}'}\cdots\delta_{n_2,n_2'}\ ,
\label{eq:iii}
\eea
where 
\bea
  &&\rho^2=\jac_N^2+\rho_{N-1}^2\ ,\quad (\rho')^2=(\jac_N')^2+\rho_{N-1}^2\ ,
   \nonumber \\     
   &&\cos\hypfi_N=\jac_N/\rho 
  \ , \quad \cos\hypfi_N'=\jac_N'/\rho'\ ,
\eea
and $G_{N-1}$ is given by Eq.~(\ref{eq:go}).
Finally, the three-dimensional integrations involved in Eq.~(\ref{eq:iii})
can be accurately performed with standard numerical techniques
(Gauss integration)~\cite{abra}. For local potentials, the
functions $V(\jac_N',\jac_N)$ reduces to
$V(\jac_N)\delta(\jac_N'-\jac_N)/\jac_N^2$.  

For the 3N interaction $W\equiv W(\jacb_N',\jacb_{N-1}',\jacb_N,\jacb_{N-1})$,
the first step is to know the projection
over three-body angular-spin states, namely 
\bea
  \lefteqn{W^{J_{N-1},\{T'\},\{T\}}_{\ell_N'S_2'j_N'\ell_{N-1}'j_{N-1}',
   \ell_N S_2 j_N \ell_{N-1}
   j_{N-1}}(\jac_N',\jac_{N-1}',\jac_N,\jac_{N-1})=}&&
   \nonumber \\     
  && \int d\hat\jac_N' d\hat\jac_N  d\hat\jac_{N-1}' d\hat\jac_{N-1}  
           \biggl[\Bigl(Y_{\ell_N'}(\hat x_N') {S_2'} \Bigr)_{j_N'}
           \Bigl ( Y_{\ell_{N-1}'}(\hat x_{N-1}') s_3 \Bigr)_{j_{N-1}'}
           \biggr]_{J_{N-1}}^\dag
   \nonumber \\     
   &\times&  
   \left\{\Bigl[ (T_2' t_3)_{T_3'}t_4\Bigr]_{T_4'}\cdots t_A\right\}^\dag_{T'T_z}
   W(\jacb_N',\jacb_{N-1}',\jacb_N,\jacb_{N-1})  
   \nonumber \\     
   &\times&    \biggl[\Bigl(Y_{\ell_N}(\hat x_N) {S_2} \Bigr)_{j_N}
           \Bigl ( Y_{\ell_{N-1}}(\hat x_{N-1}) s_3 \Bigr)_{j_{N-1}}
           \biggr]_{J_{N-1}} \nonumber \\
   &\times&
   \left\{\Bigl[ (T_2 t_3)_{T_3}t_4\Bigr]_{T_4}\cdots t_A\right\}_{TT_z}  \ ,
   \label{eq:v3npro}
\eea
Then
\bea
   \lefteqn{ <\Xi^{G'T'J\pi}_{\nu'} |W|\Xi^{GTJ\pi}_{\nu} >=}&&
   \nonumber \\     
  && \int \jac_N^2 d\jac_N \; \jac_{N-1}^2d\jac_{N-1}\; 
       (\jac_N')^2 d\jac_N' \; (\jac_{N-1}')^2d\jac_{N-1}'\; 
    (\rho_{N-2})^{D-7} d\rho_{N-2}\;
   \nonumber \\     
   &\times&  {}^{(N)}{\cal P}^{G_{N-1}',\ell_N'}_{n_N'}(\hypfi_N') 
    {}^{(N-1)}{\cal P}^{G_{N-2},\ell_{N-1}'}_{n_{N-1}'}(\hypfi_{N-1}')
     f_{l'}(\rho')    
   \nonumber \\     
   &\times& W^{J_{N-1},\{T'\},\{T\}}_{\ell_N'S_2'j_N'\ell_{N-1}'j_{N-1}',
   \ell_N S_2 j_N \ell_{N-1} j_{N-1}}(\jac_N',\jac_{N-1}',\jac_N,\jac_{N-1})
   \nonumber \\     
   &\times& {}^{(N)}{\cal P}^{G_{N-1},\ell_N}_{n_N}(\hypfi_N) 
    {}^{(N-1)}{\cal P}^{G_{N-2},\ell_{N-1}}_{n_{N-1}}(\hypfi_{N-1}) f_l(\rho) 
    \nonumber\\
   &\times& \delta_{j_{N-2},j_{N-2}'}\cdots\delta_{j_1,j_1'}
      \delta_{J_{N-1},J_{N-1}'}\cdots\delta_{J_2,J_2'}
   \nonumber\\
    &\times&  \delta_{\ell_{N-2},\ell_{N-2}'}\cdots\delta_{\ell_1,\ell_1'}
       \delta_{n_{N-2},n_{N-2}'}\cdots\delta_{n_2,n_2'}\ ,
\eea
where 
\begin{eqnarray}
         &&\rho^2=\jac_N^2+\jac_{N-1}^2+\rho_{N-2}^2\ ,
    \quad \cos\hypfi_N=\jac_N/\rho \ , \nonumber \\
  && \cos\hypfi_{N-1}=\jac_{N-1}/\sqrt{\jac_{N-1}^2+\rho_{N-2}^2} \ ,
\end{eqnarray}
and analogously for the primed quantities. Moreover,
\begin{eqnarray}
G_{N-2}&=&\sum_{i=1}^{N-2}(\ell_i+2n_i)\ , \nonumber \\
G_{N-1}&=&\ell_{N-1}+2n_{N-1}+G_{N-2}\ , \nonumber \\
G_{N-1}'&=&\ell_{N-1}'+2n_{N-1}'+G_{N-2}\ . 
\label{eq:ggn}
\end{eqnarray}

For momentum-space potentials, similar results are obtained, with 
the only change $\jacb_i\rightarrow\qb_i$, $\rho\rightarrow Q$, 
$f_l(\rho)\ra g_{l,G}(Q)$.

\section*{Acknowledgments}
This work was partially supported by the Italian {\it Ministero 
dell'Universit\`a e della Ricerca}.

\end{document}